%% file: mainV2_combined.tex
\documentclass[a4paper,11pt]{article}
\pdfoutput=1 % if you are submitting a pdflatex (i.e. if you have
\usepackage{jheppub} % for details on the use of the package, please
\usepackage{booktabs,makecell, bm}
\usepackage[font=small]{caption}
\usepackage{framed}
\usepackage{hyperref}
\usepackage[export]{adjustbox}
\usepackage{amsmath}
\usepackage{lineno}
\usepackage{ytableau}
\usepackage{subfig}
\usepackage{graphicx}
\usepackage{tikz}
\usetikzlibrary{arrows.meta,calc}
\usepackage{url}
\usepackage[numbers]{natbib}
\usepackage[colorinlistoftodos]{todonotes}

\usepackage{booktabs}
\usepackage{float}

\usepackage{xcolor}

\renewcommand{\[}{\begin{equation}}
\renewcommand{\]}{\end{equation}}
\renewcommand\arraystretch{1.25}
\makeatother

\global\long\global\long\def\Tr{\mbox{Tr}}

\def\bea{\begin{eqnarray}}
\def\eea{\end{eqnarray}}
\def\be{\begin{equation}}
\def\ee{\end{equation}}
\def\ea{\end{align}}
\def\bse{\begin{subequations}}
\def\ese{\end{subequations}}
\def\1F1{{}_1\!F_1}
\def\2F0{{}_2\!F_0}

\title{Matrix Theory from Holography}

\author[a]{Shota Komatsu,}
\author[b]{Eunwoo Lee,}
\author[c]{Chintan Patel}
\affiliation[a]{Abdus Salam Centre for Theoretical Physics, Imperial College London,\\ SW7 2AZ London, United Kingdom}
\affiliation[b]{Department of Theoretical Physics,  Tata Institute
	of Fundamental Research,\\ Homi Bhabha Rd, Mumbai 400005, India}
\affiliation[c]{Jadwin Hall, Princeton University,\\
Princeton, NJ 08540, U.S.A.}
\emailAdd{sk2925@ic.ac.uk}
\emailAdd{eunwoo.lee@tifr.res.in}
\emailAdd{cp3849@princeton.edu}
\abstract{We propose a setup that embeds Matrix Theory for M-theory into the
AdS/CFT correspondence, allowing the former to be tested using the latter. The central claim is  a triality among the BMN matrix model of rank $m$, a large-charge
monopole sector of ABJM theory with charge $J$, and M-theory on the maximally supersymmetric
eleven-dimensional pp-wave with a compact lightlike direction.  The relevant
ABJM limit sends $N,k,J\to\infty$ while keeping $N/k^2$ and
$m=J/k$ fixed. 
We provide quantitative evidence for this triality by showing that, in this triple-scaling limit, the ABJM superconformal index agrees precisely with a grand-canonical sum of Witten indices of the $U(m)$ BMN matrix model. This relation also yields a compact grand-canonical expression for the BMN index, which we use to revisit its large-$m$ behavior. Our analytic and numerical results reveal strong boson--fermion cancellations and show no evidence for the previously claimed $e^{O(m^2)}$ growth when the BMN charge scales as $Q \sim m^2$. We further study the correspondence between the BPS cohomologies of the two theories by constructing a letter-level dictionary. This leads to a new notion of BPS fortuity in ABJM theory that survives even at strict $N=\infty$: although the usual fortuity associated with finite-$N$ trace relations disappears in the triple-scaling limit, each fixed-monopole sector is governed by the effective rank $m$, and finite-$m$ trace relations generate fortuitous states that map directly to fortuitous states of the BMN matrix model. 
Finally, we discuss connections to related concepts such as large-charge matrix
models, open--closed--open triality, and worldline holography.
}

\begin{document} 

\maketitle
%\flushbottom
%%%%%%%%%%%%%%%%%%%%%%%%%%%%%%%%%%%%%%%%%

\newpage

\section{Introduction}\label{intro}
Understanding quantum gravity is one of the central goals of modern theoretical physics. Over the past
three decades, the most concrete progress in this direction has come from
holography, and in particular from the AdS/CFT correspondence \cite{Maldacena:1997re, Gubser:1998bc,Witten:1998qj}. In its standard
form, AdS/CFT asserts an equivalence between quantum gravity, or string theory,
in an asymptotically Anti-de Sitter (AdS) spacetime and a large-$N$ conformal field
theory (CFT) in one lower dimension. Bulk scattering experiments are then encoded in
the CFT as correlation functions of local operators, corresponding to small
perturbations of the large-$N$ CFT.

There is, however, another route to non-perturbative string theory and quantum
gravity, one that predates AdS/CFT but has received comparatively less
attention: the circle of ideas usually referred to as {\it Matrix Theory
conjectures} \cite{Banks:1996vh,Banks:1996my,Ishibashi:1996xs,Susskind:1997cw,Dijkgraaf:1997vv,Motl:1997th,Sen:1997we, Seiberg:1997ad}. Like AdS/CFT, these conjectures propose that certain large-$N$
quantum mechanical systems or gauge theories provide a non-perturbative
definition of string theory, M-theory, and quantum gravity. The nature of the
correspondence is, however, quite different.
\begin{enumerate}
    \item
    \textit{The asymptotic structure of spacetime is different.}
    Standard AdS/CFT defines quantum gravity in asymptotically AdS spacetime.
    By contrast, the Banks-Fischler-Shenker-Susskind (BFSS) conjecture \cite{Banks:1996vh,Susskind:1997cw}, the prototypical example of Matrix
    Theory, concerns eleven-dimensional M-theory in asymptotically flat space,
    or its compactification on a lightlike circle, known as Discrete
    Light-Cone Quantization (DLCQ).

    \item
    \textit{The degrees of freedom are localized differently.}
    In AdS/CFT, the degrees of freedom of the dual CFT live on the conformal
    boundary of AdS. In Matrix Theory, by contrast, the matrix degrees of
    freedom behave less like a boundary theory and more like a quantum
    mechanical description of particles moving in spacetime.
\end{enumerate}
These differences are precisely what make Matrix Theory valuable. It offers a
possible formulation of quantum gravity beyond asymptotically AdS spacetime,
and without relying on the existence of a conformal boundary. In particular,
the second feature is reminiscent of recent approaches to de Sitter holography,
such as static-patch holography and worldline holography, which seek to
describe de Sitter quantum gravity in terms of the quantum mechanics of an
observer in a static patch \cite{Anninos:2011af, Chandrasekaran:2022cip, Witten:2023xze}. Matrix Theory may provide a cleaner and more
controlled example of this kind of description, and may therefore
offer useful lessons for quantum gravity in spacetimes without boundaries.

At the same time, Matrix Theory differs from AdS/CFT in several technical
respects that make the conjectures substantially harder to establish.
\begin{enumerate}
\setcounter{enumi}{2}
\item
    \textit{Scattering is encoded differently.}
    In AdS/CFT, a small number of bulk particles is represented by a small
    perturbation of the boundary CFT, and bulk scattering amplitudes are
    extracted from CFT correlation functions. In Matrix Theory, by contrast,
    multi-particle states correspond to approximately block-diagonal matrix
    configurations, with each block representing an individual particle.
    Scattering is then described by interactions among these blocks, and
    therefore involves the full matrix dynamics rather than a small perturbation
    around the vacuum.

    \item
    \textit{The matrix rank plays a different role.}
    In familiar examples of AdS/CFT, the rank $N$ of the gauge group controls
    the bulk gravitational coupling, and the large-$N$ limit gives perturbative
    or semiclassical gravity. In Matrix Theory, by contrast, the matrix rank
    $m$ counts units of discretized light-cone momentum. At finite $m$, the
    theory describes the DLCQ sector with $m$ units of light-cone momentum.
    Recovering the uncompactified theory requires a strict $m\to\infty$ limit.
    Thus the matrix rank is a kinematic quantum number, not the inverse
    gravitational coupling.

    \item
    \textit{The relevant dynamical regimes are different.}
    In AdS/CFT, the standard 't~Hooft limit organizes gauge-theory diagrams by
    topology and makes the emergence of perturbative string worldsheets
    manifest \cite{tHooft:1973alw}. Matrix Theory, by contrast, generally requires
    an extreme strong-coupling regime beyond the usual 't~Hooft limit.
\end{enumerate}
These technical differences, especially the strong-coupling regime required by
the conjecture, create significant obstacles to testing and establishing Matrix
Theory conjectures. Moreover, unlike AdS/CFT, which is supported by controlled
brane-decoupling limits, a detailed operator dictionary, and many quantitative
tests, Matrix Theory conjectures are motivated by arguments \cite{Banks:1996vh,Susskind:1997cw,Seiberg:1997ad,Sen:1997we} that are compelling but less
sharp. Although there has recently been renewed interest, including studies of
BFSS scattering amplitudes \cite{Herderschee:2023pza, Miller:2022fvc,Asano:2026udd,Guevara:2026gqw}, matrix string theory \cite{Cho:2026jzs, Bajaj:2026oja}, finite-$N$ aspects of matrix models \cite{Komatsu:2024vnb}, the analysis of topologically-twisted theories \cite{Hahner:2026mfg}, a re-examination from the viewpoint of BPS decoupling limits \cite{Blair:2023noj,Blair:2024aqz}, the analysis of mass deformation by supersymmetric localization \cite{Asano:2022brd,Asano:2026hto}, and numerical bootstraps \cite{Lin:2023owt,Lin:2025srf},\footnote{See \cite{Lin:2025iir} and references therein for recent progress.} further progress is needed to put the
Matrix Theory conjectures on firmer footing.

\bigskip

The purpose of this paper is to propose a concrete setup that connects AdS/CFT to Matrix Theory and thereby provides a controlled way to study Matrix Theory conjectures. 
The central object is the Berenstein-Maldacena-Nastase (BMN) matrix model \cite{Berenstein:2002jq,Dasgupta:2002hx,Dasgupta:2002ru}, a supersymmetric mass deformation of the BFSS matrix quantum mechanics. 
A version of the Matrix Theory conjecture relates the $U(m)$ BMN matrix model to M-theory on the DLCQ of an eleven-dimensional plane-wave background, in a sector with $m$ units of momentum along the compact lightlike direction. 

\begin{figure}[t]
\centering
\resizebox{\linewidth}{!}{%
\begin{tikzpicture}[
    font=\small,
    >={Latex[length=2.3mm,width=1.6mm]},
    box/.style={
        draw,
        line width=0.75pt,
        rounded corners=5pt,
        align=center,
        inner xsep=10pt,
        inner ysep=9pt,
        outer sep=0pt,
        minimum height=1.25cm
    },
    relation/.style={
        <->,
        line width=0.8pt,
        shorten <=3pt,
        shorten >=3pt
    },
    label/.style={
        font=\small,
        inner sep=0pt
    }
]

% Nodes: use explicit line breaks, without text-width restrictions.
\node[box,draw=red,minimum width=7.2cm] (gravity) at (0,3.2)
    {\bf DLCQ of M-theory on the pp-wave\\[3pt] \bf
     with $m$ units of light-cone momentum};

\node[box,draw=blue,minimum width=4.2cm] (abjm) at (-4.6,0)
    {\bf Large-charge monopole sector\\ \bf  in triple-scaled ABJM};

\node[box,draw=green!60!black,minimum width=4.8cm] (bmn) at (4.6,0)
    {\bf $U(m)$ BMN matrix model\\[3pt]
     \bf with coupling $g_{\rm BMN}$};

% Attachment points on the lower edge of the gravity box.
% These make the diagonal arrows slope inward toward the top box.
\coordinate (gravityLeft)
    at ($(gravity.south)+(-2.35cm,0)$);
\coordinate (gravityRight)
    at ($(gravity.south)+(2.35cm,0)$);

% Diagonal relations, with labels outside the triangle.
\draw[relation]
    (abjm.north) --
    node[label,midway,above left=5pt] {\bf AdS$_4$/CFT$_3$}
    (gravityLeft);

\draw[relation]
    (bmn.north) --
    node[label,midway,above right=5pt] {\bf Matrix Theory conjecture}
    (gravityRight);

% Horizontal relation.
\draw[relation]
    (abjm.east) --
    node[label,midway,below=6pt] {\huge \bf ?}
    (bmn.west);

% Triple-scaling limit, centred beneath the ABJM box.
\node[align=center,anchor=north,inner sep=0pt]
    at ($(abjm.south)+(0,-0.25cm)$)
    {\footnotesize $ N,k,J\to\infty$\\[3pt]
     $g_{\rm BMN}^2\propto N/k^2$ fixed,\quad $m=J/k$ fixed};

\end{tikzpicture}%
}
\caption{The proposed BMN triality relating the large-charge monopole sector of ABJM theory in the triple-scaling limit, the $U(m)$ BMN matrix model, and DLCQ M-theory on the maximally supersymmetric eleven-dimensional pp-wave. In this paper, we  test the predicted field-theory duality between the triple-scaling limit of ABJM and the BMN matrix model. Together with the AdS$_4$/CFT$_3$ correspondence and its pp-wave limit, this provides an indirect test of the Matrix Theory conjecture for the BMN model.}
\label{fig:BMNtriality}
\end{figure}
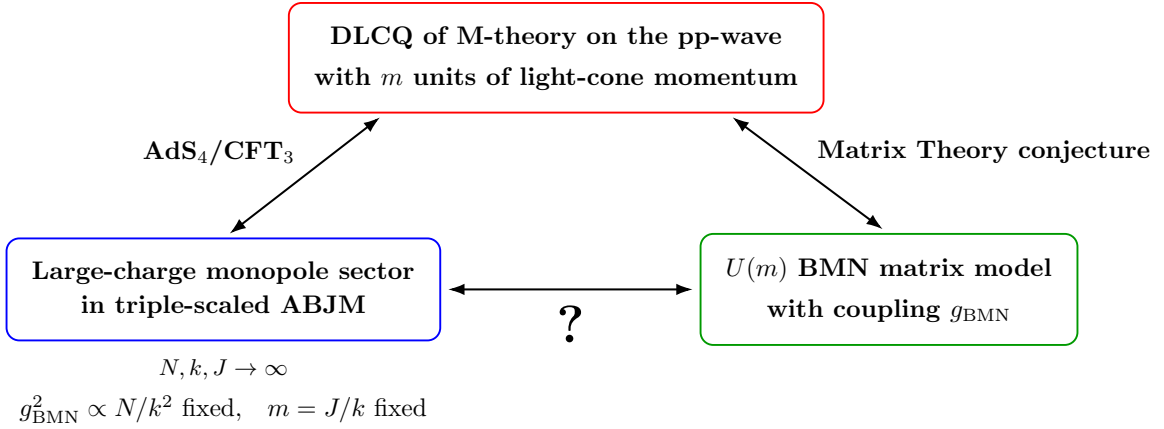

The key observation is that the same background, namely lightlike compactified
M-theory on the eleven-dimensional pp-wave, also arises as a limit of
$AdS_4\times S^7/\mathbb Z_k$ \cite{Kowalski-Glikman:1984qtj,Blau:2002dy}, whose holographic dual is ABJM
theory \cite{Aharony:2008ug}.
%\SK{Moved some parts to after eq(1.2)}
%Monopole sectors and large-charge states in ABJM theory have been studied from both field-theory and bulk perspectives \cite{Nishioka:2008gz,Gaiotto:2008cg,Kim:2009ia,Berenstein:2009sa,
%Sheikh-Jabbari:2009vjj,Kovacs:2013una}. In particular, it has been shown that the half-BPS sector of ABJM theory already exhibits the partition structure associated with the vacua of plane-wave
%matrix theory \cite{Sheikh-Jabbari:2009vjj}. 
We determine the precise limit that takes $AdS_4\times S^7/\mathbb Z_k$ to the
DLCQ pp-wave background and translate this limit carefully to the ABJM side.
This leads to a conjectural triality (see Figure \ref{fig:BMNtriality}), to be called {\it BMN triality}, among
\begin{enumerate}
    \item $U(m)$ BMN matrix model with coupling constant $g_{\rm BMN}$.
    \item Large-charge  monopole sector of $U(N)_k\times U(N)_{-k}$ ABJM theory with a global charge\footnote{The choice of the global charge $J$ is determined by the Penrose limit in the bulk. See  Sections \ref{sec:triality} and \ref{abjm bmn main} for details. 
    } $J$, 
    in a triple-scaling limit with fixed ratios among $J$, $k$, and $\sqrt N$.
    %\textbf{EL: I think we need to impose $J_1/J\to 1$ as well.}
    \item M-theory on the eleven-dimensional pp-wave background compactified along a lightlike circle of radius $R_{-}$, in a sector with lightcone momentum $-p_{-}=\frac{m}{R_{-}}$.
\end{enumerate}
The parameters of the models are related by the following dictionary,
\begin{align}
\begin{aligned}
&m=\frac{J}{k}=-R_{-}p_{-}\,,\qquad g_{\rm BMN}^2=\frac{4\pi^2}{27} \frac{N}{k^2}= \frac{R_{-}^3}{\ell_{P}^{6}\mu^3}\,,
&
\end{aligned}
\label{parameter-dictionary}
\end{align}
where $\mu$ is the strength of the four-form flux of the pp-wave background and $\ell_{P}$ is the eleven-dimensional Planck length.

Two features of \eqref{parameter-dictionary} are particularly striking. First, the rank $m$ of the BMN matrix model is determined by the charge $J$ on the ABJM side, rather than by the gauge-group rank $N$. A similar interchange between charge and matrix size appears in dual matrix-model descriptions of large-charge sectors of superconformal field theories \cite{Giombi:2018qox, Grassi:2019txd}. In those examples, however, the matrix model emerges merely as a technical rewriting of correlation functions obtained from supersymmetric localization. A closer physical analogy is provided instead by open--closed--open triality \cite{Gopakumar:2022djw,Gopakumar:2024jfq}, which has been applied to correlation functions involving determinant operators in $\mathcal N=4$ supersymmetric Yang--Mills theory \cite{Jiang:2019xdz,Jiang:2019zig,Chen:2019gsb,Budzik:2021fyh,Caron-Huot:2023wdh,Mazenc:2026yrm,Konosu:2026xxy}. There, the rank of the dual matrix model is likewise controlled by the charge of the operator insertions, but the matrix model admits a direct interpretation in terms of D-branes and open strings. As discussed in Section~\ref{oco}, our construction shares several features with this framework, suggesting that the relation between the large-charge sector of ABJM theory and the BMN matrix model may be viewed as an ``open--open'' duality. Second,  the semiclassical gravity limit, $\ell_P\to0$, corresponds to the strong-coupling limit $g_{\rm BMN}^2\to\infty$, rather than to a large-rank limit $m\to \infty$ as in the AdS/CFT correspondence. This feature has been observed already in the literature both in BFSS and BMN models \cite{Banks:1997it,Seiberg:1997ad,Komatsu:2024vnb}, but the present dictionary makes it manifest and, at the same time, explains how it is compatible with the usual AdS/CFT relation between large rank and semi-classical gravity.

\medskip

Our main quantitative test for the triality is a matching of supersymmetric indices.  After
dividing out the zero-monopole sector\footnote{Note that the formula \eqref{goal} shows that the total ABJM index factorizes into the index with zero KK momentum along the M-theory circle ($I_{\rm ABJM}^{0}$) and the sum of BMN indices. This factorization indicates a decoupling between bulk closed-string modes and open-string modes on the D0-brane worldvolume in the triple-scaling limit.}, the positive-charge part of the
large-$N$ ABJM index becomes, %at leading order 
in the triple-scaling limit\footnote{Away from the triple-scaling limit, there are corrections to the formula coming from finite-$k$, finite-$N$,
and subdominant-residue corrections. For details, see Section \ref{abjm bmn main}.},
\begin{equation}
  \frac{I_{\rm ABJM}(x,y,z,w)}{I^0_{\rm ABJM}(x,y,z,w)}
  \longrightarrow
  \sum_{m=0}^{\infty}q^{m}
  I_{{\rm BMN},U(m)}(a,b,c),
  \label{goal}
\end{equation}
where the fugacity map, defined in the main text, is $q=x^k$, $a=y/x$, $b=xz$, and $c=xw$, with $q$, $a$, $b$, and $c$ held finite. When the fugacities $a,b,c$ are turned off, \eqref{goal} reduces to the known
relation between large-charge BPS monopoles in ABJM theory and vacua of
the BMN matrix model. Monopole sectors and large-charge states in ABJM theory
have been studied from both field-theory and bulk perspectives
\cite{Nishioka:2008gz,Gaiotto:2008cg,Kim:2009wb,Kim:2009ia,Berenstein:2009sa,
Sheikh-Jabbari:2009vjj,Kovacs:2013una}. In particular, the half-BPS monopole operators are known to be in one-to-one correspondence with the BMN vacua
\cite{Sheikh-Jabbari:2009vjj}.

Our result \eqref{goal} refines this correspondence substantially. It matches not only the vacua, but also the BPS excitations around each vacuum, including their quantum numbers. 
The microscopic origin of the match is as follows. 
In the relevant limit, only monopole sectors with equal GNO fluxes in the two ABJM gauge groups contribute. 
Such flux configurations are labelled by partitions of $m$, precisely as are the vacua of the $U(m)$ BMN matrix model. 
At the level of the matrix integral of the index, the dominant residues then reproduce both the BMN single-letter index and the residual gauge integrals associated with each vacuum. 
Furthermore, we formulate a dictionary between dressed ABJM monopole operators and BMN excitations, and discuss how finite-$m$ trace relations arise from the ABJM perspective.

This in particular leads to a novel notion of BPS fortuity \cite{Chang:2024zqi} in ABJM theory\footnote{See \cite{Belin:2025hsg,Behan:2025hbx,Behan:2026her} for recent works on BPS fortuity in ABJM theory.}. In the triple-scaling limit, the microscopic ABJM rank $N$ is sent to infinity. So conventional finite-$N$ fortuity is absent. Nevertheless, each monopole sector is governed by the effective rank $m$ and finite-$m$ trace relations generate fortuitous states that map to fortuitous states of the BMN matrix model. In this sense, considering ABJM theory with different monopole charges $m$ defines a ``holographic covering" of the BMN model.
% In particular, the BMN gauge-group rank $m$ is identified with the total
% monopole charge, or equivalently the number of D0-branes in the type IIA
% description. Varying $m$ therefore defines a holographic covering \cite{Chang:2024zqi} of the BMN
% model and permits fortuitous cohomology. By contrast, the microscopic ABJM
% rank $N$ is taken to infinity, so conventional finite-$N$ fortuity is
% absent. Nevertheless, different values of $m$ label distinct superselection
% sectors of ABJM theory. BMN fortuity can thus be interpreted from the ABJM
% perspective as rank-dependent\SK{monopole-charge-dependent?} cohomology across these monopole sectors, rather
% than as fortuity with respect to $N$.

Finally, combining the relation \eqref{goal} with the $11d$ supergravity index, we obtain a closed-form grand-canonical expression for the $U(m)$ BMN index and analyze its large-charge behavior.
We find no evidence for $e^{O(m^2)}$ growth, contrary to earlier claims in the literature
\cite{Choi:2023vdm,Chang:2024lkw,Chang:2026mui}. Such growth has been interpreted as evidence for supersymmetric black
holes in the dual description, although the corresponding black-hole solutions
have not been constructed. Our result therefore calls for a reassessment of
this interpretation.

\paragraph{Relation to related works.}
The pp-wave limit of $AdS_4\times S^{7}/\mathbb{Z}_k$ and its relation to ABJM theory have been studied previously, but the relevant limits differ in fundamental ways from the one considered here.

Nishioka and Takayanagi \cite{Nishioka:2008gz} studied the ten-dimensional type IIA pp-wave obtained as the Penrose limit of $AdS_4\times \mathbb{CP}^3$ and related its string spectrum to ABJM theory. Their analysis is performed in the type IIA and 't~Hooft limits, and is therefore the ABJM analogue of the familiar planar BMN limit of $\mathcal N=4$ super Yang-Mills \cite{Berenstein:2002jq}. This is distinct from the eleven-dimensional Matrix Theory regime considered here.

More closely related is the work of Kovacs, Sato, and Shimada \cite{Kovacs:2013una}, who studied the eleven-dimensional pp-wave limit of $AdS_4\times S^{7}/\mathbb{Z}_k$ and its connection to ABJM theory and the BMN matrix model. The crucial difference is that they keep {\it $k$ fixed and finite}, rather than taking the triple-scaling limit introduced here.  On the ABJM side, they analyze fluctuations around large-charge monopole operators and compare them with the BMN model. In this limit, however, the lightlike compactification radius $R_-$ is sent to infinity, and correspondingly the BMN rank $m$ also becomes infinite. Moreover, the relation between the ABJM monopole sector and the BMN model is not realized as an exact duality in the scaling limit, but only as an approximation in the intermediate large-charge regime
$
N^{1/3}\ll J\ll N^{1/2}
$. (We briefly revisit this regime from the perspective of supersymmetric indices in Subsection~\ref{finite kk}.)

By contrast, our triple-scaling limit directly isolates a finite-rank BMN sector of ABJM while keeping the lightlike compactification radius finite. In particular, both the BMN rank $m$ and the coupling $g_{\rm BMN}$ remain finite. The resulting correspondence is therefore a genuinely DLCQ, rather than decompactified, eleven-dimensional pp-wave limit, and is technically more controlled than the large-rank relation of \cite{Kovacs:2013una}.

Conceptually, this sharpens the correspondence into a precise non-perturbative duality conjecture. Through AdS/CFT, the limiting ABJM sector provides an independent candidate definition of the same finite-momentum DLCQ sector of M-theory described by the BMN model. The finite-rank BMN conjecture can thus be formulated as a duality between two quantum field theories (see Figure \ref{fig:BMNtriality}), without relying on either a large-rank extrapolation or a weakly coupled fluctuation expansion.

Finally, the idea of generating a compact lightlike direction by taking a large
orbifold limit has appeared previously in other contexts: for pp-wave
backgrounds in type IIB string theory in \cite{Mukhi:2002ck}, and for black
hole backgrounds asymptotic to $AdS_7\times S^4$ in
\cite{Dorey:2022cfn,Dorey:2023jfw,Mouland:2023gcp}. Although these settings
are physically quite different from ours, the underlying mechanism leading to
the lightlike compactification is closely related. In particular,
\cite{Dorey:2023jfw} relates ultraspinning black holes to an ensemble of
quantum-mechanical systems, a structure reminiscent of the relation
\eqref{goal}. It would be interesting to revisit these constructions in light of our results.

\paragraph{Structure of the paper.}The remainder of this paper is organized as follows. Section~\ref{sec:triality} reviews the essential features of ABJM theory, the BMN matrix model, and the Penrose limit relating their dual geometries, and then derives the triality by translating this limit to the ABJM side. Section~\ref{review} introduces the superconformal index of ABJM theory and the Witten index of the BMN matrix model, fixing the notation and conventions used throughout the subsequent analysis.
 In Section~\ref{abjm bmn main}, we study the ABJM superconformal index in the triple-scaling limit and show that it matches the grand-canonical index of the BMN model. We also develop the correspondence between ABJM monopole sectors, BMN vacua, and their excitations. Section~\ref{dictionary} refines this dictionary at the level of BPS cohomology, with particular emphasis on how the fortuitous cohomology of the BMN model is realized on the ABJM side. Section~\ref{asymptotic behavior} analyzes the large-$m$ asymptotics of the indices and the partition functions in the regime $Q\sim m^2$ and finds no evidence for the previously proposed $e^{O(m^2)}$ growth. Finally, Section~\ref{discussion} summarizes the results and discusses future directions. Numerical data supporting the index computations are collected in Appendix~\ref{num check}.
\bigskip

\paragraph{Note Added:} After the completion of this work, while this manuscript was being prepared, \cite{Chang:2026yeg} appeared on arXiv with partial overlap with our analysis. The physical setup discussed there is {\it different} from ours: their primary focus is the finite-$k$ correspondence discussed by Kovacs, Sato, and Shimada \cite{Kovacs:2013una}, rather than the triality in the triple-scaling limit studied in this paper. Their treatment of the superconformal index also differs in  methodology. Nevertheless, the resulting behavior of the index leads to conclusions that partially overlap with ours.

\section{BMN triality}
\label{sec:triality}
In this section, we formulate a conjectural triality among ABJM theory, the BMN matrix model, and M-theory on the eleven-dimensional pp-wave with a compact lightlike direction. 
We begin by reviewing the standard dualities relating ABJM theory to M-theory on $AdS_4\times S^7/\mathbb Z_k$, and the BMN matrix
model to M-theory on the pp-wave background. We then identify the limit in
which $AdS_4\times S^7/\mathbb Z_k$ reduces to the DLCQ of
pp-wave and translate this limit into the ABJM side. The triality suggests an equivalence between the BMN matrix model and the large monopole-charge sector of ABJM theory, which we test in Section \ref{abjm bmn main} by comparing their supersymmetric indices.

\subsection{ABJM and M-theory on $AdS_4\times S^{7}/\mathbb{Z}_k$}\label{abjmrev}
ABJM theory is a three dimensional, $\mathcal{N}=6$ superconformal Chern-Simons-matter theory with the gauge group $U(N)_k \times U(N)_{-k}$, where $k$ denotes the Chern-Simons level \cite{Aharony:2008ug}. The matter content consists of four complex scalars $Y_I=(A_{1},A_2,\bar{B}_1,\bar{B}_{2})$, their fermionic superpartners $\psi^I$ and their conjugates, which transform in the bifundamental $(N, \bar{N})$ or $(\bar{N}, N)$ representations. Under the R-symmetry group $SO(6)\simeq SU(4)$, the scalars $Y_I$ transform as fundamental under $SU(4)$, and the fermions $\psi^I$ transform as antifundamental under $SU(4)$.

\paragraph{Global symmetries.}
For generic Chern--Simons level $k$, the superconformal symmetry of ABJM theory is described by $OSp(6|4)$, with conformal symmetry $SO(3,2)$ and R-symmetry $SO(6)\simeq SU(4)$. 
In addition, the two $U(N)$ gauge factors define topological currents
\begin{align}
j_1 = \frac{1}{2\pi}\ast\mathrm{Tr}(F_1) \,,
\qquad
j_2 = \frac{1}{2\pi}\ast\mathrm{Tr}(F_2) \,,
\end{align}
where $F_{1,2}$ are the field strengths of the two gauge groups. These currents are not independent: the Chern--Simons equations of motion imply that $j_-=j_1-j_2$ vanishes.\footnote{The bifundamental matter fields are neutral under the diagonal $U(1)$, generated by the sum of the two $U(1)$ charges. The equation of motion for the corresponding gauge field therefore imposes $\mathrm{Tr}(F_1-F_2)=0$.} Consequently, the total magnetic fluxes in the two gauge groups must agree. We denote their common monopole number by
\begin{equation}
m = \frac{1}{2\pi}\int_{S^2}\mathrm{Tr}(F_1)
  = \frac{1}{2\pi}\int_{S^2}\mathrm{Tr}(F_2)
  \in \mathbb Z \,.
\end{equation}

The remaining combination, $j_+=j_1+j_2$, is related to the baryonic current $j_b$ associated with the relative $U(1)$ between the $U(1)$ subgroups of both groups. Normalizing the baryonic charges of the matter fields $A_i$ and $B_i$ to be $+1$ and $-1$, respectively, the Chern--Simons equations of motion give
\begin{equation}\label{cheqom}
j_b = \frac{k}{4\pi}\ast\mathrm{Tr}(F_1+F_2)
    = \frac{k}{2}j_+ \,.
\end{equation}
It follows that the total baryonic charge is related to the monopole number by\footnote{The quantization of the monopole number follows from  the usual Dirac quantization condition.}
\begin{equation}
J \equiv \int_{S^2}\ast j_b
  = \frac{k}{2}\int_{S^2}\ast j_+
  = km \in k\mathbb Z \,.
\end{equation}
Thus the baryonic charge $J$ and the integer monopole number $m$ differ by a factor of $k$.

At $k=1,2$, the $SO(6)_R$ R-symmetry and the topological $U(1)_b$ are embedded in the enhanced $SO(8)_R$ R-symmetry. We denote its four Cartan charges by $J_i$, with $i=1,\ldots,4$. For completeness, we denote the Cartan generators of $SO(3,2)$ by $j_3,\Delta$. $j_3$ is the Cartan of $SO(3)$ and $\Delta$ is the scaling dimension. The normalization of $J_i$ is fixed by assigning the four complex scalars
\begin{equation}
Y_I=(A_1,A_2,\bar B_1,\bar B_2)
\end{equation}
the charges $J_i(Y_I)=\delta_{iI}$. 
Together with their conjugates, these scalar weights form the vector representation of $SO(8)$, while the fermion weights form a chiral spinor representation.\footnote{This is related to the fact that scalars and fermions carrying equal Baryonic charge transform under conjugate $SU(4)$ representations. }

Throughout most of the paper we consider $k>2$, where the global symmetry is $SO(6)\times U(1)_b$. Nevertheless, we continue to use the four charges $J_i$ as a basis for its Cartan subalgebra. The three $SO(6)_R$ Cartans are linear combinations of differences among the $J_i$, while the baryonic charge is their sum. The Gauss law constraint \eqref{cheqom} therefore gives
\begin{equation}
J_{\rm tot}\equiv \sum_{i=1}^4J_i=km\in k\mathbb Z \,.
\end{equation}

\paragraph{Bulk dual, $AdS_4\times S^{7}/\mathbb{Z}_k$.}
The ABJM theory at level $k$ is the boundary dual of M-theory on $\mathrm{AdS}_4\times S^7/\mathbb{Z}_k$. The corresponding eleven-dimensional supergravity background is locally described by the following metric and four-form flux:
\begin{equation}\label{back}
\begin{split}
ds^2
&=
\frac{R^2}{4}ds^2_{\mathrm{AdS}_4}
+R^2ds^2_{S^7}
\\
&=
\frac{R^2}{4}
\left(
-\cosh^2\rho\,dt^2
+d\rho^2
+\sinh^2\rho\,d\Omega_2^2
\right)
+R^2
\left(
\sum_{I=1}^4 d\mu_I^2
+\sum_{I=1}^4\mu_I^2d\phi_I^2
\right),
\\
F_4
&=
\frac{3}{8}R^3\epsilon_4
=
\frac{3}{8}R^3\cosh\rho\,\sinh^2\rho\,
dt\wedge d\rho\wedge d\Omega_2 \,,
\end{split}
\end{equation}
where $R$ denotes the radius of $S^7$, while the $\mathrm{AdS}_4$ radius is $R/2$. Here $\epsilon_4$ and $d\Omega_2$ denote the volume forms of unit-radius $\mathrm{AdS}_4$ and $S^2$, respectively. We parametrize $S^7$ using four complex coordinates,
\begin{equation}\label{s7coor}
z_I=\mu_Ie^{i\phi_I},
\qquad
\sum_{I=1}^4\mu_I^2=1.
\end{equation}
The $\mathbb{Z}_k$ quotient acts along the Hopf-fiber direction through the simultaneous identification
\begin{equation}\label{newco1}
\phi_I
\sim
\phi_I+\frac{2\pi}{k},
\qquad
I=1,\ldots,4.
\end{equation}
The parameters $N$ and $k$ of ABJM theory determine the radius $R$ in units of the eleven-dimensional Planck length $\ell_p$ through
\begin{equation}\label{eq:ABJMradiusrelation}
\frac{R}{\ell_p}=(32\pi^2Nk)^{1/6}.
\end{equation}

The Cartan charges $J_i$ of ABJM theory generate rotations in the four orthogonal two-planes parametrized by the complex coordinates $z_i$. Their sum, $J_{\rm tot}=\sum_iJ_i$, is the baryonic charge defined above and generates translations along the Hopf-fiber direction. The orbifold identification requires the corresponding angular momentum to be quantized in multiples of $k$, in agreement with the quantization condition imposed by Gauss' law in ABJM theory.

\subsection{BMN quantum mechanics and M-theory on pp-wave}\label{BMN}
BMN quantum mechanics \cite{Berenstein:2002jq} is the maximally supersymmetric mass deformation of the
BFSS matrix model \cite{Banks:1996vh}. The interacting theory preserves sixteen supercharges organized into the $SU(2|4)$ supergroup, whose bosonic symmetry contains $SO(3)\times SO(6)$.

\paragraph{Action and symmetries.}A standard way to write the Lagrangian of $U(m)$ BMN quantum mechanics is as follows \cite{Berenstein:2002jq}:
\begin{align}\label{eq:BMNMQMaction}
L = \mathrm{Tr} \Bigg[
&\frac{1}{2R_{-}} D_0 X^i D_0 X^i
+ \frac{1}{2R_{-}} D_0 X^a D_0 X^a
+ i \psi^{\dagger I\alpha} D_0 \psi_{I\alpha} \nonumber \\
&+ R_{-} \left(
  -\frac{1}{2}\left(\frac{\mu}{3R_{-}}\right)^2 (X^i)^2
  - \frac{1}{2}\left(\frac{\mu}{6R_{-}}\right)^2 (X^a)^2
  - \frac{\mu}{4R_{-}} \psi^{\dagger I\alpha} \psi_{I\alpha}
  - \frac{i\mu}{3R_{-}\ell_p^3} \epsilon_{ijk} X^i X^j X^k
\right. \nonumber \\
&\left. - \frac{1}{\ell_p^3}\psi^{\dagger I\alpha} \sigma^i{}_{\alpha}{}^{\beta} [X^i, \psi_{I\beta}]
+ \frac{1}{2\ell_p^3} \epsilon^{\alpha\beta} \psi^{\dagger\alpha I} g^a{}_{IJ} [X^a, \psi^{\dagger\beta J}]
- \frac{1}{2\ell_p^3} \epsilon^{\alpha\beta} \psi_{\alpha I} (g^{a\dagger})^{IJ} [X^a, \psi_{\beta J}]
\nonumber \right.\\
&+\left. \frac{1}{4\ell_p^6}[X^i, X^j]^2
+ \frac{1}{4\ell_p^6}[X^a, X^b]^2
+ \frac{1}{2\ell_p^6}[X^i, X^a]^2\right)
\Bigg]
\end{align}\footnote{Note that the convention for $R_-$ here is different from that in \cite{Berenstein:2002jq}, the conversion is $R_-=2R'_-$ where $R_-'$ is the compactification radius in \cite{Berenstein:2002jq}.}
Here all the fields are valued in $m\times m$ matrices and the indices $i,j$ runs over $\{1,2,3\}$ while $a,b,c$ run over $\{5,6,7,8,9,10\}$ and they transform as the vector representations of $SO(3)$ and $SO(6)$, which are bosonic subgroups of $SU(2|4)$.
The quantities $\mu^{-1}, R_{-}, \ell_p$ and $X$ all have dimensions of length, while the fermionic fields $\psi_{I\alpha}$ are dimensionless. We can also define a dimensionless combination of the parameters $\mu^{-1}, R_{-}, \ell_p$, which plays the role of the dimensionless coupling constant  of the theory:
\begin{equation}
    g_{\rm BMN}^2=\frac{R_{-}^3}{\mu^3\ell_p^6}\,.
\end{equation}

\paragraph{Fuzzy sphere vacua.} A crucial feature of the BMN model is its large set of degenerate supersymmetric vacua. The bosonic potential can be written as a sum of squares,
\begin{align}
V = \frac{R_{-}}{2} \mathrm{Tr} \Bigg[
\left( \frac{\mu}{3R_{-}} X^i + \frac{i}{2l_p^3} \epsilon_{ijk} [X^j, X^k] \right)^2
+ \frac{1}{l_p^6} (i[X^i, X^a])^2
+ \frac{(i[X^a, X^b])^2}{2l_p^6}
+ \left( \frac{\mu}{6R_{-}} X^a \right)^2
\Bigg]
\end{align}
A zero-energy configuration therefore satisfies
\begin{equation}
\begin{split}
&\frac{\mu}{3R_{-}} X^i + \frac{i}{2\ell_p^3} \epsilon_{ijk} [X^j, X^k]=0\,.\\ 
 &X^a=0\,.
\end{split}
\end{equation}
The solutions to the first equation are
\begin{equation}\label{eq:vacuasize}
    X^i=\frac{\mu \ell_p^3}{3R_-} \mathcal{J}^i
\end{equation}
where the $\mathcal J^i$ form an $m$-dimensional, not necessarily irreducible, representation of $SU(2)$.
\begin{equation}
    [\mathcal{J}^i,\mathcal{J}^j]=i\epsilon^{ijk}\mathcal{J}^k
\end{equation}
It is known that these classical configurations give the supersymmetric vacua of the quantum theory \cite{Berenstein:2002jq, Dasgupta:2002ru}.

Consequently, the BMN vacua are classified by partitions of $m$. More explicitly, the defining representation may be decomposed as \begin{equation} \mathbb C^m = \bigoplus_{s=1}^{K} \left( \mathbb C^{n_s}\otimes V_{N_s} \right), \qquad \sum_{s=1}^{K}n_sN_s=m, \end{equation} where $V_{N_s}$ denotes the $N_s$-dimensional irreducible representation of $SU(2)$, and the $N_s$ are taken to be distinct. Equivalently, each vacuum is labelled by a partition of $m$ into dimensions of $SU(2)$ irreducible representations.\footnote{From a bulk point of view, this corresponds to $K$ M2 branes with sizes 
\begin{equation}\label{dgsize}
r_s = 2\pi\sqrt{\frac{1}{N_s}\Tr_{N_s} (X^i X^i)}\sim \frac{\pi N_s \mu l_p^3}{3R_-}.
\end{equation}
where the trace is taken over the subblock of size $N_s$ and we are taking large $N_s$ limit. The factor of $2\pi$ is related to the conversion between matrix variables and supergravity coordinates(see equation 5.5 of \cite{Berenstein:2002jq}).  
These membranes are stabilised by competition between the quadratic and quartic terms in the potential, which want to contract it, and the cubic term, which wants to make it expand. }

Two limiting cases will be particularly useful. The \textit{trivial vacuum} corresponds to 
\begin{equation} 
m=1+\cdots+1, 
\end{equation} 
for which $X^i=0$ and the full $U(m)$ gauge symmetry is unbroken. The \textit{irreducible vacuum} corresponds to the single partition $m=m$ and preserves only the overall $U(1)$. More generally, for a vacuum containing $n_s$ identical irreducible blocks of dimension $N_s$, the subgroup of $U(m)$ that commutes with the $SU(2)$ embedding is\footnote{One way to see this, is that under this $U(n_k)$ transformation, the $N_k n_k$ sized block is like $\mathbf{1_{n_k}}\times \mathcal{J}^i_{N_k}$ and the $U(n_k)$ only acts on identity part and hence commutes with it.} 
\begin{equation} 
G_{\rm unbroken} = \prod_{s=1}^{K}U(n_s). \end{equation} 
Thus the same partition of $m$ that labels a BMN vacuum also determines its unbroken gauge symmetry. This partition structure will play a central role in the comparison with monopole sectors of ABJM theory below.

% The solution where we use up the whole $m$ dimensional block is called the irreducible vacuum. And the one where we split the  $m$-dimensional matrices completely into diagonal form $\{1,1,\ldots 1\}$ is called the trivial vacuum.

% In the trivial vacuum, the matrices $X^i$ do not have expectation values, and therefore they commute with all $U(m)$ generators. Hence, the trivial vacuum preserves the full $U(m)$ gauge symmetry since it is invariant under it. On the other hand, in the irreducible vacuum, the matrices $X^i$ transform under all the generators of $SU(m)$, but they still are uncharged under the $U(1)$ of $U(m)$. Hence, the irreducible vacuum preserves only $U(1)$ part of the gauge symmetry. 

% Now let us consider a vacuum with partitioning into $K$ blocks, such that there are $n_k$ blocks of size $N_k$ and 
% $$\sum_{k=1}^K n_k N_k=m$$ with all $N_k$ different. In such a vacuum, we can still perform $U(m)$ transformations which mix the $n_k$ entries of the $ m$-dimensional vector. This will mix the $n_k$ $SU(2)$ blocks.
\paragraph{BMN conjecture.} The BMN matrix model admits two conceptually distinct connections to gravity. The first is the more familiar holographic one: in an appropriate large-$m$ 't~Hooft limit ($g_{\rm BMN}^2 m$ fixed), the matrix model is dual to backreacted supergravity geometries. In particular, the geometries associated with its supersymmetric vacua were constructed in \cite{Lin:2004kw,Lin:2005nh}, and finite-temperature states are described by the corresponding black-hole geometries \cite{Costa:2014wya}.

The relation relevant for this paper is different. It is the pp-wave analogue of the Matrix Theory conjecture for the BFSS model, and we will refer to it as the \textit{BMN conjecture}. Rather than a conventional holographic limit, it asserts that the BMN matrix model in {\it non-'t Hooft limit} provides a non-perturbative description of M-theory on the maximally supersymmetric eleven-dimensional pp-wave
\begin{align}\label{eq:ppwavebkd}
\begin{aligned}
ds^2&= -2dx^+ dx^- -\left(\left(\frac{\mu}{3}\right)^2 (x^{i})^2+ \left(\frac{\mu}{6}\right)^2 (x^{a})^2\right)(dx^+)^2+ (dx^{i})^2 + (dx^{a})^2\\
    F_4&=\mu dx^+\wedge dx^1\wedge dx^2\wedge dx^3
\end{aligned}
\end{align}
As in \eqref{eq:BMNMQMaction}, the index $i$ runs over $\{1,2,3\}$ and the index $a$ runs over $\{5,6,7,8,9,10\}$.
As in BFSS Matrix Theory, it is useful to distinguish two versions of the conjecture.
\begin{itemize}
\item \textbf{Finite-rank (DLCQ) conjecture.} $U(m)$ BMN matrix model at finite $m$ is conjectured to describe the sector of M-theory on the pp-wave with $m$ units of longitudinal light-cone momentum, after compactifying 
\begin{equation} 
x^-\sim x^-+2\pi R_- . 
\end{equation} 
In our conventions, 
\begin{equation} -p_-=\frac{m}{R_-}, \qquad g_{\rm BMN}^2 = \frac{R_-^3}{\mu^3\ell_P^6}. \label{eq:couplingmapBMN} 
\end{equation} 
This is the pp-wave analogue of the finite-$N$ DLCQ interpretation of BFSS Matrix Theory proposed by Susskind \cite{Susskind:1997cw}.
\item \textbf{Large rank (decompactified) conjecture.} Uncompactified M-theory on the pp-wave is recovered by decompactifying the lightlike circle. At fixed physical light-cone momentum, this requires $m,R_-\to\infty$ with $m/R_-$ fixed. Through \eqref{eq:couplingmapBMN}, this corresponds to 
\begin{equation} m\to\infty, \qquad g_{\rm BMN}^2\to\infty, \qquad \frac{g_{\rm BMN}^2}{m^3} \quad \text{fixed}. 
\end{equation} 
Thus the decompactification limit is an intrinsically strong-coupling limit of the matrix model. This is the direct analogue of the original large-rank BFSS conjecture \cite{Banks:1996vh}.
\end{itemize}
Even in the BFSS model, for which these conjectures were originally formuated, the finite-rank conjecture is especially difficult to test. The basic obstacle is conceptual: DLCQ of M-theory is not known independently in a formulation that can be compared directly with the matrix model. In this sense, the matrix model is often best regarded as a \textit{definition} of the fixed light-cone momentum sector, rather than as one side of an independently established duality.

For the BMN model, neither of these conjectures is well-studied. Nevertheless some progress has been made recently. In particular, recent work studied the model at finite rank and strong coupling and found that its low-energy spectrum reproduces the spectrum of supergravity excitations on the pp-wave, subject to an assumption about the relevant bound states \cite{Komatsu:2024vnb}. This provides nontrivial evidence for the finite-rank conjecture, but still relies on taking a semiclassical supergravity limit.

The BMN triality proposed below is designed to overcome precisely this difficulty. The pp-wave DLCQ background will arise as a controlled limit of $AdS_4\times S^7/\mathbb Z_k$, whose M-theory description is independently provided by the ABJM theory through the AdS/CFT correspondence. This gives a second, holographic definition of the same DLCQ sector and therefore turns the BMN conjecture into a relation that can be tested quantitatively from the ABJM side. Furthermore, the connection to a non-pertubative description of M-theory is automatically guaranteed through the AdS/CFT correspondence.

\subsection{Deriving the BMN triality}
\label{penrose}
So far we have reviewed the duality between ABJM theory and M-theory on  $AdS_4\times S^7/\mathbb{Z}_k$ and the duality between the BMN model and DLCQ of M-theory on the pp-wave background.  We now describe the limit (the Penrose limit) in which $AdS_4\times S^7/\mathbb{Z}_k$ reduces to the light-like compactified pp-wave background and translate it into a triple-scaling limit of ABJM theory. This in turn allows us to relate the triple-scaling limit of ABJM to the BMN model.
\paragraph{Penrose limit of $AdS_4\times S^7/\mathbb{Z}_k$.} We take the Penrose limit of the metric in \eqref{back} along a null geodesic at the center of $AdS_4$, $\rho=0$, moving along the $\phi_1$ circle at $\mu_1=1$ ($\mu_{2,3,4}=0$), cf.~\eqref{s7coor}. Its trajectory in $AdS_4$ is given by $\phi_1=t/2$. To describe transverse displacements around this geodesic, we parametrize the coordinates in \eqref{back} and \eqref{s7coor} as\begin{align} \mu_1=\cos\theta, \qquad \mu_A=\hat n_A\sin\theta, \qquad \sum_{A=2}^4\hat n_A^2=1, \end{align} and introduce \begin{align} \rho=\frac{2r}{R}, \qquad \theta=\frac{y}{R}. \end{align} Keeping $r$ and $y$ fixed as $R\to\infty$ keeps the proper transverse distances finite. Expanding the metric gives 
\begin{align} 
ds^2 &= -\frac{R^2}{4}\,dt^2 +R^2\,d\phi_1^2 -r^2\,dt^2 -y^2\,d\phi_1^2 +d\vec r^{\,2} +d\vec y^{\,2} +O(R^{-2}).
\end{align} 
Here $d\vec r^{\,2}=dr^2+r^2d\Omega_2^2$ and $d\vec y^{\,2}=dy^2+y^2d\Omega_5^2$ are the flat metrics on $\mathbb R^3$ and $\mathbb R^6$. Hence we can introduce the Cartesian coordinates $x^{i}$ and $x^{a}$ ($i=1,\ldots, 3$, $a=5,\ldots, 10$) and write $d\vec{r}^2=(dx^i)^2$ and $d\vec{y}^2=(dx^{a})^2$.

We next choose coordinates adapted to the geodesic: \begin{align} t&=u+\frac{4v}{R^2}, & \phi_1&=\frac{u}{2}-\frac{2v}{R^2}, \nonumber\\ u&=\frac{t}{2}+\phi_1, & v&=\frac{R^2}{4}\left(\frac{t}{2}-\phi_1\right) . \label{newco2} \end{align} Here $u$ parametrizes motion along the geodesic, while $v$ measures the rescaled departure from the relation $\phi_1=t/2$. The leading $R^2du^2$ terms cancel, leaving a finite metric as $R\to\infty$: \begin{align} 
ds^2 &= -4\,du\,dv -\left((x^{i})^2+\frac{(x^{a})^2}{4}\right)du^2 +(d x^{i})^{\,2}+(dx^{a})^{\,2}+O(R^{-2}). 
\end{align} 
The four-form flux has a finite limit as well, \begin{align} 
F_4 = 3\,du\wedge dx^1\wedge dx^2\wedge dx^3. 
\end{align} 
Finally, the rescaling \begin{align} x^+=\frac{3u}{\mu}, \qquad x^-=\frac{2\mu v}{3} \end{align} brings the metric and flux to the standard pp-wave form \eqref{eq:ppwavebkd}, with transverse mass scales $\mu/3$ and $\mu/6$.

The analysis above explains formally how the pp-wave metric emerges from $AdS_4\times S^{7}/\mathbb{Z}_k$. However, to understand the physical content of this limit and translate it to the ABJM side, two further questions must be addressed: how the orbifold identification \eqref{newco1} behaves in the limit, and which states remain at finite energy after the limit is taken.

\paragraph{From orbifold to DLCQ.} The local metric alone does not determine whether the pp-wave is compactified. For this, we must also follow the $\mathbb Z_k$ identification through the limit. The quotient shifts all four angular coordinates $\phi^{I}$ by $2\pi/k$. Along the chosen geodesic, its action on 
\eqref{newco2} is \begin{align} (u,v) \sim \left( u+\frac{2\pi}{k}, v-\frac{\pi R^2}{2k} \right). 
\end{align} 
The two shifts scale differently. Taking $k\to\infty$ removes the shift in $u$, together with the rotations of the transverse coordinates. Keeping $R^2/k$ fixed, however, leaves a finite shift in $v$. The orbifold circle thus becomes a compact null direction. In the normalized coordinates, its radius is \begin{align} x^-\sim x^-+2\pi R_-, \qquad R_-=\frac{\mu R^2}{6k}, \end{align} where we have used the magnitude of the shift to define $R_->0$. The relaton to ABJM parameters \eqref{eq:ABJMradiusrelation}, $R/\ell_P=(32\pi^2Nk)^{1/6}$, then gives \begin{align} 
R_- = \frac{\mu\ell_P^2}{6}(32\pi^2)^{1/3} \left(\frac{N}{k^2}\right)^{1/3}. \end{align} Thus, at fixed $\mu$ and $\ell_P$, a finite null-circle radius requires $N/k^2$ to remain fixed. This is the geometric origin of the first scaling condition.

\paragraph{Surviving states.} We next determine which ABJM states have finite energy and momentum in the limiting geometry. With the conventions $\Delta=i\partial_t$ and $J_1=-i\partial_{\phi_1}$, the light-cone Hamiltonian is \begin{align} H = i\partial_{x^+} = \frac{\mu}{3} \left( i\partial_t+\frac{i}{2}\partial_{\phi_1} \right) = \frac{\mu}{3}\left(\Delta-\frac{J_1}{2}\right). \end{align} The pp-wave limit therefore keeps the excitation energy $\Delta-J_1/2$ finite. On the other hand, the other combination $i\partial_t-\frac{i}{2}\partial_{\phi_1} $ is related to the lightcone momentum by
\begin{align} \label{eq:pminusscaling}
-p_- = i\partial_{x^-} = \frac{6}{\mu R^2}\left(\Delta+\frac{J_1}{2}\right). \end{align}
Hence in order for states to have finite $p_{-}$ after taking the limit $R\to \infty$, $\Delta$ and $J_1$ need to individually diverge as $\propto R^2$. To translate this into ABJM, we multiply the lightlike radius $R_{-}$ to \eqref{eq:pminusscaling} and get
 \begin{align} 
 -R_-p_- = \frac{\Delta+J_1/2}{k} = \frac{J_1}{k} + \frac{\Delta-J_1/2}{k}. \end{align} 
 The second term vanishes in the limit. Hence we arrive at the second scaling condition
 \begin{align}
 J_1,k\to \infty\quad \text{with}\quad m=\frac{J_1}{k}\quad \text{fixed}\,,
 \end{align}
 where $m$ is an integer unit of lightcone momentum $m=-R_{-}p_{-}$. Together with the first scaling condition, this defines the relevant triple-scaling limit of ABJM theory.
\paragraph{Triality.} To summarize, we have translated the Penrose limit of $AdS_4\times S^{7}/\mathbb{Z}_k$ to the triple-scaling limit of ABJM theory, in which one sends $k$, $N$ and $J_1$ to infinity keeping the following combinations finite:
\begin{align}
\begin{aligned}
\frac{R_-}{\mu\ell_P^2} =  \left(\frac{4\pi^2N}{27k^2}\right)^{1/3}\,,\qquad
\lim_{k\to\infty}\frac{J_1}{k}=m\,.
\end{aligned}
\end{align}
In both equations, the left hand sides are quantities in the pp-wave background while the right hand sides are quantities in ABJM. 

Combining these with the parameter relations in the BMN conjecture --- the rank $m$ of the BMN model corresponds to the an integer unit of lightlike momentum and the coupling $g_{\rm BMN}$ is related to M-theory parameter by $g_{\rm BMN}^2=R^{3}_{-}/(\mu^3\ell_{P}^6)$ --- one can readily obtain the parameter dictionary for the triality given in \eqref{parameter-dictionary}, which we display again for convenience:
 \begin{align}
 \begin{aligned}
&\lim_{k\to\infty}\frac{J_1}{k}=m=-R_{-}p_{-}\,,\qquad g_{\rm BMN}^2=\frac{4\pi^2}{27} \frac{N}{k^2}= \frac{R_{-}^3}{\ell_{P}^{6}\mu^3}\,,
&
\end{aligned}
 \end{align}

\paragraph{Size of dual giant gravitons in the Penrose limit.} As a simple sanity check of the triality, let us discuss dual giant gravitons in $AdS_4\times S^{7}/\mathbb{Z}_k$ and see how they behave in the Penrose limit. Dual giant gravitons are  spherical M2-brane solutions in AdS and their counterparts in the pp-wave background are known to correspond to the fuzzy sphere vacua of the BMN model. 

As shown in \cite{Nishioka:2008ib}, the size of the dual giant graviton carrying $J_1=mk$ in the radial coordinate is
\begin{align}
r_{\ast}=\frac{R}{2} m\sqrt{\frac{k}{2N}}=m \ell_P\left(\frac{\pi k^2}{4N}\right)^{1/3}
\end{align}
In the triple-scaling limit, where $k\sim N^{1/2} \to \infty$, we have
\begin{align}
r_{\ast}\sim \ell_P \ll R\,.
\end{align}
Namely the M2 giant is much smaller than the size of AdS but finite in Planck units. This is in line with the fact that these states survive the triple-scaling limit and hence they should be contained in the region near the null geodesics used for the Penrose limit. 

Furthermore, by rewriting it using our dictionary \eqref{parameter-dictionary}, we obtain
\begin{align}
r_{\ast}=\frac{\pi m \ell_P}{3 g_{\rm BMN}^{2/3}}
\end{align}
which precisely argees with the size of the irreducible fuzzy sphere vacua as in  \eqref{dgsize} in the $U(m)$ BMN quantum mechanics, providing a quantitative sanity check of the triality.

\subsection{Large charge, open-closed-open triality, and worldline holography}\label{oco}

As emphasized in the Introduction, the BMN triality exhibits several striking structural features. We now place some of them in a broader context and compare them with related ideas.

\paragraph{Large-charge matrix models.}
One notable feature of the BMN triality is that the rank $m$ of the BMN matrix model is set by the charge $J_1$ of the corresponding monopole sector in ABJM theory. Similar exchanges between matrix rank and operator charge have appeared in studies of large-charge limits of superconformal field theories. For example, it was shown using supersymmetric localization in \cite{Giombi:2018qox} that correlation functions of large-charge operators on the supersymmetric Wilson loop in $\mathcal N=4$ super Yang--Mills theory admit a dual matrix-integral representation\footnote{This integral representation was first obtained using integrability in \cite{Gromov:2012eu}.} in which the matrix rank is identified with the operator charge. Analogous structures also arise for Coulomb-branch correlators in rank-one $\mathcal N=2$ superconformal field theories \cite{Grassi:2019txd}.

At the level of parameter relations, this is strikingly similar to the correspondence discussed here. The physical connection, however, is less direct. In the large-charge matrix models just described, the auxiliary matrix integral emerges from technical manipulations, such as Gram--Schmidt orthogonalization, and does not a priori come with an independent dynamical interpretation.

\begin{figure}[t]
\centering
\resizebox{\linewidth}{!}{%
\begin{tikzpicture}[
    font=\small,
    >={Latex[length=2.3mm,width=1.6mm]},
    box/.style={
        draw,
        line width=0.75pt,
        rounded corners=5pt,
        align=center,
        font=\small\bfseries,
        inner xsep=10pt,
        inner ysep=9pt,
        outer sep=0pt,
        minimum height=1.25cm
    },
    relation/.style={
        <->,
        line width=0.8pt,
        shorten <=3pt,
        shorten >=3pt
    },
    label/.style={
        font=\small\bfseries,
        align=center,
        inner sep=0pt
    }
]

% Nodes: use explicit line breaks, without text-width restrictions.
\node[box,draw=red,minimum width=4.2cm] (gravity) at (0,3.2)
    {Closed string\\[3pt]
     ($AdS_5\times S^5$)};

\node[box,draw=blue,minimum width=4.2cm] (vtype) at (-4.6,0)
    {V-type open string\\[3pt]
     ($\mathcal{N}=4$ SYM)};

\node[box,draw=green!60!black,minimum width=4.8cm] (ftype) at (4.6,0)
    {F-type open string\\[3pt]
     ($\rho$ model on giant gravitons)};

% Attachment points on the lower edge of the gravity box.
% These make the diagonal arrows slope inward toward the top box.
\coordinate (gravityLeft)
    at ($(gravity.south)+(-2.35cm,0)$);
\coordinate (gravityRight)
    at ($(gravity.south)+(2.35cm,0)$);

% Diagonal relations, with labels outside the triangle.
\draw[relation]
    (vtype.north) --
    node[label,midway,above left=5pt]
        {V-type duality\\[3pt](AdS/CFT)}
    (gravityLeft);

\draw[relation]
    (ftype.north) --
    node[label,midway,above right=5pt]
        {F-type duality}
    (gravityRight);

% Horizontal relation.
\draw[relation]
    (vtype.east) --
    node[label,midway,below=6pt] {}
    (ftype.west);

\end{tikzpicture}%
}
\caption{Relations among the three descriptions in the open--closed--open triality. Concrete realizations in $\mathcal N=4$ SYM are shown in parentheses. In the BMN triality, the V-type open-string description corresponds to triple-scaled ABJM theory, the F-type open-string description to the BMN matrix model, and the closed-string description to DLCQ M-theory on the pp-wave background.
}
\label{fig:OCO}
\end{figure}
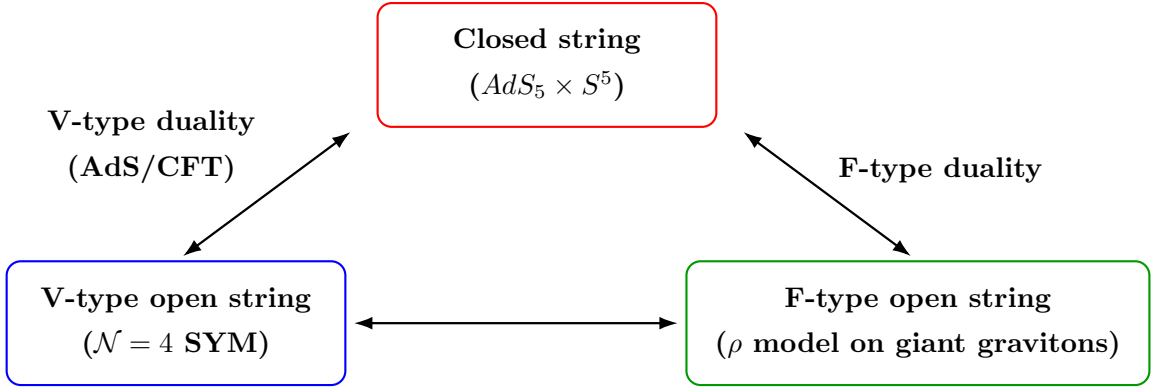

\paragraph{Open--closed--open triality.} A more physical framework for understanding such rank--charge relations is provided by the open--closed--open triality (OCO) \cite{Gopakumar:2022djw,Gopakumar:2024jfq}. Schematically, this triality relates two open-string descriptions to a common closed-string, or quantum-gravitational, description. One of the open-string descriptions, which we call the V-type dual, is the conventional AdS/CFT dual. The other, the F-type dual, is typically realized as an effective theory on probe branes propagating in the closed-string background sourced by the V-type sector. See Figure \ref{fig:OCO}.

\begin{figure}[t]
  \centering
  \resizebox{0.7\linewidth}{!}{%
  \input{oco_triality.tikz}%
  }
  \caption{Relations among the three descriptions in the
    open--closed--open triality. Starting from stacks of V-type branes and F-type branes (middle), integrating out both leads to a nontrivial closed string background with excitations (top) while integrating out only F-type branes leads to V-type open string dual with operator insertions (lower left). On the other hand, integrating out only V-type branes leads to probe F-type branes on a curved background (lower right).}
  \label{fig:derivationOCO}
\end{figure}
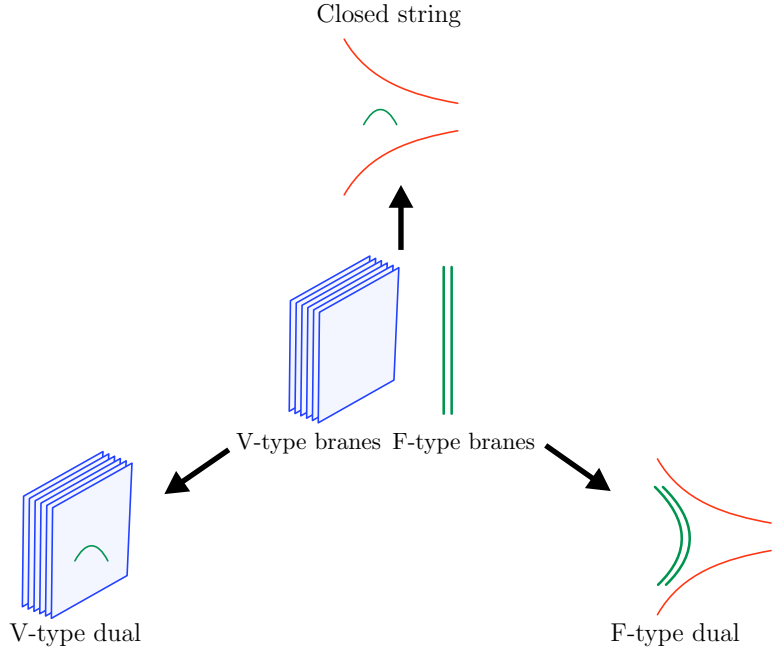

A useful way to motivate OCO is to begin with two stacks of D-branes: a large stack of V-type branes and a finite stack of F-type branes. The three descriptions arise depending on which degrees of freedom are integrated out. (See Figure \ref{fig:derivationOCO}.) Integrating out the F-type branes leaves a theory on the V-type stack, but induces deformations or operator insertions whose charges are controlled by the number of F-type branes. This gives the V-type open-string description. Conversely, integrating out the large V-type stack produces, by the usual AdS/CFT logic, a backreacted closed-string geometry. The remaining F-type branes then behave as probes in this geometry, giving the F-type open-string description. In general, their dynamics is described by an interacting open(--closed) string field theory in a curved background sourced by the V-type sector. Finally, integrating out the F-type branes as well produces excitations of the closed-string background, yielding the closed-string description.

A concrete realization appears in $\mathcal N=4$ SYM \cite{Jiang:2019xdz,Jiang:2019zig,Chen:2019gsb,Budzik:2021fyh,Caron-Huot:2023wdh,Mazenc:2026yrm,Konosu:2026xxy}. One starts with the usual stack of $N$ D3-branes together with an additional finite stack of D3-branes that become giant gravitons in $AdS_5\times S^5$. Integrating out the giant-graviton branes produces operator insertions in $\mathcal N=4$ SYM whose charges are proportional to the number of giant gravitons, giving the V-type description. If instead one integrates out the original stack of $N$ D3-branes, the giant gravitons remain as probe branes in the resulting $AdS_5\times S^5$ geometry. Their effective dynamics on the F-type branes is described by a matrix model, often referred to as the $\rho$-model in \cite{Jiang:2019xdz,Jiang:2019zig,Budzik:2021fyh,Caron-Huot:2023wdh,Mazenc:2026yrm,Konosu:2026xxy}, whose rank is therefore proportional to the charge of the corresponding operator insertion.

Our construction admits a closely analogous interpretation. Consider a stack of $N$ D2-branes at a conifold singularity with $k$ units of RR two-form flux \cite{Aganagic:2009zk}, together with $m$ D0-branes. In the appropriate low-energy limit, the V-type description is ABJM theory. Integrating out the D0-branes corresponds on the ABJM side to inserting monopole operators whose charge is set by $m$. Conversely, integrating out the D2-branes produces the holographic $AdS_4$ background, with the $m$ D0-branes remaining as probes; this is the F-type description. Finally, replacing the D0-branes by the corresponding gravitational excitations gives an M-theory description on $AdS_4\times S^7/\mathbb Z_k$ carrying $m$ units of Kaluza--Klein momentum along the M-theory circle. From this viewpoint, the BMN triality is a close analogue of OCO triality, with the BMN matrix model playing the role of the F-type open-string description.

There is, however, an essential additional ingredient in the BMN triality: we simultaneously take the Penrose limit of the $AdS_4$ geometry. This extra limit is crucial for making the triality both precise and tractable. A generic difficulty with OCO triality is that the F-type description is typically a full-fledged open(-closed) string field theory\footnote{In low-dimensional or topological settings \cite{Maldacena:2004sn,Gaiotto:2003yb,Gopakumar:2022djw,Gopakumar:2024jfq,Budzik:2021fyh}, the F-type dual may reduce to a matrix model rather than a full string field theory. This simplification reflects the restricted dynamics of such systems; in more general higher-dimensional settings, one expects the F-type description to involve a genuine open, or open--closed, string field theory.} in a curved background. Consequently, although the triality is conceptually compelling, quantitative tests are generally difficult.

The Penrose limit removes this obstacle in the present setting. In the pp-wave limit, the dynamics of the F-type branes reduces to the BMN matrix quantum mechanics, a conventional quantum-mechanical theory that can be defined and studied non-perturbatively. This limit may be viewed as a {\it second decoupling limit}: the first is the familiar decoupling limit of the V-type branes that leads to AdS/CFT, while the second simplifies the dynamics of the probe F-type branes in the backreacted geometry from string field theory to ordinary quantum field theory. The BMN triality therefore combines the OCO structure with an additional decoupling limit that renders the F-type corner independently tractable.

This perspective suggests a broader lesson. In other realizations of OCO triality, it may be possible to identify an analogous second decoupling limit in which the otherwise complicated F-type open string field theory reduces to a more conventional field theory. Such limits could provide settings in which OCO triality can be formulated more sharply and tested quantitatively beyond protected or perturbative sectors.

\paragraph{Worldline holography.}
One motivation for considering the F-type dual, in addition to the conventional V-type AdS/CFT description, is that the F-type degrees of freedom live in the bulk and therefore need not be tied to an asymptotic boundary. This feature is reminiscent of the worldline holography \cite{Anninos:2011af, Chandrasekaran:2022cip, Witten:2023xze}, which has been proposed as a framework for describing gravitational dynamics in spacetimes without boundaries, such as de Sitter space. The basic idea is to encode bulk physics in a quantum theory associated with an observer's worldline.

If such a description exists, it would considerably broaden the scope of holography beyond asymptotically AdS spacetimes. At present, however, the worldline holography is much less firmly established than standard AdS/CFT, in part because controlled top-down realizations in string theory are absent. This makes it difficult to formulate and test the proposal quantitatively.

The close relation between OCO triality and the BMN triality suggests that the BMN model may provide a useful toy model of the worldline holography in a setting that is under substantially better control. 
An intriguing aspect of this perspective is that the BMN model is an ordinary matrix quantum mechanics rather than a gravitational (or string) theory. This is especially interesting in light of recent analyses for de Sitter static patch holography which indicate the worldline theory may not be a conventional quantum-mechanical system \cite{Milekhin:2026tbi,Cui:2026bcd,Chen:2026boh,Harlow:2026pwe}. It would be interesting to understand how, and to what extent, the BMN model captures bulk quantum-gravitational dynamics, despite being an ordinary quantum mechanical system.

%%%%%%%%%%%%%%%

\section{Indices of ABJM and BMN: review}\label{review}
In this section, we review the superconformal index of ABJM theory and the supersymmetric index of BMN quantum mechanics in order to set the conventions and the notations for the following section. 
%, and the relevant Penrose--DLCQ limit of \(AdS_4\times S^7/\mathbb Z_k\) in Section~\ref{penrose}.

\subsection{Superconformal index of ABJM}\label{ABJM}

As reviewed in Subsection \ref{abjmrev}, the ABJM theory preserves $\mathcal N=6$ superconformal symmetry. The Poincare supercharges $Q$ transform as a spinor under the $SO(3)$ and as a vector under R-symmetry $SO(6)$ (or equivalently as an antisymmetric representation under $SU(4)$). Similarly, the conformal supercharges $S$ transform under spinor of $SO(3)$ and conjugate antisymmetric representation under $SU(4)$. Following the notation in \cite{Kim:2009wb}, we denote these supercharges by 
\begin{equation}
    Q_{IJ\alpha}, ~S^{IJ}_\alpha
\end{equation}
where indices $I,J$ are $SU(4)$ indices and $\alpha$ is $SO(3)$ spinor index. 

The superconformal index counts the graded sum over the states annihilated by a choice of a pair of conjugate supercharges $Q_{BPS}$ and $S_{BPS}$. The charges carried by $Q_{BPS}$ are listed in Table \ref{level k abjm}. $S_{BPS}$ carries equal and opposite charges. 

The BPS cohomology is constructed from a set of fundamental fields, or ``BPS letters,'' that are annihilated by $\delta =\{Q_{\rm BPS},S_{\rm BPS}\}$. For the above choice of the pair of supercharges, $\{Q_{\rm BPS},S_{\rm BPS}\}=\Delta-j_3-\frac{1}{2}(J_1+J_2-J_3-J_4)$\footnote{Here $j_3$ denotes the Cartan of $SO(3)\subset SO(3,2)$.}. Therefore, the states counted by the index carry zero charge under this linear combination. The BPS letters are listed in Table \ref{level k abjm}.

The superconformal index of the ABJM theory is defined as \cite{Kim:2009wb}
\begin{align}
I(p,x,y,z,w)=\Tr_{BPS}\left[(-1)^Fp^{j_3}x^{J_1}y^{J_2}z^{-J_3}w^{-J_4}\right], \quad p=xyzw.
\end{align}
To compare it with the notation in \cite{Kim:2009wb}, we write the index as follows:
\begin{equation}
I(y_0, y_1, y_2,y_3) = \mathrm{Tr}\left[(-1)^F e^{-\beta'\{Q,S\}} y_0^{(\epsilon+j_3)} y_1^{h_1}y_2^{h_2}y_3^{h_4}\right]
\end{equation}
Here, the $h_i$'s are the following linear combinations of $J_i$
\begin{align}
h_1 &= \tfrac{1}{2}(J_1 - J_2 - J_3 + J_4),\quad
h_2 = \tfrac{1}{2}(-J_1 + J_2 - J_3 + J_4) \nonumber\\
h_3 &= \tfrac{1}{2}(J_1 + J_2 - J_3 - J_4),\quad
h_4 = \tfrac{1}{2}(J_1 + J_2 + J_3 + J_4).
\end{align}
Equivalently, the map between fugacities:
\begin{align}
    x=\sqrt\frac{y_0y_1y_3}{y_2},~ y=\sqrt\frac{y_0y_2y_3}{y_1},~z=\sqrt\frac{y_0y_1y_2}{y_3},~w=\sqrt\frac{y_0}{y_1y_2y_3}.
\end{align}
so that $y_0^2=p=xyzw$.

In all the subsequent discussion, we use the basis formed by charges $J_i$.
\begin{table}[h!]
\centering
\begin{tabular}{c c c c}
\hline
Field & $\Delta$ & $(J_1,J_2,J_3,J_4)$ & Spin $j_3$ \\
\hline
$A_{1,2}$ & $1/2$ & $(1,0,0,0),(0,1,0,0)$ & $0$ \\
$B_{1,2}$ & $1/2$ & $(0,0,-1,0),(0,0,0,-1)$ & $0$ \\
$\psi_{1,2}^A$ & $1$ & $(1/2,-1/2,-1/2,-1/2)$, $(-1/2,1/2,-1/2,-1/2)$& $1/2$ \\
$\psi_{1,2}^B$ & $1$ & $(1/2,1/2,-1/2,1/2)$, $(1/2,1/2,1/2,-1/2)$ & $1/2$ \\
$\partial$ & $1$ & $(0,0,0,0)$ & $1$ \\
$Q_{\text{BPS}}$ & $1/2$ & $(1/2,1/2,-1/2,-1/2)$ & $-1/2$ \\
\hline
\end{tabular}
\caption{BPS letters that satisfy the BPS condition $\{Q_{\rm BPS},S_{\rm BPS}\}=\Delta-j_3-\frac{1}{2}(J_1+J_2-J_3-J_4)=0$.  $A_1,A_2,B_1,B_2$, and $D$ carry fugacities $x,y,z,w$, and $p$ respectively. $\psi_{1,2}^{A}$ and $\psi_{1,2}^{B}$ carry $-yzw, -xzw,-xyz,-xyw$ respectively.
}\label{level k abjm}
\end{table}

\paragraph{Large $N$ ABJM index.}

As shown in Section~3 of \cite{Kim:2009wb}, the large-$N$ ABJM index factorizes as
\begin{align}\label{abjm lNindex}
    I=I^{(0)}(y_0, y_1, y_2, y_3)I^{(+)}(y_0, y_1, y_2, y_3)I^{(-)}(y_0, y_1, y_2, y_3) 
\end{align}
where $I^{(0)}$ denotes the index contribution from single letters carrying  zero monopole charge, and $I^{(\pm)}$ denote the index contributions from single letters carrying positive/negative monopole charges. These indices are matched with zero, positive and negative KK momenta of graviton along the fiber direction from the bulk side in \cite{Kim:2009wb}.

The large $N$ index is given as
\begin{align}
    I^{(0)}=\prod_{n=1}^{\infty}\frac{(1-y_0^{2n})^2}{(1-y_0^{n}y_1^n)(1-y_0^{n}y_1^{-n})(1-y_0^{n}y_2^n)(1-y_0^{n}y_2^{-n})}
\end{align}
and
\begin{equation}
\begin{aligned}
I^{(+)} &=\sum_{M_1,M_2=0}^{\infty}
\sum_{\substack{n_1\geq\cdots\geq n_{M_1}>0\\
\tilde n_1\geq\cdots\geq\tilde n_{M_2}>0}}
y_0^{\epsilon_0}
\int dU\,d\tilde U\,
e^{ik(\sum_i n_i\alpha_i-\sum_i\tilde n_i\tilde\alpha_i)}
\\[-2pt]
&\quad\times
\exp\!\left[\sum_{r=1}^{\infty}\frac1r\sum_{i,j}
\left(f_{ij}^{\rm bif}+f_{ij}^{\rm adj}
+\tilde f_{ij}^{\rm adj}\right)\right],
\end{aligned}\label{seok}
\end{equation}
where $n_i$ and $\tilde n_i$ are the GNO charges of the two gauge groups. 
The zero-flux eigenvalues have already been integrated out at the large-$N$
saddle. 
The remaining unitary integrals run over the unbroken gauge groups in the nonzero-flux background. 
In the $r$-th plethystic term, every fugacity and holonomy appearing in the letter functions is understood to be raised to the $r$-th power.

Due to the symmetry, $I^{(-)}$ is defined in a similar way. 
\begin{align}
    I^{(-)}(y_0, y_1, y_2, y_3)=I^{(+)}(y_0,y_1,1/y_2,1/y_3).
\end{align}
For a fixed flux distribution, the Haar measure is
\begin{align}
    dU=\frac{1}{\text{(symmetry)}} \left[ \frac{d\alpha}{2\pi} \right]\prod_{i< j; n_i=n_j} \left[ 2 \sin \frac{\alpha_i - \alpha_j}{2} \right]^2
\end{align}
The measure \(d\tilde U\) is defined analogously.

The Casimir-energy shift $\epsilon_0$ is 
\begin{align}
    \epsilon_0=\sum_{i,j=1}^{N}|n_i-\tilde{n}_j|-\sum_{i<j}|n_i-n_j|-\sum_{i<j}|\tilde{n}_i-\tilde{n}_j|
\end{align}
which is zero if and only if flux distributions are aligned $\{n_i\}=\{\tilde{n}_i\}$.

The effective single-letter indices in the formula above are given by:
\begin{align}
f_{ij}^{\mathrm{bif}} &= \left(y_0^{|n_i-\tilde{n}_j|} - y_0^{|n_i|+|\tilde{n}_j|}\right) \left(f^+ e^{i(\tilde{\alpha}_j-\alpha_i)} + f^- e^{i(\alpha_i-\tilde{\alpha}_j)}\right) \nonumber \\
f_{ij}^{\mathrm{adj}} &= \left[-(1-\delta_{n_i n_j})y_0^{|n_i-n_j|} + y_0^{|n_i|+|n_j|}\right] e^{-i(\alpha_i-\alpha_j)} \nonumber \\
\tilde{f}_{ij}^{\mathrm{adj}} &= \left[-(1-\delta_{\tilde{n}_i \tilde{n}_j})y_0^{|\tilde{n}_i-\tilde{n}_j|} + y_0^{|\tilde{n}_i|+|\tilde{n}_j|}\right] e^{-i(\tilde{\alpha}_i-\tilde{\alpha}_j)}
\end{align}
where
\begin{align}
    f^+=\frac{x+y-xyz-xyw}{1-p},\quad f^-=\frac{z+w-xzw-yzw}{1-p}.
\end{align}
Here $y_0^2=xyzw$.

\subsection{Witten index of BMN quantum mechanics}
As reviewed in subsection \ref{BMN}, BMN quantum mechanics preserves $SO(3)\times SO(6)$ R symmetry. We denote the respective Cartan charges by $j_3$ and $(H_1,H_2,H_3)$ respectively, and the fermionic symmetry generators transform under $(2,4)$ and $(2,\bar 4)$ representations of the R symmetry group. The Cartans $H_i$ can be thought of as generating rotations in three orthogonal planes of $SO(6)$.
\paragraph{Witten index.}
To define BMN Witten index, we choose a pair of complex conjugate supercharges. The BPS states are annihilated by the pair, and therefore by their anticommutator. Following the conventions of \cite{Chang:2024lkw}, the anticommutator of the chosen pair of supercharges is the following: 
\begin{equation}\label{bmnanti}
    \{Q,Q^\dagger\}
    =
    H_{lc}-j_3-\frac{1}{2}(H_1+H_2+H_3)
    \geq 0,
\end{equation}

\footnote{Recall that $H_{lc}$ is the Hamiltonian of the BMN quantum mechanics.}
The BMN Witten index is defined as 
\begin{equation}
    I_{\mathrm{BMN}}(a,b,c)
    =
    \Tr\!\left[
        (-1)^F e^{-\beta\{Q,Q^\dagger\}}
        p^{j_3}a^{H_1}b^{H_2}c^{H_3}
    \right],
    \qquad p=abc,
\end{equation}
where
\begin{equation}
    p=e^{-2\omega},
    \qquad
    a=e^{-\Delta_1},
    \qquad
    b=e^{-\Delta_2},
    \qquad
    c=e^{-\Delta_3},
    \qquad
    2\omega=\Delta_1+\Delta_2+\Delta_3.
\end{equation}
Only states which are annhilated by \eqref{bmnanti} contribute to the index. 

Since, the index should be independent of coupling by the usual arguments, we can work at zero coupling and compute it. At zero coupling, expanding
around a vacuum labeled by \(\{n_k,N_k\}\) and quantizing the quadratic
fluctuations gives the BPS oscillators that are summarized in
Table~\ref{BMN letters} \cite{Dasgupta:2002hx,Dasgupta:2002ru,Kim:2002if}.

\begin{table}[h!]
\centering
\begin{tabular}{c c c}
\hline
letter & \((4\Delta_{\rm BMN},4j_3,2H_1,2H_2,2H_3)\) & fugacity weight \\
\hline
\(\beta\) & \((4j,\,4j,\,0,\,0,\,0)\) & \(e^{-2j\omega}\) \\
\hline
\(x\) & \((4j+2,\,4j,\,2,\,0,\,0)\) & \(e^{-2j\omega-\Delta_1}\) \\
    & \((4j+2,\,4j,\,0,\,2,\,0)\) & \(e^{-2j\omega-\Delta_2}\) \\
    & \((4j+2,\,4j,\,0,\,0,\,2)\) & \(e^{-2j\omega-\Delta_3}\) \\
\hline
\(\chi\) & \((4j+3,\,4j,\,1,\,1,\,1)\) &
\(e^{-2j\omega-\frac12(\Delta_1+\Delta_2+\Delta_3)}\) \\
\hline
\(\eta\) & \((4j+1,\,4j,\,1,\,1,\,-1)\) &
\(e^{-2j\omega-\frac{\Delta_1}{2}-\frac{\Delta_2}{2}+\frac{\Delta_3}{2}}\) \\
     & \((4j+1,\,4j,\,1,\,-1,\,1)\) &
\(e^{-2j\omega-\frac{\Delta_1}{2}+\frac{\Delta_2}{2}-\frac{\Delta_3}{2}}\) \\
     & \((4j+1,\,4j,\,-1,\,1,\,1)\) &
\(e^{-2j\omega+\frac{\Delta_1}{2}-\frac{\Delta_2}{2}-\frac{\Delta_3}{2}}\) \\
\hline
\(Q\) & \((1,\,-2,\,1,\,1,\,1)\) & -- \\
\hline
\end{tabular}
\caption{BPS oscillators satisfying
\(\Delta_{\rm BMN}=j_3+\frac12(H_1+H_2+H_3)\).
The bosonic oscillators are \(\beta\) and \(x\), while \(\chi\) and \(\eta\)
are fermionic. The allowed range of \(j\) depends on the pair of vacuum
blocks connected by the oscillator.}
\label{BMN letters}
\end{table}

Combining these modes, the single-letter index for fluctuations between
blocks of dimensions \(N_k\) and \(N_l\) is
\cite{Dasgupta:2002hx,Chang:2024lkw}
\begin{equation}\label{bmn single letter}
\iota_{kl}(\Delta_i)
=
-\sum_{j=\frac12|N_k-N_l|}^{\frac12(N_k+N_l)-1}
e^{-j(\Delta_1+\Delta_2+\Delta_3)}
\prod_{i=1}^{3}\left(1-e^{-\Delta_i}\right)
+\delta_{N_k,N_l}.
\end{equation}
The finite range of \(j\) reflects the truncation of matrix spherical
harmonics for finite-dimensional matrix blocks.

The Witten index for a fixed vacuum is therefore
\begin{equation}\label{bmn index}
I_{\{n_k,N_k\}}
=
\int\prod_{k=1}^{K}[dU_k]\,
\exp\!\left[
\sum_{r=1}^{\infty}\frac{1}{r}
\sum_{k,l=1}^{K}
\iota_{kl}(r\Delta_i)\,
\Tr\!\left(U_k^{\dagger r}\right)
\Tr\!\left(U_l^r\right)
\right],
\end{equation}
where \([dU_k]\) is the \(U(n_k)\) Haar measure and
\(\iota_{kl}(r\Delta_i)\equiv
\iota_{kl}(r\Delta_1,r\Delta_2,r\Delta_3)\).

\section{Testing triality by supersymmetric indices}
\label{abjm bmn main}

Having reviewed the index computations in ABJM theory and the BMN quantum mechanics, we now test
the BMN triality by matching the supersymmetric indices of the two theories.
Furthremore, we conclude the section by proposing a microscopic dictionary between ABJM monopole operators and BMN vacua, together with a map between excitations around
the corresponding backgrounds.

\subsection{Triple scaling limit of the ABJM superconformal index}

In this subsection, we evaluate the ABJM superconformal index in the
triple-scaling regime introduced above. We will show that, after normalizing
by the zero-monopole sector, the large-charge expansion of the ABJM index
organizes itself as a grand-canonical generating function for the BMN
indices. More precisely, we introduce
\begin{equation}\label{fugacity mapping}
    q=x^k,
    \qquad
    a=\frac{y}{x},
    \qquad
    b=xz,
    \qquad
    c=xw,
    \qquad
    |q|,|a|,|b|,|c|<1,
\end{equation}
and take \(k\to\infty\) while keeping \(q,a,b,c\) fixed. Thus
\(x=q^{1/k}\to1\), while sectors with angular momentum
\(J=km\) retain a finite weight \(q^m\). This identification of fugacities implies the following identification between BMN and ABJM charges:
\begin{equation}
    (H_1,H_2,H_3)=(J_2,-J_3,-J_4)
\end{equation}

With this prescription, the proposed correspondence takes the form
\begin{align}\label{BMN expansion}
    \frac{I_{\text{ABJM}}}{I^0_{\text{ABJM}}}
    \;\longrightarrow\;
    \sum_{m=0}^{\infty}
    q^m I_{\text{BMN},U(m)}(a,b,c).
\end{align}
The limit in \eqref{BMN expansion} is understood coefficientwise in the
fugacities: at any fixed order in \(q,a,b,c\), the contributions from
finite-\(k\) residues and the sectors discarded below are pushed to
parametrically higher order as \(k\to\infty\). In this sense, the
triple-scaled ABJM index provides a grand-canonical generating function
for the finite-rank BMN indices.

Operationally, we obtain \eqref{BMN expansion} by starting from the
large-\(N\) ABJM index \eqref{seok} and subsequently taking
\(k\to\infty\) at fixed \(q=x^k\). This sequential procedure probes the
parametric regime \(N\gg k\), as required by the triple scaling
\begin{equation}
    N,k,J_1\to\infty,
    \qquad
    \frac{N}{k^2}=\text{fixed},
    \qquad
    m=\frac{J_1}{k}=\text{fixed}.
\end{equation}
Strictly speaking, using the large-\(N\) result before taking the
large-\(k\) limit use the fact that the large-\(N\) expansion is sufficiently uniform in the regime \(k\ll N\) \cite{Kim:2009wb}. We will therefore interpret \eqref{BMN expansion} as the stabilized, coefficientwise statement obtained in this ordered limit.

For convenience, we rewrite the positive-monopole contribution to the large-$N$ ABJM index as
\begin{equation}
\begin{aligned}
I^{(+)}(y_0, y_1, y_2, y_3) &=  \sum_{M_1, M_2=0}^{\infty} \sum_{\substack{n_1 \geq \dots \geq n_{M_1} > 0 \\ \tilde{n}_1 \geq \dots \geq \tilde{n}_{M_2} > 0}}  y_0^{\epsilon_0} \\
& \times \int dU d\tilde{U}  e^{ik(\sum n_i \alpha_i - \sum \tilde{n}_i \tilde{\alpha}_i)}  \exp \left[ \sum_{n=1}^\infty \frac{1}{n} \left( \sum_{i,j} f^{\text{bif}}_{ij} + \sum_{i,j} f^{\text{adj}}_{ij} + \sum_{i,j} \tilde{f}^{\text{adj}}_{ij} \right) \right]
\end{aligned}
\end{equation}
where
\begin{align}
f_{ij}^{\mathrm{bif}} &= \left(y_0^{|n_i-\tilde{n}_j|} - y_0^{|n_i|+|\tilde{n}_j|}\right) \left(f^+ e^{i(\tilde{\alpha}_j-\alpha_i)} + f^- e^{i(\alpha_i-\tilde{\alpha}_j)}\right) \nonumber \\
f_{ij}^{\mathrm{adj}} &= \left[-(1-\delta_{n_i n_j})y_0^{|n_i-n_j|} + y_0^{|n_i|+|n_j|}\right] e^{-i(\alpha_i-\alpha_j)} \nonumber \\
\tilde{f}_{ij}^{\mathrm{adj}} &= \left[-(1-\delta_{\tilde{n}_i \tilde{n}_j})y_0^{|\tilde{n}_i-\tilde{n}_j|} + y_0^{|\tilde{n}_i|+|\tilde{n}_j|}\right] e^{-i(\tilde{\alpha}_i-\tilde{\alpha}_j)}
\end{align}
and
\begin{align}
    f^+=\frac{x+y-xyz-xyw}{1-p},\quad f^-=\frac{z+w-xzw-yzw}{1-p}, \quad y_0^2=p=xyzw.
\end{align}

Taking the large-$k$ limit further simplifies the index. For fixed
fugacities, sectors in which the GNO-charge
multisets of the two gauge groups are not aligned carry suppressing factors
of order $y_0^{\mathcal{O}(k)}$ and therefore vanish as $k\to\infty$
\cite{Kim:2009wb}. Thus, up to Weyl permutations, the condition
\begin{equation}
    \{n_i\}=\{\widetilde n_i\}
\end{equation}
is not imposed separately but emerges automatically from the large-$k$
limit: only the aligned monopole sectors survive at leading order.

Moreover, the negative-monopole contribution $I^{(-)}$ decouples.
Negative-monopole states carry leading fugacity weights of order $z^k$ or
$w^k$. Since $\lvert b\rvert,\lvert c\rvert<1$ while $x\to1$, these
contributions vanish as $k\to\infty$, and hence $I^{(-)}\to1$.
Consequently, the large-$N$ index reduces to
\begin{equation}
    I_{\mathrm{ABJM}}
    \longrightarrow
    I_{\mathrm{ABJM}}^{(0)}I_{\mathrm{ABJM}}^{(+)}.
\end{equation}
In view of \eqref{BMN expansion}, we may therefore restrict
$I^{(+)}$ to the automatically selected aligned sectors
$\{n_i\}=\{\widetilde n_i\}$.

Next, we integrate out the eigenvalues $\alpha_i$ by applying Cauchy's theorem to the poles strictly bounded within the unit circle, $\vert{}e^{i\alpha_i}\vert{} < 1$. 
These poles arise either from the bifundamental interactions, which yield poles of the form
\begin{align}
    e^{i\alpha_i} \in \left(y_0^{\vert{}n_i-\tilde{n}_j\vert{}} - y_0^{\vert{}n_i\vert{}+\vert{}\tilde{n}_j\vert{}}\right)f^+ e^{i\tilde{\alpha}_j}
\end{align}
or from the adjoint interactions, which yield poles of the form
\begin{align}
    e^{i\alpha_i} = y_0^{\vert{}n_i\vert{}+\vert{}n_j\vert{}} e^{i\alpha_j}
\end{align}

In the large $k$ limit, the integral is heavily controlled by the phase factor $e^{ik\sum_i n_i(\alpha_i-\tilde{\alpha}_i)}$.
Consequently, only a specific subset of poles contributes to the index, as the remaining poles are exponentially suppressed.
Since $x$ is the dominant fugacity in $f^{+}$ under the scaling
\eqref{fugacity mapping}, the leading interaction term
$\sum_{i,j} x e^{i(\widetilde{\alpha}_j-\alpha_i)}$ dominates the
holonomy integral and aligns the eigenvalues of the two gauge groups:
\begin{align}
    e^{i\alpha_j} = x e^{i\tilde{\alpha}_j}
\end{align}
Any additional bifundamental contributions rely on subleading fugacities and are therefore heavily suppressed, scaling as $\mathcal{O}(\vert{}y/x\vert{}^{k}) \ll 1$. In a similar spirit, contributions from the adjoint poles, as well as mixed residues involving both types of poles, are exponentially suppressed.

We therefore evaluate the physical integrand at the bifundamental pole,
\begin{align}
    \alpha_j = \tilde{\alpha}_j - i \ln x.
\end{align}
We show below that, if one integrates out the $\alpha_i$ eigenvalues using only the bifundamental poles, one reproduces the BMN index. 
\begin{itemize}
\item \textbf{Contour residue ($i=j$ subtraction):} We pick up the simple pole from the $i=j$ term in the $f^+$ bifundamental Plethystic Exponential. Taking the leading term $x$ from $f^+$, the PE expands into an infinite product:
$$\exp\left[ \sum_{m=1}^\infty \frac{1}{m} \sum_{i,j} x^m e^{im(\tilde{\alpha}_j - \alpha_i)} \right] = \prod_{i,j} \frac{1}{1 - x e^{i(\tilde{\alpha}_j - \alpha_i)}}$$
Notice that this product does contain the $i=j$ diagonal terms. To evaluate the integral over ${\alpha}_i$, we pick up the simple pole from the $i=j$ pairing This yields the diagonal subtraction: $-\delta_{n_i,n_j}\delta_{i,j}$.
\item \textbf{The phase factor} evaluated at the localized saddle:
$$\exp\left[ik \sum_i n_i (\alpha_i - \tilde{\alpha}_i)\right] \longrightarrow \exp\left[k \ln x \sum_i n_i\right] = x^{k \sum_i n_i}$$
\item \textbf{The $f^+$ bifundamental:}
$$f_{ij}^+ e^{i(\tilde{\alpha}_j - \alpha_i)} \longrightarrow \frac{f_{ij}^+}{x} e^{i(\tilde{\alpha}_j - \tilde{\alpha}_i)}$$
\item \textbf{The $f^-$ bifundamental:}
$$f_{ij}^- e^{i(\alpha_i - \tilde{\alpha}_j)} \longrightarrow x f_{ij}^- e^{i(\tilde{\alpha}_i - \tilde{\alpha}_j)}$$
By swapping dummy indices $i \leftrightarrow j$, this equates to $x f_{ji}^- e^{i(\tilde{\alpha}_j - \tilde{\alpha}_i)}$.
\item \textbf{The $f^{\text{adj}}$ 1-loop determinant:}
Because $\tilde{f}^{\text{adj}}$ couples $\tilde{\alpha}_i$ to $\tilde{\alpha}_j$, the uniform shift $+i \ln x$ cancels:
$$f^{\text{adj}}_{ij} e^{-i({\alpha}_i - {\alpha}_j)} \longrightarrow f^{\text{adj}}_{ij} e^{-i(\tilde{\alpha}_i - \tilde{\alpha}_j)} = f^{\text{adj}}_{ji} e^{i(\tilde{\alpha}_j - \tilde{\alpha}_i)}$$
\item \textbf{Additional Haar measure ($i \neq j$ subtraction):} The Haar measure of the second gauge group $d\tilde{U}$ identically reduces to $dU$. Exponentiating this surviving measure elevates it into the Plethystic Exponential, acting as a fermionic ghost that subtracts the off-diagonal modes within the unbroken blocks: $-\delta_{n_i,n_j}(1-\delta_{i,j})$.
\end{itemize}

Assembling the full effective action requires summing the $f^{\pm}$ matter, $f^{\text{adj}}$, $\tilde{f}^{\text{adj}}$, and combining the exact shifts from the measure and residue, $\left[-\delta_{n_i,n_j}(1-\delta_{i,j})\right] + \left[-\delta_{n_i,n_j}\delta_{i,j}\right] = -\delta_{n_i,n_j}$. This reconstructs the identity matrix subtraction for the broken subgroup:
\begin{align}
    f_{ij}^{\text{eff}} = \left(y_0^{|n_i-n_j|} - y_0^{|n_i|+|n_j|}\right)\left( \frac{f_{ij}^+}{x}  + x f_{ji}^- \right)- \delta_{n_i, n_j} + 2f_{ij}^{\text{adj}}
\end{align}
where flux alignment enforces $f_{ij}^{\text{adj}} = \tilde{f}_{ij}^{\text{adj}}$.
The effective single letter index after integrating out $\alpha_i$ is 
\begin{align}
    &\left(y_0^{|n_i-n_j|} - y_0^{|n_i|+|n_j|}\right)\left(\frac{1+y/x-yz-yw}{1-p}+\frac{x(z+w-xzw-yzw)}{1-p}\right)-\delta_{n_i,n_j}\nonumber\\
    &+\left[-(1-\delta_{n_i n_j})y_0^{|n_i-n_j|} + y_0^{|n_i|+|n_j|}\right]+\left[-(1-\delta_{{n}_i {n}_j})y_0^{|n_i-{n}_j|} + y_0^{|{n}_i|+|{n}_j|}\right]\nonumber\\
    =&\delta_{n_i,n_j}-\left(y_0^{|n_i-{n}_j|} - y_0^{|n_i|+|{n}_j|}\right)\left(-\frac{1+y/x-yz-yw}{1-p}-\frac{x(z+w-xzw-yzw)}{1-p}+2\right)\nonumber\\
    = &\delta_{n_i,n_j}-\frac{y_0^{|n_i-{n}_j|} - y_0^{|n_i|+|{n}_j|}}{1-p}(1-a)(1-b)(1-c)\,.
\end{align}
Remarkably, this  coincides with the single-letter index of the BMN matrix model \eqref{bmn single letter} under the identification $a=y/x$, $b=xz$, and $c=xw$, as advertised in the introduction.

Summing over the monopole sectors, the ABJM index in the triple-scaling limit becomes\footnote{Strictly speaking, the derivation only requires $N,k\to\infty$ with $N\gg k$.}
\begin{align}
\frac{I_{\text{ABJM}}}{I^0_{\text{ABJM}}}\longrightarrow\sum_{\{n_l\}~\text{finite}}q^{\sum_l ln_l} \int \prod_{l:n_l>0} [dU_l] \exp \left( \sum_{r=1}^\infty \frac{1}{r} \sum_{l,l'=1}^K f_{ll'}^{\rm eff}(r\Delta) \text{Tr} U_{l}^{\dagger r} \text{Tr} U_{l'}^r \right),
\end{align}
The right-hand side agrees precisely with the grand-canonical sum of the BMN indices in \eqref{bmn index}. This equality constitutes the main technical result of this paper. In the following subsections, we examine the microscopic origin of the correspondence by studying simplifying limits and explain how this agreement emerges at the level of the underlying degrees of freedom.

\subsection{Half-BPS states of ABJM theory and BMN vacua}

In this subsection, we restrict our attention to half-BPS states in ABJM theory and establish a relation between its half-BPS monopole operators and the vacua of BMN quantum mechanics. As we argue below, the BMN vacua correspond one-to-one to the ABJM monopole sectors whose two gauge groups carry identical GNO charges.

As reviewed in Section \ref{BMN}, the vacua of the \(U(m)\) BMN matrix model are labelled by the partitions of $m$. According to our conjecture, the \(U(m)\) BMN
quantum mechanics arises as an effective description of the ABJM sector with
total monopole flux $m$ in each gauge group. From the ABJM point of view, a
partition of $m$ is realized by distributing the total flux among the
eigenvalues of the field strength. Thus a flux sector is specified by ordered
nonzero GNO charges
\begin{equation}
  N_1\geq N_2\geq \cdots \geq N_s>0,
  \qquad
  \sum_{a=1}^{s} N_a=m .
\end{equation}
The remaining eigenvalues carry zero flux. The choice of which eigenvalues carry which charges are gauge equivalent, and the only gauge invariant data is the set of integers $N_i$.

The same data may be represented by a Young tableau with \(N_a\) boxes in the
\(a\)-th row. The corresponding bare monopole operator is not gauge invariant
by itself. Because of the Chern--Simons Gauss law, a monopole carrying total
flux $m$ also carries baryonic charge \(km\) \cite{Aharony:2008ug}. Therefore
it must be dressed by matter fields with total baryonic charge \(km\).

More precisely, a monopole with GNO charges \(N_a\) transforms under gauge
transformations in a representation whose Young tableau has \(kN_a\) boxes in
the \(a\)-th row \cite{Klebanov:2009sg}. To obtain a gauge-invariant operator,
one must dress the monopole by matter fields transforming in the conjugate
representation. Since in this paper we consider a boost along the \(J_1\)
direction of \(S^7/\mathbb{Z}_k\), we use the scalar field \(A_1\), which carries
one unit of \(J_1\) charge and hence one unit of baryonic charge.

Because the operator creates magnetic flux, it is most naturally described as a
dressed 't Hooft disorder operator rather than as an ordinary local polynomial
operator. In the sense explained in \cite{Aharony:2008ug}, we may write it
schematically as a monopole operator dressed by suitable powers of \(A_1\).

To make contact with the BMN matrix model, suppose that the equal GNO charges
have multiplicities \(n_\ell\), so that \(n_\ell\) eigenvalues carry flux
\(\ell\). Then
\begin{equation}
  \sum_{\ell\geq 1} \ell\, n_\ell = m .
\end{equation}
This flux background breaks the relevant gauge symmetry into a subgroup $\prod_l U(n_l)\times U(n_l)$. Schematically, the vacuum operator may be
written as
\begin{equation}\label{vacuumop}
  T_{\{n_\ell\}}
  \prod_{\ell\geq 1}
  \left[
    \det\!\left(A_1\right)_{n_\ell\times n_\ell}
  \right]^{kl} .
\end{equation}

\footnote{More appropriately, since the matter fields are bifundamental under the unbroken gauge group, the determinant is of the following form:
$$\epsilon_{a_1,\ldots a_{n_l}}^{\hat b_1\ldots \hat b_{n_l}}A^{a_1}_{\hat b_1}\ldots A^{a_{n_l}}_{b_{n_l}}$$} 
Here \(T_{\{n_\ell\}}\) denotes the 't Hooft disorder operator creating the
corresponding flux sector. The determinant factors make the operator invariant
under the unbroken block gauge group.\footnote{The fully gauge-invariant
operator is obtained by averaging this representative over the full
\(U(N)\times U(N)\) gauge orbit.}

We identify this ABJM dressed monopole operator with the BMN vacuum labelled by
the same partition of $m$. The unbroken gauge group on the ABJM side though is square of the unbroken gauge group in the corresponding BMN vacuum.

For example, if one eigenvalue carries all $m$ units of flux and the remaining
eigenvalues carry zero flux, the partition is \((m)\). This corresponds to the
irreducible BMN vacuum. In ABJM the monopole background isolates a single
fluxed eigenvalue, leaving a \(U(1)\times U(1)\) factor associated with the fluxed block.

On the other hand, if $m$ eigenvalues each carry one unit of flux, the
partition is \((1^m)\). This corresponds to the trivial BMN vacuum. The ABJM
operator preserves a \(U(m)\times U(m)\) block associated with the fluxed eigenspace.

In the bulk description, the GNO data encode groups of M2-brane dual giants in
\(AdS_4\). A part \(N_a\) of the partition corresponds to a bound state carrying
\(N_a\) units of dual-giant charge. This agrees with the BMN picture, where the
same partition labels the grouping of fuzzy spheres in the pp-wave vacuum.

To quantify the above, let us consider the half-BPS index of ABJM theory \cite{Sheikh-Jabbari:2009vjj}. At large $N$, the positive-charge single-particle contribution is
\begin{align}
I_{sp}=\frac{x^k}{1-x^k}.
\end{align}
where $x$ is the fugacity for charge $J_1$. The states contributing to this index are precisely the half-BPS monopole operators considered above. 

Multi-particling the single-particle index by applying the plethystic exponential to this specific term reproduces the half BPS index.
One can organize it by the monopole charge $m$ as follows:
\begin{equation}
 \operatorname{PE}\!\left[I_{\rm sp}^{1/2\text{-BPS}}\right]
 =\prod_{n=1}^{\infty}\frac{1}{1-x^{kn}}
 =\sum_{m=0}^{\infty}p(m)x^{km},
\end{equation}
where \(p(m)\) is the number of integer partitions of $m$. This agrees with the number of vacua of the \(U(m)\) BMN matrix model, where $I_{\text{BMN},U(m)}$ in \eqref{BMN expansion} is identified as $p(m)$.
\subsection{Two-fugacity index}

Let us now consider a more refined version of the ABJM index to get a better idea about the dictionary between the operators of ABJM theory and excitations of BMN in the given vacuum.

We retain two of the four fugacities of the fully refined index. 
We may retain either \((x,y)\), which have the same baryonic charge, or
\((x,w)\), which have opposite charges. These choices excite
\((A_1,A_2)\) and \((A_1,B_2)\), respectively.\footnote{The first
specialization preserves \(1/3\) of the supersymmetry and the second
\(1/6\).} All contributing letters in these specializations are bosonic, so
the index equals the corresponding partition function.

For the first case, where the excitations are restricted to $A_1$ and $A_2$, the total number of $A_1$ and $A_2$ must be a multiple of $k$. Consequently, the single-particle index for the $\mathbb{Z}_k$ orbifold projection is given by
\begin{align}
    I_{\text{sp},\mathbb{Z}_k} = \sum_{\substack{i,j \ge 0 \\ i+j \in k\mathbb{Z}> 0}} x^i y^j.
\end{align}
Taking the plethystic exponential of this single-particle index yields the full multi-particle index:
\begin{align}
    I_{\text{ABJM}}=\prod_{i,j\text{~and~} i+j\in k\mathbb{Z}{> 0}}\frac{1}{1-x^iy^j}.
\end{align}
This can be reorganized as
\begin{align}\label{indexg1}
    I_{\text{ABJM}}=\prod_{n=1}^{\infty}\prod_{j=0}^{kn}\frac{1}{1-x^{kn}(y/x)^j}.
\end{align}
For \(|x|>|y|\) and \(k\to\infty\), this expression counts fluctuations
around the large \(A_1\) background. Their relative fugacity is \(a=y/x\).

Note that the BMN index with the fugacity $a$ is given as
\begin{align}\label{bmnindex2}
\sum_{m=0}^\infty x^{km} I_{\text{BMN}}(m; a) &= \sum_{m=0}^{\infty} x^{km}\sum_{\sum n_l N_l=m}\left(\prod_{l=1}^K \prod_{j=1}^{n_l} \frac{1}{1 - a^j} \right)\nonumber\\
&=\prod_{n=1}^\infty\prod_{j=0}^\infty \frac{1}{1-x^{kn} a^j}.
\end{align}

Similarly, in the second case where we consider $A_1$ and $B_2$ excitation,
the multi-particle index becomes
\begin{align}\label{indexg2}
    I_{\text{ABJM}}=\prod_{j=1}^{\infty}\frac{1}{1-(xw)^j}\prod_{n=1}^{\infty}\prod_{j=0}^{\infty}\frac{1}{1-x^{kn}(xw)^j}\prod_{n=1}^{\infty}\prod_{j=0}^{\infty}\frac{1}{1-w^{kn}(xw)^j}.
\end{align}
For $\vert{}x\vert{} > \vert{}w\vert{}$, the third factor in \eqref{indexg2} is exponentially suppressed as $k \to \infty$, while the second reduces to \eqref{bmnindex2} upon identifying $a = xw$. The first factor equals the contribution to the index arising from the zero-monopole single-letter partition function.
This is consistent with the identification \(c=xw\) used in equation \eqref{fugacity mapping}, because the BMN
index is symmetric under permutations of \(a,b,c\).

What these indices reveal is that, despite the two ABJM limits preserving different subgroups of the global symmetry and yielding distinct finite-$k$ partition functions, they both converge to the identical BMN expansion in the infinite-$k$ limit. 
This convergence is an algebraic manifestation of the global symmetry enhancement from the $U(3)\subset SU(4)$ symmetry of the ABJM theory to the full $SO(6)$ R-symmetry of the BMN matrix model. 
Physically, this enhancement emerges because the large-$k$ limit provides an infinitely large background charge.
Consequently, one can insert arbitrarily many relative fluctuation factors, such as $A_1^{-1}A_2$ (in the first case) or $A_1 B_2$ (in the second case), without incurring a topological penalty, thereby restoring the unbroken rotational symmetries among the matrix model scalars.

\subsection{Dictionary of excitations}\label{exc}
Based on the discussions in the preceding subsections,  we now propose a dictionary between the operators that correspond to excitations on top of the monopole operators and the BMN excitations. 

The excitations should be invariant under the unbroken part of the gauge group and respect the Gauss law condition (total baryon charge should still be $mk$). Since the determinant operator above already carries $mk$ baryonic charge, the excitations should be neutral. For concreteness, we specialize to the trivial BMN vacuum in this paper but generalization to the other representations is straightforward.

\subsubsection*{$SO(6)$ Scalars}

The ABJM scalars $A_{1,2}$ and $\bar B_{1,2}$ carry positive baryon charge and their complex conjugates carry negative baryon charge. 
Consider the following six operators:

\begin{equation}\begin{split}
   &  (A_2)^{a}_{\hat b} (A_1^{-1})^{\hat b}_{c},~  (\bar B_{1,2})^a_{\hat b}(A_1^{-1})^{\hat b}_c \\
& (A_1)^a_{\hat b} (\bar A_2)^{\hat b}_c, ~(A_1)^a_{\hat b} (B_{1,2})^{\hat b}_c
\end{split}    
\end{equation}

These carry no baryon charge.
We propose that these six fields can be identified with six scalars of BMN quantum mechanics. A gauge invariant operator can be formed by taking sums of powers of traces obtained by these operators. Note that the operators are invariant under the right half of the unbroken gauge group since the hatted indices are contracted and hence behave like adjoint under the left half of the unbroken gauge group, same as in BMN quantum mechanics. Note that at finite $k$, the excitations in upper line behave very differently from those in the lower line, because for the excitations in the first line, there is an upper cutoff on how many time one can excite them, since they use up $A_1$ insertions from the determinant. However this cutoff goes to infinity in large $k$ limit, making all the six scalars identical. This is also the reason for the enhancement of R symmetry from $U(3)\subset SU(4)$ to $SO(6)$ in this sector. 

\subsubsection*{$SO(3)$ scalars}

The three $SO(3)$ scalars of BMN quantum mechanics are proposed to be dual to 
\begin{equation}
    (D_i A_1) A_1^{-1}
\end{equation}
where $D_i$ are the three covariant derivatives and the hatted indices are contracted just as above. 

\subsubsection*{Fermions}

Recall that fermions in ABJM theory transform in fundamental irrep of $SU(4)$ and their conjugates in antifundamental irrep of $SU(4)$. One difference from bosons being that fundamental fermions carry negative baryonic charge as opposed to fundamental bosons carrying positive baryonic charge\footnote{This is what makes them chiral spinor of $SO(8)$ when there is symmetry enhancement as opposed to bosons which become vectors of $SO(8)$.}. The $SU(4)$ fundamental fermions are labelled as $\psi_I$ in our notation. Under the enhanced $SO(8)$ symmetry, all these behave as an antichiral spinor of $SO(8)$.

Since the fundamental fermions $\psi_I$ carry negative baryon charge, their cartans under $SO(8)$ would be:
$$\left(\frac{-1}{2},\frac{-1}{2},\frac{-1}{2},\frac{1}{2}\right),\left(\frac{-1}{2},\frac{-1}{2},\frac{1}{2},\frac{-1}{2}\right),\left(\frac{-1}{2},\frac{1}{2},\frac{-1}{2},\frac{-1}{2}\right),\left(\frac{1}{2},\frac{-1}{2},\frac{-1}{2},\frac{-1}{2}\right)$$

Similarly $\bar \psi^I$ would carry negative cartan charges to above.

We propose that the BMN fermions can be obtained from ABJM fermions by taking traces of the following:
\begin{equation}
    A_1 \psi_I,~ \bar\psi^I A_1^{-1}
\end{equation}

These fermions rearrange into chiral and antichiral fermions of BMN quantum mechanics as follows. The fermions with $J_1$ Cartan charge negative tranform in the chiral representation of $SO(6)$ and the ones with positive $J_1$ in the antichiral of $SO(6)$ in BMN quantum mechanics. Similar to bosons, these fermions also behave identically in the large $k$ limit, indication both, the enhancement of global symmetry to $SO(6)$ and enhancement of supersymmetry as well.

\section{Correspondence of cohomology classes}\label{dictionary}

In this section, we relate the classical $Q$-cohomology of the BMN matrix model to that of ABJM theory in a fixed large-charge monopole sector. When we say $Q$-cohomology, we refer to the cohomology generated by a special choice of supercharges in both theories, introduced in section \ref{review}. 

In Subsection \ref{Qcor}, we relate the BPS letters of the two theories and compare the respective actions of supercharges on these letters.
Then, in Subsection \ref{monback}, we describe the covering complexes for half-BPS monopole backgrounds.
In Subsection \ref{finm}, we explain finite-rank BMN cohomology using ABJM covering complexes.

\subsection{BPS letters and $Q$-action correspondence}\label{Qcor}
We focus on the trivial BMN vacuum preserves \(U(m)\) gauge group. Following subsection \ref{exc}, the proposed correspondence between BMN BPS
oscillators, listed in Table \ref{BMN letters}, and finite excitations in the ABJM monopole background is
\begin{align}\label{dic}
    x_I
    &\sim
    A_1B_1,\ A_1B_2,\ A_2A_1^{-1},
    \nonumber\\
    \eta^I
    &\sim
    -\psi_2^BA_1^{-1},\ \psi_1^BA_1^{-1},\ A_1\psi_1^A,
    \nonumber\\
    \beta
    &\sim
    (DA_1)A_1^{-1}.
\end{align}
Here, we write \(x_I\) for \(x_{0I}\) and \(\eta^I\) for
\(\eta^I_{1/2}\), since in the trivial BMN vacuum \(x_I\) can only have
\(j=0\), whereas \(\eta^I\) can only have \(j=1/2\). Matrix products and
commutators in \eqref{dic} act on the effective monopole block, and
\(A_1^{-1}\) is understood in the determinant-hole sense introduced in \ref{exc}.

We can explicitly check that the map \eqref{dic} intertwines $Q$-actions on both sides. Let us call them $Q_{\rm ABJM}$ and $Q_{\rm BMN}$.
% \begin{equation}
%     \Psi^2=\psi_1^A,
%     \qquad
%     \Psi^3=-\psi_2^B,
%     \qquad
%     \Psi^4=\psi_1^B.
% \end{equation}
The tree-level ABJM supersymmetry transformations following from the superpotential are
\begin{align}
    Q_{\rm ABJM}A_i&=Q_{\rm ABJM}B_i=0,
    \nonumber\\
    Q_{\rm ABJM}\psi_i^A
    &=\frac{4\pi i}{k}\epsilon_{rs}B_rA_iB_s,
    \nonumber\\
    Q_{\rm ABJM}\psi_i^B
    &=-\frac{4\pi i}{k}\epsilon_{rs}A_rB_iA_s,
\end{align}
where \(\epsilon_{12}=1\). Using the map \eqref{dic}, it is easy to see that
\begin{equation}
    Q_{ABJM}x_I=0.
\end{equation}
For the fermionic letters, one finds, up to a common normalization, we find
\begin{align}
&Q_{\rm ABJM}\left(-\psi_2^BA_1^{-1}\right)
\propto [A_1B_2,A_2A_1^{-1}],\nonumber\\
&Q_{\rm ABJM}\left(\psi_1^BA_1^{-1}\right)\propto 
[A_2A_1^{-1},A_1B_1]\nonumber\\
&Q_{\rm ABJM}\left(A_1\psi_1^A\right)\propto [A_1B_1,A_1B_2]
\end{align}
This reproduces the $Q$ action on the fermionic BMN oscillators $\eta^I$:
\begin{equation}
Q_{\rm BMN}\eta^I = \frac12\epsilon^{IJK}[x_J,x_K],
\end{equation}
where the overall factors are absorbed into the normalization of $\eta^I$.

The transformation of \(\beta\) can be checked similarly. The \(Q\)-action
on a derivative letter is
\begin{align}
Q_{\rm ABJM}(DA_i)=\frac{8\pi}{k}\Big(
&A_{[1}\psi^A_{2]}A_i
-A_i\psi^A_{[2}A_{1]}
+\psi^B_{[1}B_{2]}A_i
-A_iB_{[2}\psi^B_{1]}
\Big),
\end{align}
where the brackets denote antisymmetrization of the \(SU(2)\) indices.
Setting \(i=1\), the terms containing \(\psi_2^A\) cancel, leaving
\begin{align}
Q_{\rm ABJM}(DA_1)=\frac{8\pi}{k}\Big(
&-A_2\psi^A_{1}A_1+A_1\psi^A_{1}A_2
+\psi^B_{1}B_2A_1-\psi_2^{B}B_1A_1
\nonumber\\
&-A_1B_2\psi_{1}^B+A_1B_1\psi_2^{B}
\Big).
\end{align}
% On the other hand, defining
% \begin{equation}
%     \mathcal S\equiv\sum_{I=1}^3[x_I,\eta^I],
% \end{equation}
% the dictionary gives
% \begin{align}
% \mathcal S A_1
% ={}&
% A_1B_1\Psi^3-\Psi^3B_1A_1
% +A_1B_2\Psi^4-\Psi^4B_2A_1
% \nonumber\\
% &+A_2\Psi^2A_1-A_1\Psi^2A_2.
% \end{align}
% Comparison with the preceding expression yields
% \begin{equation}
%     Q(DA_1)
%     =
%     -\frac{8\pi}{k}\,\mathcal S A_1.
% \end{equation}
% Since \(Q_{\rm ABJM}A_1=0\), the determinant-hole bookkeeping gives
% \(Q_{\rm ABJM}(A_1^{-1})=0\), and hence
% \begin{equation}
%     Q_{\rm ABJM}\left((DA_1)A_1^{-1}\right)
%     =
%     -\frac{8\pi}{k}
%     \sum_{I=1}^3[x_I,\eta^I].
% \end{equation}
This reproduces the $Q$ action on the $\beta$ oscillator after an overall rescaling:
\begin{equation}
    Q_{\rm BMN}\beta=[x_I,\eta^I].
\end{equation}

Let $\lambda_{\rm triv}=(1^m)$ denote the trivial vacuum of the $U(m)$
BMN matrix model, and let
$\mathcal C^{\rm BMN}_{m,\lambda_{\rm triv}}$ denote its excitation
complex. On the ABJM side, let
$\widetilde{\mathcal C}^{\rm ABJM}_{\lambda_{\rm triv}}$ denote the
intrinsic stabilized complex of finite modifications of the corresponding
$A_1$-polarized monopole background, as defined in
Subsection~\ref{monback}. The dictionary \eqref{dic} defines a map
\begin{equation}
    \Phi_m:
    \mathcal C^{\rm BMN}_{m,\lambda_{\rm triv}}
    \longrightarrow
    \widetilde{\mathcal C}^{\rm ABJM}_{\lambda_{\rm triv}}.
\end{equation}
The preceding calculation establishes
\begin{equation}\label{cochain-map}
    Q_{\rm ABJM}\circ\Phi_m
    =
    \Phi_m\circ Q_{\rm BMN},
\end{equation}
so that $\Phi_m$ is a cochain map.

At the level of the localized stabilized letter algebra, $\Phi_m$ is
injective onto the subalgebra generated by the composites in
\eqref{dic}. Indeed, once $A_1$ is treated as invertible in the
determinant-hole sense, the dictionary can be inverted on its image.
Thus, modulo the matrix identities common to the two $U(m)$
descriptions, the BMN letters obey no additional relations after being
mapped to ABJM. The image of $\Phi_m$ therefore forms a
$Q_{\rm ABJM}$-closed subcomplex of
$\widetilde{\mathcal C}^{\rm ABJM}_{\lambda_{\rm triv}}$.

The agreement of the refined ABJM and BMN indices in the triple-scaling limit, together with the identification of the corresponding vacuum sectors, led us to conjecture that $\Phi_m$ is a quasi-isomorphism:
\begin{equation}
    H^\bullet_{Q_{\rm BMN}}
    \left(
        \mathcal C^{\rm BMN}_{m,\lambda_{\rm triv}}
    \right)
    \simeq
    H^\bullet_{Q_{\rm ABJM}}
    \left(
        \widetilde{\mathcal C}^{\rm ABJM}_{\lambda_{\rm triv}}
    \right).
\end{equation}
We assume this conjecture in the remainder of this section. We emphasize, however, this chain-level injectivity does not by itself establish the quasi-isomorphism on $Q$-cohomology.
Also, even if the proposed triality holds at the quantum level, it does not by itself guarantee that the classical $Q$-cohomologies of the two descriptions coincide. We leave a direct verification of this conjecture for future work.

\subsection{Monopole-background covering complexes}\label{monback}

The total monopole number $m$ is a conserved charge, so the Hilbert space decomposes into fixed-$m$ sectors. The triple-scaling limit is taken within one such sector and retains states whose deviation from a chosen $A_1$-polarized monopole background involves only an $\mathcal O(1)$ number of constituent letters. 

As explained in \eqref{vacuumop}, the corresponding monopole dressed background operators, polarized along the $A_1$ direction, are schematically written as: 
\begin{equation}\label{A1-background}
    \mathcal B_{\lambda,k}^{(1)}
    =
    T_{\{n_\ell\}}
    \prod_{\ell\geq1}
    \left[
        \det(A_1)_{n_\ell\times n_\ell}
    \right]^{\ell k}.
\end{equation}
Here, the partitions $\lambda\vdash m$ label the distinct vacua of the $U(m)$ BMN matrix model. At weak coupling, it is natural to organize the Hilbert space around a given background $\lambda$ into products of neutral multi-particle graviton states and background-specific dressing excitations. By an excitation of $\mathcal B_{\lambda,k}^{(1)}$, we mean a modification that involves only a finite number of constituent letters in the $k\to\infty$ limit. Such modifications include replacing a finite number of $A_1$ fields with other bosonic or fermionic letters, inserting derivatives, or removing a finite number of $A_1$ insertions from the determinant background—the latter operation being denoted symbolically by $A_1^{-1}$.

This notion of excitation is local in the space of the backgrounds \eqref{A1-background}.
For example, consider the operator which is made up of $A_2$ instead of $A_1$:
\begin{equation}\label{A2-background}
    \mathcal B_{\lambda,k}^{(2)}
    =
    T_{\{n_\ell\}}
    \prod_{\ell\geq1}
    \left[
        \det(A_2)_{n_\ell\times n_\ell}
    \right]^{\ell k}.
\end{equation}
It has the same GNO charges as \eqref{A1-background}, but converting
\(\mathcal B_{\lambda,k}^{(1)}\) into
\(\mathcal B_{\lambda,k}^{(2)}\) requires replacing $mk$
copies of \(A_1\) by \(A_2\). This is an \(\mathcal O(k)\), rather than an
\(\mathcal O(1)\), modification. Consequently,
\(\mathcal B_{\lambda,k}^{(2)}\) is not contained in the finite-excitation
space constructed around \(\mathcal B_{\lambda,k}^{(1)}\). Namely infinite $k$ limit produces a local Fock-space patch around
\(\mathcal B_{\lambda,k}^{(1)}\), while the \(A_2\)-polarized state is
pushed to infinite excitation number, and thus not belong to the spectrum.

We separate the universal neutral-graviton complex from the
background-specific dressing complex. In what follows,
$\widetilde{\mathcal C}^{\rm ABJM}_\lambda$ denotes only the intrinsic
stabilized complex of finite modifications of
$\mathcal B_{\lambda,k}^{(1)}$, after the separately factorized neutral
graviton sector has been removed. Correspondingly,
$\mathcal C^{\rm ABJM}_{N,k;\lambda}$ below denotes its finite-$(N,k)$
realization.
Denoting the intrinsic cohomology by
\begin{equation}
    \widehat{\mathcal H}_{\lambda}^{(Q)}
    \equiv
    H_Q\!\left(\widetilde{\mathcal C}^{\rm ABJM}_\lambda\right),
\end{equation}
we expect the decoupling of the spectrum:
\begin{equation}
    \mathcal H_{m,\lambda}^{(Q)}
    \simeq
    \mathcal H_0^{(Q)}
    \otimes
    \widehat{\mathcal H}_{\lambda}^{(Q)},
\end{equation}
where \(\mathcal H_0^{(Q)}\) is the neutral graviton cohomology. Summing
over the partitions of $m$, we obtain the conjectural decomposition
\begin{equation}\label{cohomology-factorization}
    \mathcal H_m^{(Q)}
    \simeq
    \mathcal H_0^{(Q)}
    \otimes
    \bigoplus_{\lambda\vdash m}
    \widehat{\mathcal H}_{\lambda}^{(Q)}.
\end{equation}
The tensor product in \eqref{cohomology-factorization} is understood to
preserve the grading by all protected quantum numbers.

Indeed, it follows that the index is factorized as
\begin{equation}
    I_m
    =
    I_0
    \sum_{\lambda\vdash m}\widehat I_\lambda
    \equiv
    I_0\,\widehat I_m.
\end{equation}
As discussed around \eqref{BMN expansion}, \(\widehat I_m\) agrees with the
index of the \(U(m)\) BMN matrix model, including the sum over its different vacua.
One can interpret the factorization above as decoupling of closed strings and open strins on D0 branes. This is very plausible in the triple-scaling limit because $g_s$ goes to zero and also the excited states of open strings decouple since 'tHooft coupling goes to infinity, and so does string tension.

The realization of this intrinsic stabilized complex at finite $N$ and
$k$ defines a projection
Schematically, the realization of the covering complex at finite $N$ and
\(k\) defines a projection
\begin{equation}
    \Pi_{N,k;\lambda}:
    \widetilde{\mathcal C}^{\rm ABJM}_\lambda
    \longrightarrow
    \mathcal C^{\rm ABJM}_{N,k;\lambda},
    \qquad
    \mathcal J_{N,k;\lambda}
    \equiv
    \ker\Pi_{N,k;\lambda}.
\end{equation}
Universal relations of the determinant-modification algebra are already
included in \(\widetilde{\mathcal C}^{\rm ABJM}_\lambda\), whereas
\(\mathcal J_{N,k;\lambda}\) contains the additional relations that arise
upon realizing the background at finite $N$ and \(k\). Since the
projection is a cochain map, it induces
\begin{equation}
    (\Pi_{N,k;\lambda})_*:
    H_Q\!\left(\widetilde{\mathcal C}^{\rm ABJM}_\lambda\right)
    \longrightarrow
    H_Q\!\left(\mathcal C^{\rm ABJM}_{N,k;\lambda}\right).
\end{equation}
A class in the image of this map has a \(Q\)-closed lift in the stabilized
monopole-background covering complex and is a monotone. By contrast, if a lift \(\widetilde{\mathcal O}\) satisfies
\begin{equation}
    Q\widetilde{\mathcal O}\neq0,
    \qquad
    Q\widetilde{\mathcal O}\in\mathcal J_{N,k;\lambda},
\end{equation}
then its image is \(Q\)-closed only after imposing finite-\(N\) or finite-\(k\) relations, providing a diagnostic for fortuity.\footnote{This condition is not by itself sufficient: the projected class must also be nontrivial and must not admit any \(Q\)-closed lift to the stabilized
covering complex.}

In the standard holographic covering formulation, an ABJM cohomology class at finite $N$ is fortuitous if its \(Q\)-closure
relies on trace relations of the \(U(N)\times U(N)\) gauge group and
therefore fails to remain \(Q\)-closed at sufficiently large $N$
\cite{Chang:2024zqi}. Such classes are known to exist in finite-rank ABJM
theory, including in nontrivial monopole sectors
\cite{Belin:2025hsg,Behan:2025hbx}.

This mechanism is absent in the triple-scaling sector considered here, where
\begin{equation}
    N\gg k\gg1,
    \qquad
    J=mk=\mathcal O(k)\ll N,
\end{equation}
with $m$ held fixed. A finite excitation modifies the
\(\mathcal O(k)\) constituent fields of the monopole background by only
\(\mathcal O(1)\). Consequently, at every fixed order in the excitation
fugacities, these operators remain parametrically below the degree at which
trace relations of the microscopic \(U(N)\times U(N)\) gauge group become
relevant. Their \(Q\)-closure therefore cannot rely on finite-$N$ ABJM
trace relations. 
\footnote{The construction of covering space is reminiscent of the baryonic covering spaces introduced in Ref.~\cite{Choi:2026faq}. There, baryonic monotones are defined as finite modifications of baryonic backgrounds, possibly dressed by neutral graviton states. 
Because a baryonic operator has conformal dimension of order \(\mathcal O(N)\), it does not belong to the standard fixed-charge holographic covering, and the notion of monotonicity must be generalized accordingly.
In the present setting, by contrast, the monopole-dressed background carries charge \(J=mk=\mathcal O(k)=\mathcal O(\sqrt N)\ll N\). It therefore remains in the ordinary monotone sector with respect to the microscopic rank $N$;
indeed, perturbative supergravity states may themselves carry monopole, or equivalently Kaluza--Klein momentum, charge.}

It follows that the cohomology classes retained in the triple-scaling limit
are monotones with respect to the ambient ABJM rank $N$.
There are no fortuitous ABJM cohomologies, in the standard finite-$N$
sense, within this limiting sector. This statement is understood
coefficientwise at fixed excitation charges; it does not apply to
excitations whose charges are themselves scaled with $N$.

This conclusion does not exclude fortuitous cohomology in the corresponding
BMN matrix model. The effective BMN rank \(m=J/k\) remains finite, and the
trace relations of the \(U(m)\) matrix algebra therefore remain present.
We now use the cochain correspondence constructed in
Subsection~\ref{Qcor} to describe this distinct finite-$m$ effect from
the ABJM perspective.

\subsection{Finite-$m$ BMN fortuity from the ABJM perspective}\label{finm}

We now return to the finite-$m$ trace relations. Consider a rank-independent covering algebra for the BMN excitation complex and its realization at
rank $m$,
\begin{equation}
    \pi_m:
    \widetilde{\mathcal C}_{\rm BMN}
    \longrightarrow
    \mathcal C_{\rm BMN}^{(m)}
    \simeq
    \widetilde{\mathcal C}_{\rm BMN}/\mathfrak I_m,
    \qquad
    \mathfrak I_m\equiv\ker\pi_m.
\end{equation}
The ideal \(\mathfrak I_m\) contains the trace relations of the effective
\(U(m)\) matrix algebra. If a lift
\(\widetilde{\mathcal O}\in\widetilde{\mathcal C}_{\rm BMN}\) satisfies
\begin{equation}
    Q_{\rm BMN}\widetilde{\mathcal O}\neq0,
    \qquad
    Q_{\rm BMN}\widetilde{\mathcal O}\in\mathfrak I_m,
\end{equation}
then \(\pi_m(\widetilde{\mathcal O})\) is \(Q_{\rm BMN}\)-closed only
because of the finite-$m$ trace relations. If it is a nontrivial cohomology, it is a fortuitous cohomology of the \(U(m)\) BMN matrix model.

The simplest known example occurs in the \(U(2)\) BMN model in the trivial vacuum \cite{Chang:2022mjp,Choi:2022caq,Choi:2023znd}. More precisely, the nontrivial cohomology class belongs to the interacting \(SU(2)\) sector, after factoring out the decoupled center-of-mass \(U(1)\) multiplet.
Accordingly, the matrix-valued letters appearing below are understood to be traceless. The rank-two fortuitous operator is
\begin{align}\label{O0}
    \mathcal O_{\rm f}^{(2)}
    =
    \epsilon_{p_1p_2p_3}
    \Tr(x_r\eta^{p_1})
    \Tr(x_s\eta^{p_2})
    \Tr\!\left(\eta^{p_3}[\eta^r,\eta^s]\right).
\end{align}
Here, $\Tr$ is the trace over the $m\times m$ matrix space.

Let \(\widetilde{\mathcal O}_{\rm f}\) denote the corresponding word in the
rank-independent BMN covering algebra. Its \(Q\)-closure at rank two is
equivalently expressed as
\begin{equation}
    Q_{\rm BMN}\widetilde{\mathcal O}_{\rm f}\neq0,
    \qquad
    Q_{\rm BMN}\widetilde{\mathcal O}_{\rm f}
    \in\mathfrak I_2,
\end{equation}
so that
\begin{equation}
    \mathcal O_{\rm f}^{(2)}
    =
    \pi_2\!\left(\widetilde{\mathcal O}_{\rm f}\right)
    \in H_{Q_{\rm BMN}}.
\end{equation}
When the same covering-space word is realized at a larger rank, the
required trace identity is no longer available. For example,
\begin{equation}
    Q_{\rm BMN}
    \pi_4\!\left(\widetilde{\mathcal O}_{\rm f}\right)
    \neq0.
\end{equation}

To express this result in ABJM variables, define the trivial-vacuum monopole
background by
\begin{equation}
    \mathfrak M_m
    \equiv
    \mathcal B_{(1^m),k}^{(1)}
    =
    T_{\{n_1=m\}}
    [\det(A_1)_{m\times m}]^k.
\end{equation}
The ABJM representative corresponding to
\(\mathcal O_{\rm f}^{(2)}\) is schematically
\begin{equation}
    \mathcal O_{{\rm f},{\rm ABJM}}^{(2)}
    =
    \mathfrak M_2\,
    \Phi_2\!\left(\mathcal O_{\rm f}^{(2)}\right),
\end{equation}
where the BMN letters are replaced according to \eqref{dic}. Since
\(\Phi_2\) is a cochain map, this is \(Q_{\rm ABJM}\)-closed. By contrast,
the analogous excitation of the \(m=4\) monopole background,
\begin{equation}
    \mathcal O_{{\rm f},{\rm ABJM}}^{(4)}
    =
    \mathfrak M_4\,
    \Phi_4\!\left(
        \pi_4(\widetilde{\mathcal O}_{\rm f})
    \right),
\end{equation}
is not \(Q_{\rm ABJM}\)-closed.

The rank dependence in this example concerns the effective BMN rank $m$,
not the microscopic ABJM rank $N$. At \(m=2\), the relevant trace identity
is a structural relation of the two-dimensional fluxed block and remains
valid as the ambient rank $N$ is taken to infinity. Consequently,
\(\mathcal O_{{\rm f},{\rm ABJM}}^{(2)}\) is not fortuitous with respect to
$N$. It belongs to the stabilized charge-two monopole-background
cohomology and is therefore a generalized monotone in the ABJM covering
described above.

For a general partition \(\lambda=\{n_\ell\}\), the corresponding BMN
vacuum has unbroken gauge group
\begin{equation}
    G_\lambda=\prod_{\ell\geq1}U(n_\ell),
\end{equation}
and the relevant finite-rank relations depend on the multiplicities
\(n_\ell\). Developing the corresponding covering algebras and extending
the cochain map beyond the trivial vacuum would provide a
systematic framework for classifying BMN cohomology from the ABJM
perspective. We leave this problem for future work.

\section{Asymptotic behavior of BMN index}\label{asymptotic behavior}

In this section, we study the large-$m$ behavior of several index-like quantities associated with the BMN BPS spectrum and examine their physical implications. Our primary focus is the regime in which the excitation charges scale as $\mathcal{O}(m^2)$. At zero coupling, exciting the $\mathcal{O}(m^2)$ free BPS oscillators leads to an expected entropy of order $\mathcal{O}(m^2)$. In what follows, we investigate quantities that probe this spectrum at strong coupling and demonstrate that all of them exhibit severe cancellations, failing to grow as $\mathcal{O}(m^2)$.
\begin{itemize}
\item In Subsection~\ref{sugra bmn}, we study the asymptotic growth of the full BMN index at large $m$ with charges of order $m^2$. Using the proposed triality, we obtain a closed-form expression for the index from the supergraviton index on $\mathrm{AdS}_4 \times S^7/\mathbb{Z}_k$ in the sector carrying $m$ units of Kaluza--Klein momentum. We derive the asymptotic growth of the index analytically and verify the result against direct numerical evaluations of the same quantity. 
%In addition, an independent numerical check confirms that the large-$k$ graviton index matches the explicit BMN index summed over all vacua. 

\item In Subsection \ref{sec:multi_graviton_entropy}, we compute the unsigned partition function of free BPS gravitons in the same positive-momentum sector, and derive its large-$m$ entropy analytically. It corresponds to the BMN partition function in the strong-coupling regime.

\item In Subsection \ref{comp ent}, we compare various quantities computed in the previous two subsections, along with the comparison with the sum over absolute values of the BMN indices in various BMN vacua. This is to get a quantitative understanding of the cancellations in the index among various vacua. The comparison is depicted in a plot in Figure \ref{fig:bmn-graviton-comparison}.

\item In Subsection~\ref{finite kk}, we turn to a parameter regime distinct from the one studied in the rest of the paper. We investigate the ABJM index at finite $k$ and large charge, comparing its behavior with the large-$m$ BMN index. This comparison is physically motivated by the fact that the appropriate large-charge Penrose limit of the unorbifolded $\mathrm{AdS}_4 \times S^7$ geometry yields the uncompactified eleven-dimensional pp-wave background, which is conjectured to be dual to BMN quantum mechanics in a suitable large-rank limit. This is the regime studied by Kovacs, Sato, and Shimada \cite{Kovacs:2013una}. We test this holographic conjecture by comparing their indices.

\end{itemize}

\subsection{Growth of the BMN index}\label{sugra bmn}

As established previously, the BMN index maps to the ABJM index in the triple-scaling limit. At large $N$, the ABJM index agrees order by order with the multiparticle supergravity index on \(AdS_4\times S^7/\mathbb Z_k\) \cite{Kim:2009wb}. 
Assuming this matching extends to the triple-scaling limit, the positive-charge graviton index, after dividing out the neutral factor, should reduce to the $U(m)$ BMN index, so that
\begin{align}
\frac{I_{\rm grav}}{I_{\rm grav}^0}=\sum_{m=0}^{\infty} q^m I_{\textrm {BMN}, U(m)},
\end{align}
where $I_{\rm grav}$ is a supergraviton index on \(AdS_4\times S^7/\mathbb Z_k\).

Building on this, in this subsection we derive a closed-form expression for the full $U(m)$ BMN index and analyze its asymptotic growth at large $m$ for charges of order $m^2$. As a cross-check, we numerically verify that the large-$k$ graviton index carrying KK momentum $m$ matches the complete $U(m)$ BMN index, with the corresponding data presented in Appendix~\ref{num1}. We note that isolating the individual fuzzy sphere vacuum contributions from the ABJM graviton index is nontrivial and lies beyond the scope of this work.

Our analytical results, supported by numerical checks, show that the indicial entropy grows as $\mathcal O(m)$, reflecting substantial boson--fermion cancellations. This differs from the $\mathcal O(m^2)$ behavior reported in earlier studies.

\subsubsection{Closed form expression for the BMN index}\label{abjm grav}

We begin by constructing the graviton index in AdS$_4\times S^7/\mathbb{Z}_k$. To this end, we perform a $\mathbb{Z}_k$ projection on the single-particle index in AdS$_4\times S^7$ \cite{Bhattacharya:2008zy,Bhattacharya:2008bja,Kim:2009wb}:
\begin{equation}\label{spgi}
I_{sp}(x,y,z,w) = \frac{(1-xyz)(1-yzw)(1-zwx)(1-wxy)}{(1-x)(1-y)(1-z)(1-w)(1-p)^2}
- \frac{1-p+p^2}{(1-p)^2}.
\end{equation}

Here $p=xyzw$. We assign a $U(1)_b$ charge of $+1$ to the variables $x, y$ and a charge of $-1$ to $z, w$. The goal is to compute the $\mathbb{Z}_k$ orbifold projection of this index, which restricts the spectrum to states where the total $U(1)$ charge $\ell = N_x + N_y - N_z - N_w$ is an integer multiple of $k$.
The projected single-particle index is formally defined via a discrete Fourier transform over the $k$-th roots of unity, $\omega = e^{2\pi i / k}$:\footnote{This can be seen by noting that when the Baryonic charge is not a multiple of $k$, the sum over $j$ vanishes since the sum over all units of identity is zero.}
\begin{equation}
    I_{\mathbb{Z}_k}(x,y,z,w) = \frac{1}{k} \sum_{j=0}^{k-1} I_{sp}\left(\omega^j x, \omega^j y, \omega^{-j} z, \omega^{-j} w\right) = \sum_{m=-\infty}^{\infty} H_{mk} \,,
\end{equation}
where $H_\ell$ is the generating function restricted to the fixed-charge sector $\ell >0$.
\footnote{To obtain $H_\ell$, we can scale the fugacities as $x,y\to qx, qy$, $z,w\to z/q, w/q$ and do the following contour integral:
$$
H_\ell(x,y,z,w)=\frac{1}{2\pi i}\oint dq\,q^{-\ell-1}I_{sp}(qx,qy,z/q,w/q).$$ It is easy to see that picking up poles $q=1/x$ and $q=1/y$ gives rise to positive powers of Baryon charge and $q=z$ and $q=w$ give rise to negative powers of baryon charge.}
\begin{equation}
\label{eq:masterHc}
    H_\ell = \frac{1}{1-p} \Big[ (1-p^2)K_\ell - O_1 K_{\ell-1} - O_{-1}K_{\ell+1} \Big]\,,
\end{equation}
where $K_\ell$ is the fixed-charge coefficient of the free denominator $[(1-x)(1-y)(1-z)(1-w)]^{-1}$, and
\begin{equation}
    O_1 = xy(z+w) - p(x+y) \,, \quad O_{-1} = zw(x+y) - p(z+w) \,.
\end{equation}

For $\ell$ positive, the coefficients $K_\ell$ are given by the positive poles of the free generating function:
\begin{equation}\label{Kc}
    K_\ell^{(+)} = \frac{x^{\ell+1}}{(x-y)(1-xz)(1-xw)} + \frac{y^{\ell+1}}{(y-x)(1-yz)(1-yw)} \,.
\end{equation}

Now let us take $k\to\infty$ with $q=x^k$ and $a=y/x$, $b=xz$, $c=xw$ fixed inside the unit disk, so $x\to1$ as we did in Section \ref{abjm bmn main}. Since $l$ is multiplet of $k$, the second term is exponentially suppressed and vanishes due to $(\vert{}y\vert{}/\vert{}x\vert{})^{l} \to 0$. From \eqref{eq:masterHc} and \eqref{Kc}, we obtain
\begin{equation}\label{graviton Hc}
H_\ell \approx x^\ell \frac{(1-ab)(1-bc)(1-ca)}{(1-p)(1-a)(1-b)(1-c)},
\end{equation}
where $a = y/x$, $b = xz$, and $c = xw$, so $p=abc$.
The formula for negative charges $\ell < 0$ follows immediately via reflection. Setting $\ell = -\vert{}\ell\vert{}$, we find:
\begin{equation}
H_{-|\ell|}(x,y,z,w) = H_{|\ell|}(z,w,x,y).
\end{equation}
Furthermore, contributions from negative charges vanish in the $k \to \infty$ limit as $w^{\vert{}l\vert{}}$ and $z^{\vert{}l\vert{}}$ approach zero.

For the zero-charge sector, $H_0$ is given by
\begin{align}
    H_0 
    =\frac{wx}{1-wx}+\frac{wy}{1-wy}+\frac{xz}{1-xz}+\frac{yz}{1-yz}-2\frac{xyzw}{1-xyzw}
\end{align}
which reproduces the zero-monopole graviton index \cite{Bhattacharya:2008bja}.

Finally, the multi-particle index is computed using the plethystic exponential:
\begin{align}\label{abjm k}
I_{\text{grav}} = \text{PE}[I_{\mathbb{Z}k}] \xrightarrow{k\to \infty} I^0_{\text{grav}}I^+_{\text{grav}}.
\end{align}

Equation \eqref{abjm k} can be re-expressed as
\begin{align}\label{grav gen}
\frac{I_{\text{grav}}}{I^0_{\text{grav}}} \longrightarrow \frac{1}{\prod_{n=1}^{\infty}(1-q^n)}
\prod_{n=1}^{\infty}\prod_{i=1}^{\infty}
\frac{(1-q^n (abc)^i)}
{(1-q^n a^i)(1-q^n b^i)(1-q^n c^i)} = \sum_{m=0}^{\infty} q^m \tilde{I}_m,
\end{align}
where $x^k=q$. The following large-$m$ analysis is performed after taking this limit.
Equating \eqref{grav gen} to \eqref{BMN expansion}, the index $I_{\text{BMN},~U(m)}$ is identified with $\tilde{I}_m$.

\subsubsection{Asymptotic entropy growth}\label{index_entropy}

We now determine the scaling of the coefficient $\tilde I_m$ charge of order $m^2$ with $m\to \infty$. 

To this end, we define $F_{grav}$ as follows:
\begin{align}
F_{grav}=    \log \frac{I_{\text{grav}}}{I^0_{\text{grav}}} &= \sum_{j=1}^{\infty} \frac{1}{j} \sum_{n=1}^{\infty} q^{nj} + \sum_{n=1}^{\infty} \sum_{i=1}^{\infty} \sum_{j=1}^{\infty} \frac{q^{nj}}{j} \left(a^{ij} + b^{ij} + c^{ij} - (abc)^{ij}\right) \nonumber\\
    &= \sum_{j =1}^{\infty} \sum_{n=1}^{\infty} \frac{q^{nj}}{j} f(a^j, b^j, c^j) \,,
\end{align}
where the single-particle index is identified as:
\begin{equation}\label{def f}
    f(a,b,c) = 1 + \frac{a}{1-a} + \frac{b}{1-b} + \frac{c}{1-c} - \frac{abc}{1-abc} \,.
\end{equation}

For simplicity, we introduce a uniform chemical potential $\Delta$ by setting $a=b=c=s=e^{-\Delta}$; $Q$ denotes the exponent of $s$ (half the exponent if $s=t^2$). We evaluate the single-particle index exactly at this uniform fugacity:
\begin{equation}
    f(\Delta) = 1 + \frac{3e^{-\Delta}}{1-e^{-\Delta}} - \frac{e^{-3\Delta}}{1-e^{-3\Delta}} \,.
\end{equation}
In the high-temperature limit ($\Delta\to0$), $f(\Delta)=C/\Delta+\mathcal O(\Delta^3)$, where $C=8/3$. With $q=e^{-\omega}$, define
\begin{equation}
 A(q)=\sum_{r=1}^{\infty}\frac{q^r}{r^2(1-q^r)},
 \qquad B(q)=q\partial_q A(q).
\end{equation}
At fixed $0<q<1$, $F_{grav}\sim CA(q)/\Delta$, and the saddle equations give
\begin{equation}
 m\sim\frac{CB(q)}{\Delta},\qquad
 Q\sim\frac{CA(q)}{\Delta^2},\qquad
 \kappa=\frac{A(q_\kappa)}{CB(q_\kappa)^2}.
\end{equation}
Thus, at fixed $\kappa>0$,
\begin{equation}
 S(m,\kappa m^2)
 =m\left[-\log q_\kappa+\frac{2A(q_\kappa)}{B(q_\kappa)}\right]+o(m)
 =\mathcal O(m).
\end{equation}
In conclusion, for $Q\sim \mathcal O(m^2)$, the entropy grows as $\mathcal O(m)$. 

% Here the equal-fugacity product \eqref{grav gen} has nonnegative coefficients: its numerator cancels one denominator factor at each internal level divisible by three.

In the regime $Q = \kappa m^2$ with $\kappa \gg 1$, further simplifications occur. Deriving the explicit formula below requires taking an additional dilute limit where $q\ll 1$ (or $\omega\gg 1$), under which expanding the exact charge equation yields
\begin{align}
 m&=\sum_{r=1}^{\infty}\sum_{n=1}^{\infty}nq^{nr}f(r\Delta)
 \nonumber\\
 &=qf(\Delta)+q^2\bigl[2f(\Delta)+f(2\Delta)\bigr]+\mathcal O(q^3).
\end{align}
As we will see below, fixing the total charge $Q$ imposes $\Delta = m/Q$. Therefore, the fugacity behaves as:
\begin{equation}
    q \approx \frac{m}{f(\Delta)} = \frac{m\Delta}{C} \sim \frac{m^2}{Q} \,.
\end{equation}
For $q$ to be infinitesimally small, we must enforce a charge scaling of $Q \gg m^2$. If this restriction is met, the higher momentum modes ($n\ge2$) and statistical corrections ($r\ge2$) are suppressed by powers of $q\sim m^2/Q$, leaving the leading Maxwell--Boltzmann term:
\begin{equation}
    \log \frac{I_{\text{grav}}}{I^0_{\text{grav}}} \approx q f(\Delta) = e^{-\omega} \frac{C}{\Delta} \,.
\end{equation}
This gives $\tilde{I}_m(\Delta)$ the classical ideal gas form:
\begin{equation}
    \tilde{I}_m(\Delta) \approx \frac{1}{m!} (f(\Delta))^m \implies \log \tilde{I}_m(\Delta) \approx m \log\left(\frac{C}{\Delta}\right) - m \log m + m \,.
\end{equation}

\begin{figure}[t]
    \centering
    \includegraphics[width=0.9\textwidth]{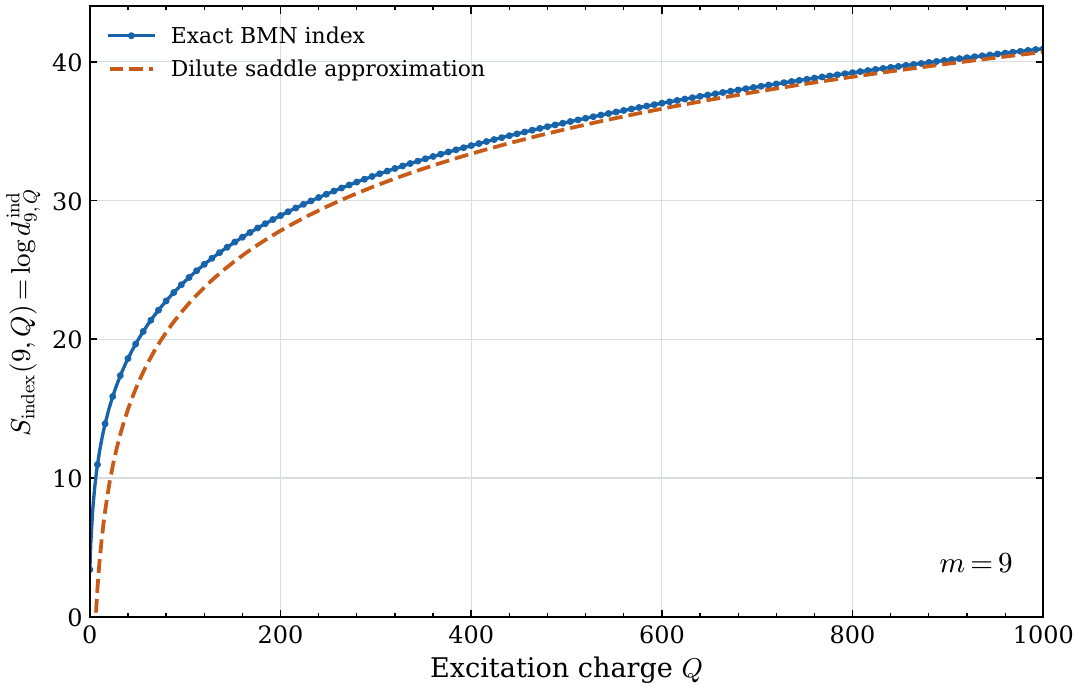}
    \caption{BMN indicial entropy
    $S_{\mathrm{index}}(m,Q)=\log d^{\mathrm{ind}}_{m,Q}$ at $m=9$,
    where $a=b=c=s=t^2$ and $d^{\mathrm{ind}}_{m,Q}$ denotes the coefficient of $q^ms^Q$.
    Blue points are obtained by exact coefficient extraction from \eqref{grav gen}.
    The dashed orange curve is the dilute approximation
    $m\log\!\bigl(8Q/(3m^2)\bigr)+2m-\log(2\pi Q)$.}
    \label{fig:bmn-index-entropy}
\end{figure}

Finally, we perform a Legendre transform on $\log \tilde{I}_m(\Delta)$ with respect to $\Delta$ to fix the total charge $Q$, :
\begin{equation}
    S(m, Q) \simeq \min_{\Delta} \Big( \log \tilde{I}_m(\Delta) + Q \Delta \Big) \,.
\end{equation}
Taking the derivative with respect to $\Delta$:
\begin{equation}
    -\frac{m}{\Delta} + Q = 0 \implies \Delta = \frac{m}{Q} \,.
\end{equation}
Substituting this back into the entropy expression gives the dilute approximation
\begin{equation}
    S(m, Q) = m \log\left( \frac{C Q}{m} \right) - m \log m + m + Q \left(\frac{m}{Q}\right) = m \log\left( \frac{C Q}{m^2} \right) + 2m \,.
\end{equation}
The Gaussian correction within this dilute model is $-\log(2\pi Q)$. Omitted tower and statistical terms contribute $\mathcal O(m^3/Q)$, which can exceed this logarithm; for example, $Q\gg m^3$ makes them negligible. Figure~\ref{fig:bmn-index-entropy} compares the dilute approximation with the index.

\subsection{Graviton partition function and asymptotic entropy}
\label{sec:multi_graviton_entropy}

In this subsection, we perform an analysis analogous to that of the previous subsection. However, instead of computing the graviton index on the ABJM side, we evaluate the BPS partition function and consider its large-$k$ limit. We define this BPS partition function as a trace over BPS graviton states subject to the same chemical potential constraints as the index, but without the fermion parity insertion $(-1)^F$, thereby preventing boson--fermion cancellations. This quantity can be interpreted as the BMN partition function evaluated at strong coupling. 
It is very important to keep in mind, that, unlike index, this quantity is not protected against change in coupling constant, and hence we should not expect that this matches the similar quantity explicitly computed in BMN quantum mechanics at weak coupling.
% We therefore view this analysis as an exploratory step toward a more quantitative understanding of the boson--fermion cancellations in the graviton index, rather than a concrete check of the proposed triality. 

The graviton partition function can be evaluated using the single-graviton spectrum tabulated in \cite{Bhattacharya:2008zy}. As demonstrated below, the entropy of the graviton partition function at charge $Q \sim m^2$ scales as $\mathcal{O}(m \log m)$. Although this is significantly larger than the index entropy ($\sim \mathcal{O}(m)$), it remains much smaller than the entropy of the free matrix model ($\sim \mathcal{O}(m^2)$).

We begin by constructing the graviton partition function. The single-particle graviton partition function on $\text{AdS}_4 \times S^7$, evaluated on states with $\delta=\{Q,S\}=0$, is given by
\begin{equation}
    Z_{sp}(x,y,z,w) = \frac{\mathcal{N}(x,y,z,w)}{(1-x) (1-y) (1-z) (1-w)(1-xyzw)}\,,
\end{equation}
where the numerator is defined as
% \begin{align}
%     \mathcal{N}(x,y,z,w) &= 2 w^2 x^2 y^2 z^2-w^2 x^2 y^2 z-w^2 x^2 y z^2\nonumber\\
%     &+w^2 x^2 y z-w^2 x y^2 z^2+w^2 x y^2 z+w^2 x y z^2-w x^2 y^2 z^2\nonumber\\
%     &+w x^2 y^2 z+w x^2 y z^2+w x y^2 z^2-2 w xy z+2 w x y+2 w x z-w x\nonumber\\
%     &+2 w y z-w y-w z+w+2 x y z-x y-x z+x-y z+y+z.
% \end{align}
\begin{align}
    \mathcal{N}(x, y, z, w) &= \sum_{i} x_i - \sum_{i<j} x_i x_j + 2 \sum_{i<j<k} x_i x_j x_k - 2p \nonumber \\
    &\quad + p \left( \sum_{i<j} x_i x_j \right) - p \left( \sum_{i<j<k} x_i x_j x_k \right) + 2p^2 \,,
\end{align}
Here $(x_1,x_2,x_3,x_4)=(x,y,z,w)$ and $p=xyzw$. As in the previous subsection, the projected single-particle partition function is formally defined via a discrete Fourier transform over the $k$-th roots of unity, $\omega = e^{2\pi i / k}$:
\begin{equation}
    Z_{\mathbb{Z}_k}(x,y,z,w) = \frac{1}{k} \sum_{j=0}^{k-1} Z_{sp}\left(\omega^j x, \omega^j y, \omega^{-j} z, \omega^{-j} w\right) = \sum_{m=-\infty}^{\infty} G_{mk} \,,
\end{equation}
where $G_\ell$ is the generating function restricted to charge $\ell$.

Now let us take the same $k\to\infty$ limit with $q=x^k$ and $a,b,c$ fixed. At fixed transverse fugacities in the convergence domain, the dominant positive-charge residue gives
\begin{equation}
    x^{-\ell}G_\ell \longrightarrow
    \frac{(1+ab)(1+bc)(1+ca)}
    {(1-a)(1-b)(1-c)(1-abc)}
    \equiv g(a,b,c)
    \qquad (\ell\to\infty).
\end{equation}
where $a, b, c$ are defined as in \eqref{graviton Hc}.

The positive-mode $\mathbb{Z}_k$ orbifold projection of the single-particle partition function is obtained by summing over the $mk$ sectors. Substituting $q = x^k$, we evaluate the geometric sum:
\begin{equation}\label{plus}
    Z^+_{\mathbb{Z}_k} = \sum_{m=1}^{\infty} G_{mk} = \sum_{m=1}^{\infty} q^m g(a,b,c) = \frac{q}{1-q} g(a,b,c) \,.
\end{equation}
For the unsigned multiparticle function, bosons and fermions obey different statistics \cite{Bhattacharya:2008zy,Chang:2024lkw}. 
The unsigned and signed single-particle functions satisfy
$g=g_B+g_F$ and $f=g_B-g_F$. Explicitly,
\begin{equation}
    g_B=\frac{1+abc(a+b+c)}
    {(1-a)(1-b)(1-c)(1-abc)},\qquad
    g_F=\frac{ab+bc+ca+(abc)^2}
    {(1-a)(1-b)(1-c)(1-abc)}.
\end{equation}
After removing the neutral sector, Bose statistics for bosons and Fermi statistics for fermions give
\begin{align}\label{eq:grav-bose-fermi}
 \log\frac{Z_{\text{grav}}}{Z^0_{\text{grav}}}
 &=\sum_{r=1}^{\infty}\frac{q^r}{r(1-q^r)}
 \bigl[g_B(a^r,b^r,c^r)+(-1)^{r+1}g_F(a^r,b^r,c^r)\bigr].
\end{align}

\subsubsection*{Asymptotic entropy growth for $Q =\kappa m^2$}

At equal fugacities $a=b=c=e^{-\Delta}$, \eqref{eq:grav-bose-fermi} becomes
\begin{align}
 \log\frac{Z_{\text{grav}}}{Z^0_{\text{grav}}}
 &=\sum_{\substack{r\ge1\\r\,\mathrm{odd}}}
 \frac{q^r g(r\Delta)}{r(1-q^r)}
 +\sum_{\substack{r\ge1\\r\,\mathrm{even}}}
 \frac{q^r f(r\Delta)}{r(1-q^r)}.
\end{align}
We denote the fixed-momentum graviton coefficient by $Z_m^{\mathrm{grav}}$:
\begin{align}
 \frac{Z_{\text{grav}}}{Z^0_{\text{grav}}}
 =\sum_{m=0}^{\infty}q^m Z_m^{\mathrm{grav}},\qquad Z_0^{\mathrm{grav}}=1.
\end{align}

\begin{figure}[t]
    \centering
    \includegraphics[width=0.9\textwidth]{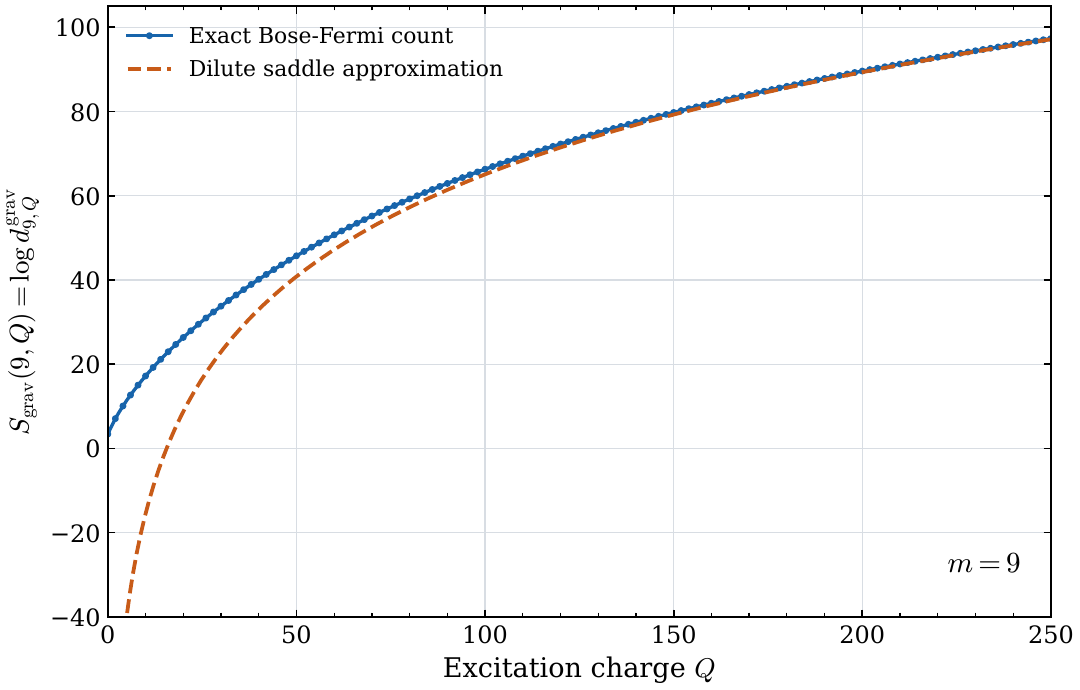}
    \caption{Graviton entropy $S_{\mathrm{grav}}(m,Q)=\log d^{\mathrm{grav}}_{m,Q}$
    at $m=9$, with $a=b=c=s=t^2$ and
    $d^{\mathrm{grav}}_{m,Q}$ denotes the coefficient of $q^ms^Q$.
    Blue points are computed from \eqref{eq:grav-bose-fermi}, with Bose statistics for bosons and
    Fermi statistics for fermions. The dashed orange curve is the dilute saddle
    approximation $m\log\!\bigl(Q^4/(96m^5)\bigr)+5m-\log(\pi Q)$.}
    \label{fig:graviton-entropy}
\end{figure}

To analyze the entropy growth for $Z_m^{\mathrm{grav}}$ at large $m \gg 1$ with a $U(1)$ charge scaling of $Q \sim \mathcal{O}(m^2)$, we evaluate the asymptotic density of states using a thermodynamic saddle-point approximation. 

As we have done in the previous subsection, we introduce a uniform chemical potential $\Delta$ for the charges, setting $a = b = c = e^{-\Delta}$. In the large-charge (high-temperature) limit, $\Delta \to 0$. We expand the single-particle partition function $g(\Delta)$:
\begin{equation}
    g(\Delta) = \frac{(1+e^{-2\Delta})^3}{(1-e^{-\Delta})^3(1-e^{-3\Delta})} \approx \frac{8}{\Delta^3 (3\Delta)} = \frac{8}{3\Delta^4} \,.
\end{equation}
Letting $C = 8/3$, the single-particle partition function behaves as $g(\Delta)=C/\Delta^4+\mathcal O(\Delta^{-2})$.

Next, we extract the coefficient of $q^m$ at large $m$. Let $\omega$ be the chemical potential conjugate to $m$ (so $q = e^{-\omega}$). The D0-brane charge $m$ is fixed by the saddle-point equation:
\begin{align}
 m&=q\partial_q\log\frac{Z_{\text{grav}}}{Z^0_{\text{grav}}}
 \nonumber\\
 &=qg(\Delta)+q^2\bigl[2g(\Delta)+f(2\Delta)\bigr]+\mathcal O(q^3).
\end{align}

Let \(Q=\kappa m^2\), with fixed \(\kappa>0\), and take \(m\to\infty\). As shown below, the charge saddle then gives \(\Delta=4/(\kappa m)\), so that
\begin{equation}
    g(\Delta)\simeq \frac{C}{\Delta^4}
    =\frac{C\kappa^4}{4^4}m^4.
\end{equation}
The momentum saddle consequently yields
\begin{equation}
    m \simeq q\,g(\Delta)
    \qquad\Longrightarrow\qquad
    q \simeq \frac{4^4}{C\kappa^4m^3}
    =\frac{96}{\kappa^4m^3}\ll1.
\end{equation}
Thus, unlike the index calculation, no additional large-$\kappa$ limit is required to reach the dilute regime. The statistical terms (\(r\ge2\)) and higher-momentum contributions (\(n\ge2\)) are therefore suppressed by powers of \(q\). To leading order, the grand canonical potential reduces to the Maxwell--Boltzmann term
\begin{equation}
    \log \frac{Z_{\text{grav}}}{Z^0_{\text{grav}}}
    \simeq q\,g(\Delta)
    =e^{-\omega}\frac{C}{\Delta^4}.
\end{equation}
It follows that \(Z_m^{\mathrm{grav}}(\Delta)\) takes the classical ideal-gas form,
\begin{equation}
    Z_m^{\mathrm{grav}}(\Delta)
    \simeq \frac{1}{m!}\bigl(g(\Delta)\bigr)^m,
\end{equation}
and hence
\begin{equation}
    \log Z_m^{\mathrm{grav}}(\Delta)
    \simeq
    m\log\left(\frac{C}{\Delta^4}\right)
    -m\log m+m.
\end{equation}

Finally, we perform a Legendre transform on $\log Z_m^{\mathrm{grav}}(\Delta)$ with respect to $\Delta$:
\begin{equation}
    S(m, Q) \simeq \min_{\Delta} \Big( \log Z_m^{\mathrm{grav}}(\Delta) + Q \Delta \Big) \,.
\end{equation}
Taking the derivative with respect to $\Delta$ gives the saddle-point equation:
\begin{equation}
    -\frac{4m}{\Delta} + Q = 0 \implies \Delta = \frac{4m}{Q} \,.
\end{equation}
Since $Q \sim m^2$, we verify that $\Delta \sim m^{-1} \to 0$, validating our high-temperature approximation. Substituting $\Delta$ back into the entropy expression yields:
\begin{equation}
    S(m, Q) \simeq m \log\left( \frac{C Q^4}{256 m^4} \right) - m \log m + m + Q \left(\frac{4m}{Q}\right) = m \log\left( \frac{C Q^4}{256 m^5} \right) + 5m \,.
\end{equation}
Inserting the charge scaling $Q = \kappa m^2$:
\begin{align}
    S(m,\kappa m^2)=3m\log m+m\left[\log\left(\frac{\kappa^4}{96}\right)+5\right]+\mathcal O(\log m).
\end{align}
The Gaussian correction is $-\log(\pi Q)$ in this dilute graviton model.

\subsection{Comparison of entropy diagnostics}
\label{comp ent}

In this subsection, we summarize the entropy estimates for the various partition functions and indices in the regime
$Q=\kappa m^2$,
with fixed \(\kappa>0\). 

\begin{table}[H]
\centering
\renewcommand{\arraystretch}{1.25}
\begin{tabular}{lc}
\toprule
\textbf{Quantity} & \textbf{Entropy at \(Q\sim m^2\)} \\
\midrule
Free BPS BMN partition function
& \(\mathcal{O}(m^2)\) \\

ABJM multiparticle graviton partition function
& \(\mathcal{O}(m\log m)\) \\

Coefficientwise absolute sum of vacuum-sector indices
& \(S_{\mathrm{index}}\leq S_{\mathrm{abs}}
   \stackrel{?}{\lesssim}S_{\mathrm{grav}}\) \\

BMN index
& \(\mathcal{O}(m)\) \\
\bottomrule
\end{tabular}
\caption{Summary of the entropy diagnostics in the regime \(Q\sim m^2\). The evaluation of the BMN index assumes the relation \eqref{grav gen}. The tentative inequality \(S_{\mathrm{abs}}\lesssim S_{\mathrm{grav}}\), with \(S_{\mathrm{grav}}=\mathcal{O}(m\log m)\), is suggested by the available finite-\(m\) data but has not been established analytically.}
\label{tab:entropy-summary}
\end{table}

Let us first contrast the free BPS BMN partition function with the fully summed BMN index. Since the free matrix model contains \(\mathcal{O}(m^2)\) oscillators, its unsigned BPS entropy at \(Q\sim m^2\) grows as \(\mathcal{O}(m^2)\). By contrast, the BMN index at the same charge grows only as \(\mathcal{O}(m)\), as shown in Section~\ref{index_entropy}. This large difference indicates substantial boson--fermion cancellations.

The multiparticle graviton partition function computed in Section~\ref{sec:multi_graviton_entropy} provides an intermediate diagnostic. Its entropy grows as \(\mathcal{O}(m\log m)\), which is parametrically smaller than the free BPS entropy but larger than the graviton index. Under the proposed triality, it is natural to interpret this quantity as describing the supergravity subsector of the strongly coupled BMN theory. The difference between the unsigned graviton partition function and the graviton index exhibits large boson--fermion cancellations within the supergravity spectrum.

A complementary measure of cancellations is provided by the coefficientwise absolute sum of the vacuum-sector indices. Writing the $U(m)$ BMN index as a summation of the vacuum sector indices, 
\begin{equation}
I_m(s)=\sum_{P\vdash m}I_P(s),\qquad
    I_P(s)=\sum_Q d_{P,Q}s^Q,
\end{equation}
we define
\begin{equation}
    A_{m,Q}=\sum_{P\vdash m}\lvert d_{P,Q}\rvert,
    \qquad
    S_{\mathrm{abs}}(m,Q)=\log A_{m,Q}.
\end{equation}
This quantity removes cancellations between different vacuum sectors, although it retains boson--fermion cancellations within each individual sector.

Recent non-renormalization results show, subject to the stated normalizability assumptions, that BPS multiplicities remain constant between finite nonzero couplings
\cite{Chang:2026ccg,Colin-Ellerin:2026eta}. The strict free limit used to evaluate the matrix integrals is an endpoint and requires separate control, so these results do not by themselves establish that \(A_{m,Q}\) is coupling independent. Nevertheless, \(A_{m,Q}\) remains a useful diagnostic of cancellations among the sector indices.

The triangle inequality gives
\begin{equation}
    \left\lvert\sum_{P\vdash m}d_{P,Q}\right\rvert
    \leq A_{m,Q},
\end{equation}
and therefore
\begin{equation}
    S_{\mathrm{index}}(m,Q)\leq S_{\mathrm{abs}}(m,Q).
\end{equation}
Moreover, every individual vacuum sector, including the trivial vacuum \(P=(1^m)\), obeys
\begin{equation}
    \lvert d_{P,Q}\rvert\leq A_{m,Q}.
\end{equation}
Thus, \(A_{m,Q}\) also provides an upper bound on the magnitude of the trivial-vacuum index coefficient.

\begin{figure}[t]
    \centering
    \includegraphics[width=0.9\textwidth]{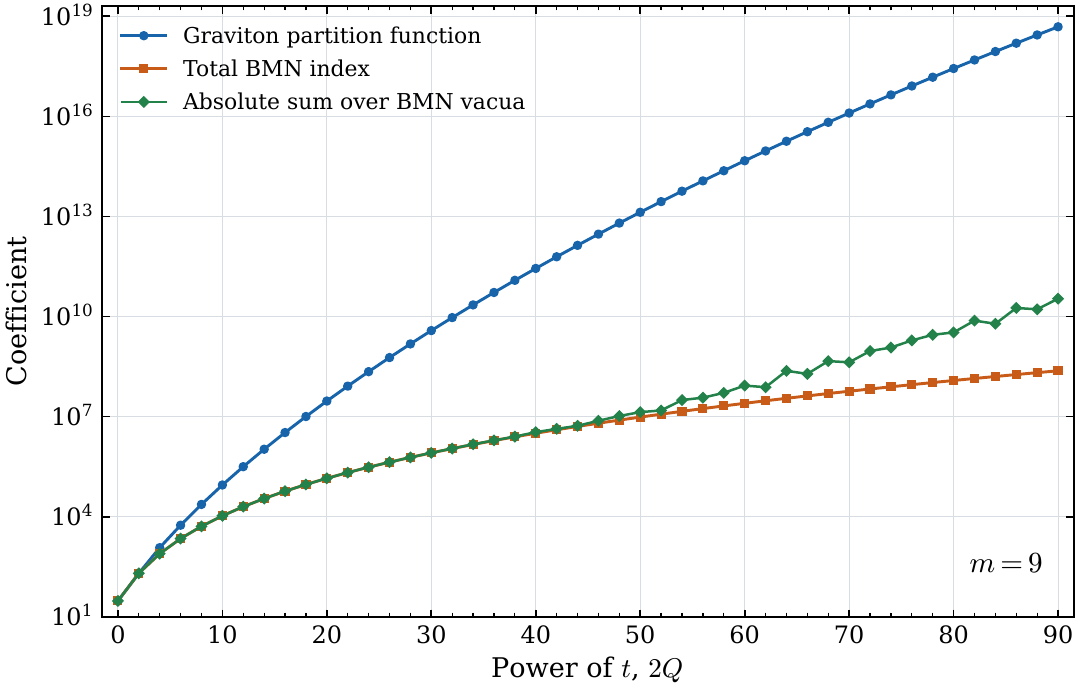}
    \caption{Comparison of coefficients at \(m=9\), with the fugacity specialization \(a=b=c=s=t^2\). Blue circles represent the unsigned graviton coefficients \(d^{\mathrm{grav}}_{9,Q}\), defined as the coefficients of \(q^9s^Q\) in \(Z_{\mathrm{grav}}/Z_{\mathrm{grav}}^0\) and previously shown in Figure~\ref{fig:graviton-entropy}. Orange squares represent the total BMN index coefficients \(d^{\mathrm{ind}}_{9,Q}\), while green diamonds represent the coefficientwise absolute sum over vacuum sectors,
    \(A_{9,Q}=\sum_{P\vdash9}\lvert d_{P,Q}\rvert\).
    The horizontal axis shows the power of \(t\), namely \(2Q\), and the vertical axis is logarithmic.}
    \label{fig:bmn-graviton-comparison}
\end{figure}

As shown in Figure~\ref{fig:bmn-graviton-comparison}, the available \(m=9\) data place \(A_{m,Q}\) well below the unsigned graviton coefficients over the displayed charge range. Since \(A_{m,Q}\) bounds each individual sector, these data suggest that even the trivial-vacuum index does not display an \(\mathcal{O}(m^2)\) enhancement in this regime. If this ordering persists in the large-\(m\) limit at fixed \(Q/m^2\), it would imply
\begin{equation}
    S_{\mathrm{triv}}
    \leq S_{\mathrm{abs}}
    \lesssim S_{\mathrm{grav}}
    =\mathcal{O}(m\log m).
\end{equation}
% At present, however, this is only a finite-\(m\) observation and not an asymptotic bound. In particular, no general inequality relating \(A_{m,Q}\) to the physical graviton degeneracy is known.

These observations motivate a reassessment of previous arguments for \(\mathcal{O}(m^2)\) entropy in the trivial-vacuum BMN index, as well as more recent finite-rank evidence for similar growth after summing over vacuum sectors
\cite{Choi:2023vdm,Chang:2024lkw,Chang:2026mui}. In fact, the latter asymptotic conclusion is directly contradicted by the analytic computation in Subsection~\ref{abjm grav}, which gives an \(\mathcal{O}(m)\) entropy for the fully summed index at fixed \(Q/m^2\), assuming the relation \eqref{grav gen}.

The analytic small-charge saddle relevant to the trivial-vacuum BMN index applies to
\begin{equation}
    Q=\epsilon m^2,\qquad \epsilon\ll1,
\end{equation}
with \(\epsilon\) fixed as \(m\to\infty\)
\cite{Choi:2023vdm}. It gives $S(Q)\propto \epsilon^{3/2}m^2$,
and therefore predicts an \(\mathcal{O}(m^2)\) entropy. Although the leading eigenvalue distribution explicitly solves the saddle-point equation, the calculation controls only the leading small-\(\epsilon\) behavior. Moreover, a complete contour analysis establishing that this saddle contributes to the relevant index coefficient has not been given.

Numerically, the comparatively small value of \(A_{m,Q}\), which bounds every individual vacuum-sector coefficient, suggests that the tension may be present. Resolving it will require either larger-rank calculations at fixed \(Q/m^2\) or a reassessment of the saddle using the same charge refinement and integration contour. We leave it to the future study.

\subsection{Growth of finite-$k$ ABJM index}\label{finite kk}
Throughout this paper, our main focus has been the finite-rank BMN conjecture: the proposed equivalence between DLCQ M-theory on the eleven-dimensional pp-wave and the finite-rank BMN matrix model. In this subsection, we instead turn to the large-rank BMN conjecture, which relates M-theory on the decompactified pp-wave background to BMN matrix quantum mechanics at large rank.

Concretely, we study the large-$N$ ABJM graviton index at fixed $k$ in the limit $m\to\infty$. For simplicity, we specialize to $k=1$, where the structure is particularly transparent. We show that, in an appropriate charge regime, the leading grand-canonical behavior of the ABJM index agrees with that of the large-rank BMN matrix model.

The motivation is straightforward from the gravitational perspective. At $k=1$, taking the large-$N$, large-$m$ limit of ABJM theory corresponds to the Penrose limit of unorbifolded $AdS_4\times S^7$. The resulting geometry is the noncompact eleven-dimensional pp-wave, whose Matrix Theory description is conjectured to be given by the BMN model at large rank. This is precisely the regime considered by Kovacs, Sato, and Shimada \cite{Kovacs:2013una}. Our analysis provides a quantitative test of this correspondence at the level of supersymmetric indices.

To quantify this claim, we compare the large $N$, large $m$ index of ABJM theory at finite $k$ with large rank BMN index. Note that since one needs to take large $N$ limit first, we can directly work with graviton index in $AdS_4\times S^7$.

The multi-graviton index in $AdS_4\times S^7$ reduces to the plethystic
exponential of the single-particle graviton index.
To use the monopole charge as the expansion parameter, we keep
$a=y/x$, $b=xz$, and $c=xw$ fixed, as in the previous subsection,
and define
\begin{equation}
    \mathcal I_{sp}(x;a,b,c)
    \equiv I_{sp}(x,ax,b/x,c/x),
    \qquad
    I_{\mathrm{ABJM}}=\operatorname{PE}[\mathcal I_{sp}].
\end{equation}
In these variables, $x$ is conjugate to the monopole charge
$J=km=m$. The Laurent expansion contains both positive and
negative charges and is taken in the chamber
$\max(|b|,|c|)<|x|<\min(1,|a|^{-1})$.
We remove the neutral single-particle factor
\begin{equation}
    I_{\mathrm{ABJM}}^0
    =\operatorname{PE}[h_0(a,b,c)].
\end{equation}
where $h_0$ is coefficient of $x^0$ in $\mathcal I_{sp}(x;a,b,c).$

The dominant positive-charge pole is located at $x=1$, and
\begin{equation}
    \mathcal I_{sp}(x;a,b,c)-h_0(a,b,c)
    =
    \frac{x}{1-x}f(a,b,c)+R(x;a,b,c),
\end{equation}
where $f$ is defined in \eqref{def f} and $R$ is regular at $x=1$. Consequently,
\begin{align}
    \log\frac{I_{\mathrm{ABJM}}}{I_{\mathrm{ABJM}}^0}
    &=
    \sum_{r=1}^{\infty}
    \frac{x^r}{r(1-x^r)}f(a^r,b^r,c^r)
    +\log Z_R .
\end{align}
The first term is precisely the BMN generating function upon
identifying $x=q$. For fixed internal fugacities inside the
convergence domain, it grows as $\mathcal O(\omega^{-1})$ as
$x=e^{-\omega}\to1$, whereas $\log Z_R=\mathcal O(1)$.
Thus, the finite-$k$ corrections are subleading in the
large-Hopf-charge limit.

If the internal fugacities,
$a=b=c=e^{-\Delta}$, also approach one, the same conclusion holds provided
\begin{equation}
    \omega\ll\Delta.
\end{equation}
In the simultaneous small-$(\omega,\Delta)$ regime, this condition
corresponds to $Q\ll m$. When $\omega$ and $\Delta$ become
comparable, the remaining residues are no longer subleading and
the finite-$k$ index can deviate from the BMN generating function
at leading order.

This restriction has a natural bulk interpretation. The Penrose limit retains states whose transverse excitation charges are parametrically smaller than the large Hopf-fiber momentum \(J=m\). When \(Q\) becomes comparable to \(m\), these states are no longer localized near the null geodesic defining the Penrose limit and begin to resolve the finite curvature and global structure of \(AdS_4\times S^7\). Understanding the crossover regime \(Q/m=\mathcal{O}(1)\) is left for future work.

\section{Discussion}\label{discussion}

In this paper, we have proposed a concrete route from a conventional AdS/CFT duality to a  Matrix Theory conjecture. The starting point is the observation that the Penrose limit of $AdS_4\times S^7/\mathbb{Z}_k$, combined with an appropriate scaling of the orbifold identification, produces the DLCQ of the eleven-dimensional pp-wave geometry. On the ABJM side, this limit isolates monopole sectors in the triple-scaling regime introduced in this paper. On the gravitational side, the same sector is characterized by finite light-cone momentum and is conjectured to be described by the finite-rank BMN matrix model. 

The central claim of the paper is therefore a triality between triple-scaled ABJM theory, DLCQ M-theory on the eleven-dimensional pp-wave, and the finite-rank BMN matrix model. The main nontrivial evidence for this proposal comes from supersymmetric indices. We showed that the ABJM superconformal index in the triple-scaling limit precisely reproduces the grand-canonical generating function of BMN Witten indices, with the ABJM monopole charge mapped to the rank of the BMN gauge group.

We conclude with some observations and directions for future study. First, several natural questions arise directly from our analysis of the index.

\paragraph{Index cancellations and comparison with earlier results.} The coefficient formula \eqref{grav gen} implies that the total signed index has entropy of order \(\mathcal O(m)\) at fixed \(Q/m^2\), while the multiparticle graviton partition function gives an unsigned entropy of order \(\mathcal O(m\log m)\). These results motivate a reassessment of earlier arguments for \(\mathcal O(m^2)\) index entropy \cite{Choi:2023vdm,Chang:2024lkw,Chang:2026mui}. It would be valuable to revisit the proposed large-rank saddles and their contributions to the index, together with numerical computations at increasing rank and fixed \(Q/m^2\), to determine the origin of the apparent discrepancy.

\paragraph{``D0-brane expansion".}
The expansion in \eqref{BMN expansion} resembles the giant-graviton
expansions of finite-$N$ holographic indices
\cite{Imamura:2021ytr,Gaiotto:2021xce}.
Here \(q^{m}\) grades the positive monopole charge, or equivalently
the DLCQ momentum, and the coefficient is the index of the \(U(m)\) matrix
model. In the type-IIA frame, it is natural to call this D0-brane
expansion. However, unlike the giant-graviton expansion, there is no alternating sign in expansions because this is an ordinary decomposition into charge sectors and there is no tachyonic fluctuations to be analytically continued. By contrast, signs in giant-graviton expansions arise from the wrapped-brane fluctuation index and from the analytic continuation used to reorganize finite-$N$ corrections \cite{Gaiotto:2021xce,Lee:2022vig,Imamura:2022aua}. It would be interesting to see if there are other setups\footnote{See \cite{Dorey:2023jfw} for related discussions for ultraspinning black holes.} for superconformal indices that give rise to these D0-brane-type expansions.

\paragraph{Taking further limits.} Our triality relates a particular large-$N$ limit of ABJM theory to the finite-rank $U(m)$ BMN matrix model. 
Conversely, suitable large-$m$ limits of the BMN model are known to produce a variety of theories associated with M2- and M5-branes, including six-dimensional $(2,0)$ theory \cite{Lin:2005nh,Asano:2017xiy,Asano:2017nxw}, little string theory \cite{Lin:2005nh,Asano:2022brd,Asano:2026hto}, and finite-rank ABJM theory \cite{Lin:2005nh,Chang:2024lkw}. In a sense, these relations run in the opposite direction to our triality: they embed other brane theories into the BMN model, whereas our triality embeds the BMN model into a large-charge sector of ABJM theory. Combining the two suggests a nested, or ``Matryoshka-like,'' structure of dualities: theories associated with M2- and M5-branes arise as limits of the BMN model, which itself arises as a subsector of large-$N$ ABJM theory. It would be interesting to test this hierarchy directly at the level of protected quantities, by taking the corresponding limits of the ABJM superconformal index. 

A related question is to study the limit in which the BMN model reduces to the BFSS matrix model. Understanding the corresponding limit of the ABJM index may provide a new route to exploring the original Matrix Theory conjecture by BFSS \cite{Banks:1996my}.
\vspace{20pt}

\noindent There are also several broader questions concerning the proposed BMN triality.

\paragraph{Dynamical tests.}
A stronger test of the proposed triality should go beyond protected counting.
One approach is to quantize ABJM theory on \(S^2\times\mathbb R\) around
a dressed-monopole background, identify the fluctuations that remain light
in the triple-scaling limit, and determine their dynamics to see if it agrees with that of the BMN model.
It is known in the literature that Higgsing around a half-BPS background yields maximally supersymmetric
Yang--Mills theory on \(\mathbb R\times S^2\)
\cite{Ezhuthachan:2011kf}, while large-angular-momentum analyses have
matched leading excitation spectra with the pp-wave matrix model at
fixed \(k\) \cite{Kovacs:2013una}.
These results motivate, but do not establish, the corresponding
interacting reduction in our scaling.
Deriving the BMN masses, quartic scalar interaction, Myers coupling,
fermion interactions, and coupling dictionary
\eqref{parameter-dictionary}, together with a controlled decoupling
argument, would provide an action-level derivation of the limiting theory.
Complementary tests include matching ABJM excitation energies with BMN
energy shifts and comparing normalized three-point functions.
A longer-term goal would be to relate matrix-model correlators to bulk
transition amplitudes, drawing on BFSS amplitude calculations\footnote{A conventional scattering interpretation would
require further care because the pp-wave background does not allow for standard asymptotic states.}
\cite{Herderschee:2023pza} and strong-coupling studies of the BMN spectrum
\cite{Komatsu:2024vnb}.
The dressed-operator construction of Section~\ref{dictionary} may
provide a starting point for these comparisons.
\paragraph{Open-closed-open triality.} A related direction is to sharpen the connection with the open--closed--open (OCO) triality discussed in Subsection~\ref{oco}. It would be particularly useful to start directly from a D2--D0 brane system and show that, by integrating out the appropriate brane sector and taking suitable decoupling limits, one recovers both ABJM theory with monopole insertions and the BMN matrix model. Such a derivation would provide a complementary, microscopic route to the BMN triality.

An instructive precedent is the D3--D($-1$) system, where integrating out different brane sectors leads to instanton physics in $\mathcal N=4$ SYM and provides a string-theoretic realization of the ADHM construction \cite{Billo:2002xy,Ferrari:2012nw}. The relation between ABJM monopole operators and the BMN model is suggestively reminiscent of the Nahm construction of monopoles. It would therefore be interesting to understand whether techniques and lessons from the D3--D($-1$) system admit a direct analogue in the present setting.

More broadly, the BMN triality may point to a useful refinement of OCO triality itself. As discussed in Subsection~\ref{oco}, the F-type dual is generically expected to be a full-fledged open-string field theory. In the present construction, however, the triple-scaling limit isolates a regime in which the F-type description reduces to an ordinary matrix quantum mechanics. This makes the triality considerably sharper and more amenable to quantitative tests. It would be valuable to search for analogous decoupling limits in other realizations of OCO triality, where the F-type sector may likewise simplify to a conventional field theory.
\paragraph{Other examples of triality.}
Our proposed triality may extend to other  holographic
theories admitting a large-charge Penrose limit with a compact null
direction. In three dimensions, ADHM-type theories associated with D2--D6 systems are natural
candidates \cite{Ferrara:1998vf,Gomis:1998xj,Itzhaki:1998uz,Pelc:1999ms},
since an appropriate Penrose limit yields the maximally supersymmetric
eleven-dimensional plane wave. Their indices and protected correlators
provide observables for testing the resulting matrix-model description
\cite{Bobev:2022wem,Chester:2014mea,Mezei:2017kmw,Gaiotto:2020vqj}.

In the forthcoming work \cite{KLPT:toappear}, we show that the ADHM
superconformal index also reproduces the BMN index in the triple-scaling
limit. We further establish a correspondence between protected
Coulomb-branch correlators computed using supersymmetric localization
\cite{Dedushenko:2017avn,Dedushenko:2018icp} and the protected BMN correlators obtained by
Asano et al.~\cite{Asano:2017nxw}\footnote{We would like to thank Henry Lin for a useful discussion on this.} in the same limit.

The triality may also exist in type IIB string theory.
A suitable large-orbifold Penrose limit of
\(AdS_5\times S^5/\mathbb Z_k\) is known to yield a plane wave background with a compact
lightlike direction \cite{Mukhi:2002ck}.
One may then seek to match the large-charge superconformal index
of the dual four-dimensional quiver gauge theory with the index of
level-one mass-deformed ABJM theory on \(\mathbb R\times T^2\),
a proposed DLCQ description of the IIB plane wave \cite{Gomis:2008vc}.
The known agreement of supersymmetric vacuum counting
\cite{Kim:2010mr} motivates extending this comparison to excited
BPS states and their refined index.

Instead of finding and testing the triality, one may also be able to use the triality to make novel predictions for field theories. For instance, the indices of recently-constructed BMN-like matrix models \cite{Lee:2026xoo} may be used to constrain the large-charge limit of  putative three-dimensional
parent theory.\footnote{
A caveat is that a Penrose limit can enhance supersymmetry \cite{Itzhaki:2002kh}, producing
protected states not captured by the parent index. The latter therefore need not determine the
full refined matrix-model index.}
\paragraph{Matrix Theory conjectures and worldline holography.}
Finally, the BMN triality may provide a useful framework for sharpening our understanding of Matrix Theory conjectures themselves. In particular, it would be valuable to use our setup of the BMN triality to clarify  precisely how the Matrix-Theory degrees of freedom encode the underlying quantum-gravitational dynamics.

A related direction is to further develop the connection to worldline holography discussed in Subsection~\ref{oco}. Because the BMN triality provides a controlled setting in which bulk quantum-gravitational physics is captured by a tractable quantum-mechanical theory living on observer's worldline, it may serve as a useful toy model for exploring which aspects of worldline holography can be formulated and tested beyond purely qualitative arguments.

% Study non-BPS spectrum - computing anomalous dimensions of various operators in ABJM/BMN or directly compare ABJM/BMN actions.
% Scattering like quantities in plane-wave background and compare it to correlators in BMN.

% Equation \eqref{BMN expansion} looks like giant graviton expansion. Indeed, can be thought of as a D0 brane expansion. (but without minus sign! maybe because unlike D3 brane wrapping S3 in S5, no tachyonic directions?)

% Begin with other 3d holographic theories and taking the similar limit ($k\to \infty$, $x\to1$...) to obtain $U(N)$ adjoint matrix integral formula. This will allow us to guess the form of the matrix model. (However, there can be supersymmetry enhancement and emergent BPS states can appear from the Penrose limit (cf. AdS$_5\times T^{1,1}$ \cite{Itzhaki:2002kh}). In this case we won't be able to see the refined information.) 

% Or the other way around. Begin with BMN-like matrix model, write down the index, and then reverse engineer the superconformal index of 3d CFT.
% So there will be a map between 3d CFTs and matrix models (doesn't have to be 1-to-1).

% It would be interesting to study ADHM theory, as a Penrose limit will give us the maximally symmetric 11d plane-wave geometry. One can also study the penrose limit of orbifolded $AdS_5\times S^5$ and try to match the superconformal index of the dual quiver gauge theories to index of a theory dual to 10d Type IIB pp-wave(massive ABJM theory at level 1 defined on 2-torus looks like a good candidate). 

\section*{Acknowledgement}
We thank the 20th Asian Winter School
on Strings, Particles and Cosmology in IISER Bhopal, where the collaboration was initiated.
We thank Shivansh Tomar for valuable discussions and collaboration on related topics. We would also like to thank Elisabetta Armanini, Jaehyeok Choi, Seok Kim, Henry Lin, Juan Maldacena, Shiraz Minwalla, Mukund Rangamani, Hidehiko Shimada and Arkady Tseytlin for valuable discussions at various stages of this project.  The work of E.L. was supported by the Infosys Endowment for the study of the Quantum Structure of Spacetime. The work of C.P was supported in part by the NSF grant PHY-2609862, the Simons Foundation grant 917464, and the Simons Foundation International grant SFI-MPS-QCD-00014867-01 (Simons Collaboration on Confinement and QCD Strings).

\vspace{1cm}

\appendix

\section{Numerical data}\label{num check}

In this appendix, we list the numerical results pertaining to various quantities computed in Section \ref{asymptotic behavior}.

\subsection{BMN index from ABJM graviton index}\label{num1}
In this subsection, we numerically compute the complete BMN index using the large $k$ limit of the graviton index in $AdS_4\times S^7/\mathbb Z_k$. 

Expanding the generating function \eqref{grav gen}, we obtain the following $U(m)$ BMN indices at $a=b=c=t^2$.
We display representative results for $m=2,3,4,5,6,9$; the same procedure extends directly to higher ranks and expansion orders.

$U(2)$:
\[
\begin{aligned}
&2 + 6 t^{2} + 12 t^{4} + 13 t^{6} + 18 t^{8}\\
&{}+ 21 t^{10} + 25 t^{12} + 27 t^{14} + 33 t^{16} + 35 t^{18}\\
&{}+ 39 t^{20} + \mathcal O(t^{22})
\end{aligned}
\]

$U(3)$:
\[
\begin{aligned}
&3 + 12 t^{2} + 27 t^{4} + 45 t^{6} + 63 t^{8}\\
&{}+ 87 t^{10} + 115 t^{12} + 141 t^{14} + 174 t^{16} + 213 t^{18}\\
&{}+ 249 t^{20} + \mathcal O(t^{22})
\end{aligned}
\]

$U(4)$:
\[
\begin{aligned}
&5 + 21 t^{2} + 57 t^{4} + 105 t^{6} + 177 t^{8}\\
&{}+ 261 t^{10} + 376 t^{12} + 507 t^{14} + 672 t^{16} + 857 t^{18}\\
&{}+ 1086 t^{20} + \mathcal O(t^{22})
\end{aligned}
\]

$U(5)$:
\[
\begin{aligned}
&7 + 36 t^{2} + 105 t^{4} + 224 t^{6} + 405 t^{8}\\
&{}+ 666 t^{10} + 1016 t^{12} + 1473 t^{14} + 2058 t^{16} + 2790 t^{18}\\
&{}+ 3687 t^{20} + \mathcal O(t^{22})
\end{aligned}
\]

$U(6)$:
\[
\begin{aligned}
&11 + 57 t^{2} + 186 t^{4} + 429 t^{6} + 846 t^{8}\\
&{}+ 1479 t^{10} + 2424 t^{12} + 3702 t^{14} + 5454 t^{16} + 7735 t^{18}\\
&{}+ 10692 t^{20} + 14385 t^{22} + 19029 t^{24} + 24663 t^{26} + 31545 t^{28}\\
&{}+ 39753 t^{30} + 49566 t^{32} + 61074 t^{34} + 74623 t^{36} + 90273 t^{38}\\
&{}+ 108447 t^{40} + 129245 t^{42} + 153096 t^{44} + 180129 t^{46} + 210869 t^{48}\\
&{}+ 245391 t^{50} + 284313 t^{52} + 327760 t^{54} + 376344 t^{56} + 430248 t^{58}\\
&{}+ 490195 t^{60} + 556269 t^{62} + 629334 t^{64} + 709551 t^{66} + 797727 t^{68}\\
&{}+ 894105 t^{70} + 999654 t^{72} + 1114458 t^{74} + 1239654 t^{76} + 1375452 t^{78}\\
&{}+ 1522881 t^{80} + 1682253 t^{82} + 1854817 t^{84} + 2040651 t^{86} + 2241210 t^{88}\\
&{}+ 2456760 t^{90} + \mathcal O(t^{92})
\end{aligned}
\]

$U(9)$:
\[
\begin{aligned}
&30 + 201 t^{2} + 780 t^{4} + 2210 t^{6} + 5166 t^{8}\\
&{}+ 10638 t^{10} + 20004 t^{12} + 35070 t^{14} + 58209 t^{16} + 92478 t^{18}\\
&{}+ 141696 t^{20} + 210648 t^{22} + 305167 t^{24} + 432309 t^{26} + 600576 t^{28}\\
&{}+ 820057 t^{30} + 1102665 t^{32} + 1462452 t^{34} + 1915713 t^{36} + 2481276 t^{38}\\
&{}+ 3180987 t^{40} + 4039795 t^{42} + 5086113 t^{44} + 6352446 t^{46} + 7875474 t^{48}\\
&{}+ 9696453 t^{50} + 11862042 t^{52} + 14424364 t^{54} + 17441478 t^{56} + 20978460 t^{58}\\
&{}+ 25107409 t^{60} + 29907966 t^{62} + 35468667 t^{64} + 41886938 t^{66} + 49269654 t^{68}\\
&{}+ 57734808 t^{70} + 67411514 t^{72} + 78440514 t^{74} + 90976224 t^{76} + 105186751 t^{78}\\
&{}+ 121254453 t^{80} + 139378338 t^{82} + 159774069 t^{84} + 182674530 t^{86} + 208332627 t^{88}\\
&{}+ 237021346 t^{90} + \mathcal O(t^{92})
\end{aligned}
\]

We also numerically check that the explicit vacuum-sector BMN indices computed from \eqref{bmn index} reproduce the results of Subsection~\ref{num1} upon summation, hence providing a sanity check of the proposal of using the graviton index to compute the BMN index. 

\subsection{Absolute value sum of each vacuum BMN index}

To quantify cancellations between vacua, we list the sums of the absolute values of the sector coefficients at each power of $t$ in this subsection. 

$U(2)$:
\[
\begin{aligned}
&2 + 6 t^{2} + 12 t^{4} + 13 t^{6} + 18 t^{8}\\
&{}+ 21 t^{10} + 25 t^{12} + 33 t^{14} + 33 t^{16} + 35 t^{18}\\
&{}+ 45 t^{20} + \mathcal O(t^{22})
\end{aligned}
\]

$U(3)$:
\[
\begin{aligned}
&3 + 12 t^{2} + 27 t^{4} + 45 t^{6} + 63 t^{8}\\
&{}+ 87 t^{10} + 115 t^{12} + 147 t^{14} + 174 t^{16} + 213 t^{18}\\
&{}+ 267 t^{20} + \mathcal O(t^{22})
\end{aligned}
\]

$U(4)$:
\[
\begin{aligned}
&5 + 21 t^{2} + 57 t^{4} + 105 t^{6} + 177 t^{8}\\
&{}+ 261 t^{10} + 376 t^{12} + 513 t^{14} + 672 t^{16} + 889 t^{18}\\
&{}+ 1104 t^{20} + \mathcal O(t^{22})
\end{aligned}
\]

$U(5)$:
\[
\begin{aligned}
&7 + 36 t^{2} + 105 t^{4} + 224 t^{6} + 405 t^{8}\\
&{}+ 666 t^{10} + 1016 t^{12} + 1479 t^{14} + 2058 t^{16} + 2790 t^{18}\\
&{}+ 3705 t^{20} + \mathcal O(t^{22})
\end{aligned}
\]

$U(6)$:
\[
\begin{aligned}
&11 + 57 t^{2} + 186 t^{4} + 429 t^{6} + 846 t^{8}\\
&{}+ 1479 t^{10} + 2424 t^{12} + 3708 t^{14} + 5454 t^{16} + 7751 t^{18}\\
&{}+ 10710 t^{20} + 14391 t^{22} + 19211 t^{24} + 24927 t^{26} + 31575 t^{28}\\
&{}+ 42313 t^{30} + 51864 t^{32} + 61980 t^{34} + 82917 t^{36} + 102933 t^{38}\\
&{}+ 130179 t^{40} + 130327 t^{42} + 231030 t^{44} + 243243 t^{46} + 336035 t^{48}\\
&{}+ 246141 t^{50} + 485421 t^{52} + 717920 t^{54} + 1208064 t^{56} + 471450 t^{58}\\
&{}+ 1987547 t^{60} + 1532097 t^{62} + 4822632 t^{64} + 3575635 t^{66} + 7823289 t^{68}\\
&{}+ 7869063 t^{70} + 15178952 t^{72} + 19446558 t^{74} + 27173094 t^{76} + 38867862 t^{78}\\
&{}+ 47511093 t^{80} + 82811727 t^{82} + 71427463 t^{84} + 149628531 t^{86} + 151855116 t^{88}\\
&{}+ 227929136 t^{90} + \mathcal O(t^{92})
\end{aligned}
\]

$U(9)$:
\[
\begin{aligned}
&30 + 201 t^{2} + 780 t^{4} + 2210 t^{6} + 5166 t^{8}\\
&{}+ 10638 t^{10} + 20004 t^{12} + 35076 t^{14} + 58209 t^{16} + 92478 t^{18}\\
&{}+ 141714 t^{20} + 210660 t^{22} + 305323 t^{24} + 432321 t^{26} + 601230 t^{28}\\
&{}+ 822275 t^{30} + 1103571 t^{32} + 1483758 t^{34} + 1936507 t^{36} + 2524602 t^{38}\\
&{}+ 3432861 t^{40} + 4304923 t^{42} + 5314533 t^{44} + 7532520 t^{46} + 10478570 t^{48}\\
&{}+ 13641849 t^{50} + 15346788 t^{52} + 31460380 t^{54} + 37046742 t^{56} + 51653106 t^{58}\\
&{}+ 85405251 t^{60} + 76259478 t^{62} + 236216625 t^{64} + 191396006 t^{66} + 462733146 t^{68}\\
&{}+ 424324974 t^{70} + 926448210 t^{72} + 1178659638 t^{74} + 1927745664 t^{76} + 2841888437 t^{78}\\
&{}+ 3348573423 t^{80} + 7541535372 t^{82} + 6084457929 t^{84} + 18078235224 t^{86} + 16495562931 t^{88}\\
&{}+ 34436411990 t^{90} + \mathcal O(t^{92})
\end{aligned}
\]

\subsection{Graviton partition function}

In this subsection, we list the numerically computed graviton partition function in the large $k$ limit, extracting fixed KK momentum contribution. 

Using the Bose--Fermi prescription in Section~\ref{sec:multi_graviton_entropy}, we expand the unsigned BPS graviton partition function after dividing out the neutral factor.
The series below correspond to fixed momentum $m$, with $a=b=c=s=t^2$ and $Q$ equal to the exponent of $s$. The results shown against label $U(m)$ refer to the graviton partition function contribution with KK momentum $m$.

$U(2)$:
\[
\begin{aligned}
&2 + 6 t^{2} + 24 t^{4} + 67 t^{6} + 180 t^{8}\\
&{}+ 435 t^{10} + 983 t^{12} + 2079 t^{14} + 4161 t^{16} + 7913 t^{18}\\
&{}+ 14397 t^{20} + \mathcal O(t^{22})
\end{aligned}
\]

$U(3)$:
\[
\begin{aligned}
&3 + 12 t^{2} + 51 t^{4} + 171 t^{6} + 513 t^{8}\\
&{}+ 1413 t^{10} + 3635 t^{12} + 8805 t^{14} + 20304 t^{16} + 44797 t^{18}\\
&{}+ 94971 t^{20} + \mathcal O(t^{22})
\end{aligned}
\]

$U(4)$:
\[
\begin{aligned}
&5 + 21 t^{2} + 99 t^{4} + 357 t^{6} + 1185 t^{8}\\
&{}+ 3567 t^{10} + 10040 t^{12} + 26607 t^{14} + 67080 t^{16} + 161809 t^{18}\\
&{}+ 375456 t^{20} + \mathcal O(t^{22})
\end{aligned}
\]

$U(5)$:
\[
\begin{aligned}
&7 + 36 t^{2} + 177 t^{4} + 692 t^{6} + 2439 t^{8}\\
&{}+ 7848 t^{10} + 23544 t^{12} + 66585 t^{14} + 179136 t^{16} + 461288 t^{18}\\
&{}+ 1142817 t^{20} + \mathcal O(t^{22})
\end{aligned}
\]

$U(6)$:
\[
\begin{aligned}
&11 + 57 t^{2} + 300 t^{4} + 1239 t^{6} + 4626 t^{8}\\
&{}+ 15705 t^{10} + 49702 t^{12} + 148032 t^{14} + 419220 t^{16} + 1135935 t^{18}\\
&{}+ 2960850 t^{20} + 7454589 t^{22} + 18190615 t^{24} + 43141209 t^{26} + 99673587 t^{28}\\
&{}+ 224788433 t^{30} + 495699216 t^{32} + 1070432988 t^{34} + 2266543563 t^{36} + 4711203645 t^{38}\\
&{}+ 9622931433 t^{40} + \mathcal O(t^{42})
\end{aligned}
\]

$U(9)$:
\[
\begin{aligned}
&30 + 201 t^{2} + 1182 t^{4} + 5576 t^{6} + 23418 t^{8}\\
&{}+ 89376 t^{10} + 316790 t^{12} + 1055490 t^{14} + 3338067 t^{16} + 10089998 t^{18}\\
&{}+ 29311170 t^{20} + 82188642 t^{22} + 223238825 t^{24} + 589091391 t^{26} + 1513994712 t^{28}\\
&{}+ 3797583655 t^{30} + 9313584183 t^{32} + 22368379230 t^{34} + 52681378483 t^{36} + 121817872896 t^{38}\\
&{}+ 276862079295 t^{40} + \mathcal O(t^{42})
\end{aligned}
\]

\bibliography{biblio}

\end{document}

%% file: oco_triality.tikz
% Preamble requirements: \usepackage{tikz} and \usetikzlibrary{arrows.meta}
\begingroup
\definecolor{Vblue}{RGB}{35,66,255}
\definecolor{Fgreen}{RGB}{0,148,79}
\definecolor{Gred}{RGB}{255,55,19}
\begin{tikzpicture}[
  x=1cm, y=1cm,
  font=\large,
  line cap=round,
  line join=round,
  vbrane/.style={
    draw=Vblue,
    fill=Vblue!5!white,
    fill opacity=1,
    line width=0.75pt
  },
  fbrane/.style={draw=Fgreen, line width=1.2pt},
  insertion/.style={draw=Fgreen, line width=0.75pt},
  geometry/.style={draw=Gred, line width=0.75pt},
  map/.style={
    draw=black,
    line cap=butt,
    line width=2.5pt,
    -{Triangle[length=3.8mm,width=3.8mm]}
  },
  brane label/.style={font=\normalsize},
  % Draw the rear brane first and the front brane last.
  % Each opaque face hides the portions of the branes behind it.
  pics/vstack/.style={code={
    \foreach \i in {0,...,5} {
      \begin{scope}[shift={({0.095*\i},{-0.038*\i})}]
        \path[vbrane]
          (0,0) -- (1.24,0.70) -- (1.32,2.54)
          -- (0.02,1.81) -- cycle;
      \end{scope}
    }
  }},
  % The two red curves representing the closed-string geometry.
  pics/throat/.style={code={
    \draw[geometry]
      (0,2.55) .. controls (0.40,1.85) and (1.13,1.63)
               .. (1.86,1.50);
    \draw[geometry]
      (0,0) .. controls (0.40,0.70) and (1.13,0.92)
            .. (1.86,1.05);
  }}
]

% ------------------------------------------------------------------
% Central configuration: V-type and F-type branes.
% ------------------------------------------------------------------
\pic at (-1.50,3.64) {vstack};
\draw[fbrane] (1.03,3.60) -- (1.03,6.00);
\draw[fbrane] (1.16,3.60) -- (1.16,6.00);

\node[brane label] at (-1.18,3.12) {V-type branes};
\node[brane label] at ( 1.36,3.12) {F-type branes};

% The three arrows.
\draw[map] ( 0.33,6.28) -- ( 0.33,7.35);
\draw[map] (-2.49,3.01) -- (-3.55,2.28);
\draw[map] ( 2.70,3.08) -- ( 3.77,2.33);

% ------------------------------------------------------------------
% Top: closed-string description.
% ------------------------------------------------------------------
\pic at (-0.60,7.18) {throat};
\draw[insertion]
  (-0.28,8.33) .. controls (-0.10,8.66) and (0.09,8.66)
              .. (0.26,8.33);
\node at (0.13,10.12) {Closed string};

% ------------------------------------------------------------------
% Bottom left: V-type open-string description.
% ------------------------------------------------------------------
\begin{scope}[shift={(-5.87,0.44)}]
  \pic {vstack};
  % Draw the green excitation after the stack, on its front face.
  \draw[insertion]
    (0.86,0.75) .. controls (1.04,1.08) and (1.23,1.08)
                .. (1.40,0.75);
\end{scope}
\node at (-5.00,-0.02) {V-type dual};

% ------------------------------------------------------------------
% Bottom right: F-type open-string description.
% ------------------------------------------------------------------
\pic at (4.53,0.31) {throat};
\draw[fbrane]
  (4.49,2.40) .. controls (5.04,1.86) and (5.09,1.33)
              .. (4.54,0.80);
\draw[fbrane]
  (4.62,2.40) .. controls (5.17,1.86) and (5.22,1.33)
              .. (4.67,0.80);
\node at (4.82,-0.02) {F-type dual};

\end{tikzpicture}
\endgroup